\pdfoutput=1
\documentclass[reprint,amsmath,amssymb,prb]{revtex4-2}

\usepackage{graphicx}
\usepackage{dcolumn}
\usepackage{bm}
\usepackage{braket}
\usepackage{siunitx}
\usepackage{xcolor}

\DeclareSIUnit\angstrom{\text {Å}}
\DeclareSIUnit\rydberg{\text {Ry}}
\DeclareSIUnit\bohr{\text {Bohr}}

\newcommand*{\ch}{\ensuremath{\hat{c}^{\phantom{\dagger}}}}
\newcommand*{\cc}{\ensuremath{\hat{c}^{\dagger}}}
\newcommand*{\cb}{\ensuremath{\hat{c}^{\phantom{\dagger}}}}
\newcommand*{\cbc}{\ensuremath{\hat{c}^{\dagger}}}
\newcommand*{\ct}{\ensuremath{\hat{c}^{\phantom{\dagger}}}}
\newcommand*{\ctc}{\ensuremath{\hat{c}^{\dagger}}}
\newcommand*{\ah}{\ensuremath{\hat{a}^{\phantom{\dagger}}}}
\newcommand*{\ac}{\ensuremath{\hat{a}^{\dagger}}}
\newcommand*{\ab}{\ensuremath{\hat{a}^{\phantom{\dagger}}}}
\newcommand*{\abc}{\ensuremath{\hat{a}^{\dagger}}}
\newcommand*{\at}{\ensuremath{\hat{a}^{\phantom{\dagger}}}}
\newcommand*{\atc}{\ensuremath{\hat{a}^{\dagger}}}

\newcommand*{\kb}{\ensuremath{\mathbf{k}}}
\newcommand*{\qb}{\ensuremath{\mathbf{q}}}
\newcommand*{\Rb}{\ensuremath{\mathbf{R}}}
\newcommand*{\Gb}{\ensuremath{\mathbf{G}}}
\newcommand*{\Db}{\ensuremath{\mathbf{D}}}
\newcommand*{\Dbb}{\ensuremath{\mathbf{\bar{D}}}}
\newcommand*{\hb}{\ensuremath{\hbar}}
\newcommand*{\ka}{\ensuremath{\kappa}}
\newcommand*{\al}{\ensuremath{\alpha}}

\DeclareMathOperator{\Tr}{Tr}

\usepackage[hidelinks]{hyperref}

\begin{document}

\preprint{PRB/123-QED}

\title{Fully coupled electron-phonon transport in two-dimensional-material-based devices using efficient FFT-based self-energy calculations}

\author{Rutger Duflou}
\email{rutger.duflou@imec.be}
\author{Gautam Gaddemane}
\author{Michel Houssa}
\author{Aryan Afzalian}
\affiliation{
imec, Kapeldreef 75, Leuven, 3001,Belgium}
\affiliation{Semiconductor Physics Laboratory, KU Leuven, Celestijnenlaan 200 D, Leuven, 3001, Belgium}

\date{\today}

\begin{abstract}
Self-heating effects can significantly degrade the performance in nanoscale devices. We investigate self-heating effects in such devices based on two-dimensional materials using ab-initio techniques. A new algorithm was developed to allow for efficient self-energy computations, achieving a $\sim$500 times speedup. It is found that for the simple case of free-standing MoS$_2$ without explicit metal leads, the self-heating effects do not result in significant performance degradation.
\end{abstract}

\maketitle


\section{\label{sec:introduction}Introduction}
The last two decades, two-dimensional (2D) materials have risen in interest as candidates for next-generation devices. They are predicted to show excellent electrostatic control, reducing short-channel effects, and to suffer less from device variation \cite{Schwierz2015,Logoteta2014,Chhowalla2016,Kang2014}.
Ab-initio methods, such as Density Functional Theory (DFT) and the Non-Equilibrium Green's Function (NEGF) formalism \cite{Afzalian2020}, have been helpful tools in the research on 2D materials \cite{Schwierz2015}.
It has been shown that, to capture the correct behavior for 2D-material-based devices with NEGF, it is of great importance to incorporate electron-phonon interactions \cite{Afzalian2020,Duflou2021}.
Additionally, it has been shown for silicon nanowires that neglecting self-heating effects can alter the behavior of coupled electron-phonon systems \cite{Rhyner2014}. We therefore aim to investigate the influence of self-heating on electron-phonon scattering in 2D-material-based devices.
Our DFT-NEGF quantum transport solver, ATOMOS, allows for the simulation of devices with electron-phonon scattering by phonons at an equilibrium temperature \cite{Afzalian2020,Afzalian2021} and for the simulation of ballistic phonon transport \cite{Duflou2023SISPAD}.
In this work, we extend ATOMOS to allow for the simulation of fully coupled electron-phonon transport for 2D materials. Additionally, we provide a new algorithm based on the Fast-Fourier-Transform (FFT) technique to greatly reduce the computational cost of the self-energy calculation required for such a coupled electron-phonon transport.
In Section \ref{sec:theory}, we discuss the theoretical foundations of our NEGF implementation of the electron and phonon Green's function.
In Section \ref{sec:methods}, we describe the methods used to perform a device simulation and elaborate on the FFT-based self-energy calculation. In Section \ref{sec:model_testing}, we discuss the errors introduced by the approximations within this FFT-based self-energy computation and provide estimates for the gain in computational efficiency.
Finally, in Section \ref{sec:self_consist_device}, we show the results of the simulation of a fully coupled 2D-material-based device with self-heating.

\section{\label{sec:theory}Theory}

\subsection{\label{secsec:Hamiltonian}The DFT Hamiltonian}

It can be shown that within the DFT formalism, a material is described by the following Hamiltonian \cite{Giustino2017},

\begin{align}
\begin{split}
    \hat{H} =& \sum_{n\kb} e^{\,}_{\kb n}\cc_{\kb n}\ch_{\kb n} + \sum_{\nu \qb}\hb\omega^{\,}_{\qb\nu}(\ac_{\qb\nu}\ah_{\qb\nu}+\frac{1}{2}) +\\ &N_p^{-\frac{1}{2}}\sum_{\substack{\kb \qb\\mn\nu}} g_{mn\nu}(\kb,\qb)\cc_{\kb+\qb m}\ch_{\kb n}(\ah_{\qb\nu}+\ac_{-\qb\nu}).
\end{split}
\label{eq:hamiltonian_rec}
\end{align}

The description is in reciprocal space, with electronic band energies $e_{\kb n}$, phonon energies $\hb\omega_{\qb\nu}$ and electron-phonon interaction parameters $g_{mn\nu}(\kb,\qb)$. Here, $n$ ($\nu$) denotes the band index (phonon mode) and $\kb$ ($\qb$) the k-point in the Brillouin zone for the electrons (phonons). $\cc_{\kb n}$ and $\ch_{\kb n}$ ($\ac_{\qb\nu}$ and $\ah_{\qb\nu}$) are the corresponding electron (phonon) creation and annihilation operators. However, these operators create and annihilate particles in reciprocal space. Device simulations typically require a real space description, which can be achieved by transforming the reciprocal space operators to real space, using a Wannier transformation \cite{Marzari2012},

\begin{equation}
    \ctc_{\Rb_e m} = \frac{1}{\sqrt{N_e}}\sum_{n\kb}e^{-i\kb\cdot\Rb_e}U^{\,}_{nm,\kb}\cc_{\kb n},
    \label{eq:el_rec2rel_cre}
\end{equation}

where $m$ denotes a Wannier function index, $\Rb_e$ denotes the primitive cell lattice point, $N_e$ is the number of k-points and $U_{nm,\kb}$ is a matrix built to maximize the real space localization of the electron. A similar real space transformation can be achieved for the localization of phonons \cite{Giustino2007},

\begin{equation}
    \atc_{\Rb_p\ka\al} = \frac{1}{\sqrt{N_p}}\sum_{\nu\qb}e^{-i\qb\cdot\Rb_p}e_{\ka\al\nu,\qb}^*\ac_{\qb \nu},
    \label{eq:ph_rec2rel_cre}
\end{equation}

where $e_{\ka\al\nu,\qb}$ is the eigenvector of the dynamical matrix and $\kappa$, $\alpha$ and $\Rb_p$ denote an atom index, its polarization direction and its primitive cell lattice point, respectively. Note that there are minor differences compared to the conventions in Ref. \cite{Giustino2007}. 

\begin{figure}[ht]
    \centering
    \includegraphics[width=\linewidth]{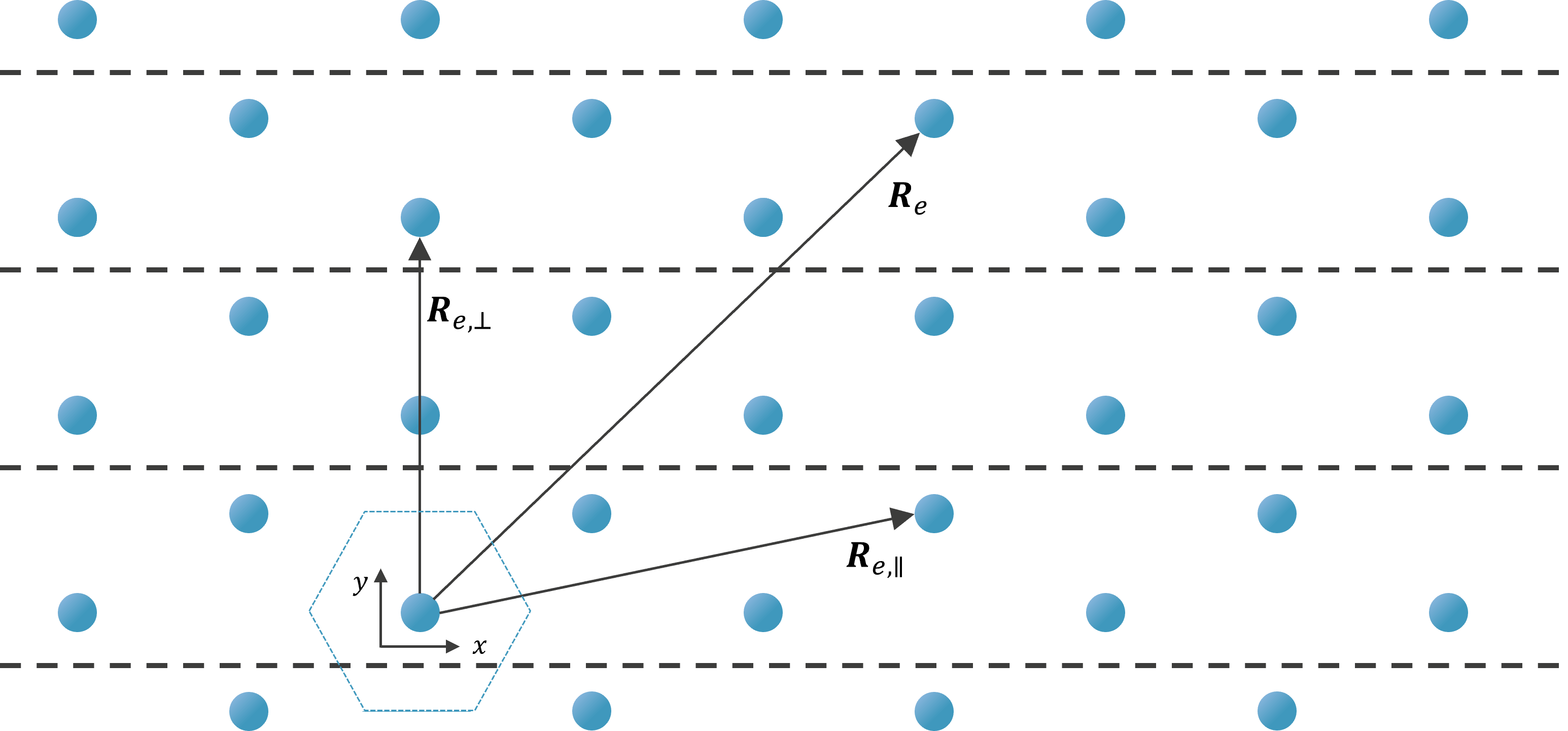}
    \caption{Subdivision of the real space lattice in a part along the transport direction and a periodic part orthogonal to the transport direction, which can be transformed to reciprocal space through a Fourier transformation.}
    \label{fig:mixed_space}
\end{figure}

In some cases, a mixed space description is beneficial. An exemplary case is a planar transistor, e.g., made from a 2D material. The transport direction requires a real space description to allow for the insertion of carriers at the source and their extraction at the drain. The out-of-plane direction, orthogonal to both the 2D material plane and the transport direction, is non-periodic and thus implies a real space description as well. The third direction, however, is typically very homogeneous and can thus be considered periodic. This periodicity allows for the subdivision of the system in a transport part and a periodic part which can be Fourier transformed to reciprocal space, as shown in Fig. \ref{fig:mixed_space}. 

This concept can be extended to other numbers of periodic directions, from 0 for nanowires to 2 for resistors or diodes. A more complete discussion of the transformations used is given in Appendix \ref{app:transformations}. The final result is the following mixed space Hamiltonian

\begin{align}
    \hat{H} =
    &\sum_{nn'\kb_{\perp}}
\bar{h}_{\substack{nn'\\\kb_{\perp}}}
 \cbc_{\kb_{\perp}n}\cb_{\kb_{\perp}n'} + \sum_{\nu\nu'\qb_{\perp}}
 \bar{d}_{\substack{\nu\nu'\\\qb_{\perp}}}
     \abc_{\qb_{\perp}\nu}\ab_{\qb_{\perp}\nu'}.
\label{eq:hamiltonian_mix}\\ 
 &+N_{\perp}^{-\frac{1}{2}}\sum_{\substack{n n'\nu\\\kb_{\perp}\qb_{\perp}}}
    \bar{g}_{\substack{nn'\nu\\\kb_{\perp}\qb_{\perp}}}
    \cbc_{\kb_{\perp}+\qb_{\perp}
    n}\cb_{\kb_{\perp}n'}(\ab_{\qb_{\perp}\nu}+\abc_{-\qb_{\perp}\nu}),\nonumber
\end{align}

where the indices $n$ and $\nu$ should not be confused with the band indices in \eqref{eq:hamiltonian_rec}, but represent a grouping of the indices $\left(m,\Rb_{e,\parallel}\right)$ and $\left(\kappa,\alpha,\Rb_{p,\parallel}\right)$ defined above and in Fig. \ref{fig:mixed_space}. $N_{\perp}$ is equal to the number of orthogonal k-points. The third term in \eqref{eq:hamiltonian_mix} represents the electron-phonon interactions and will be referred to as $\hat{H}_I$.

\subsection{\label{secsec:Greensfunctions}The NEGF formalism}

The NEGF formalism relies on the definition of an electron and phonon Green's function \cite{Fetterch3,Maciejko2007}

\begin{align}
    iG_{\substack{n,m\\\kb_{\perp}}}(t,t') &= \frac{1}{\hb}\big<T_c\big[\cb_{\kb_{\perp}n}(t)\cbc_{\kb_{\perp}m}(t')\big]\big>, \label{eq:el_cont_G}\\
    iD_{\substack{\nu,\mu\\\qb_{\perp}}}(t,t') &= \frac{1}{\hb}\big<T_c\big[(\ab_{\qb_{\perp}\nu}(t)+\abc_{-\qb_{\perp}\nu}(t))\nonumber\\
    &\qquad\qquad(\abc_{\qb_{\perp}\mu}(t')+\ab_{-\qb_{\perp}\mu}(t'))\big]\big>, \label{eq:ph_cont_D}
\end{align}

where the creation and annihilation operators are given in the Heisenberg picture and are ordered on a two-branch contour. The averaging over the states is determined by a non-equilibrium occupation \cite{Maciejko2007}. The indices $n$ and $m$ ($\nu$ and $\mu$) can be understood as row and column indices, defining the Green's functions as matrices, $\bm{G_{\kb_{\perp}}}(t,t')$ ($\bm{D_{\qb_{\perp}}}(t,t')$). Although we refer to $\bm{D_{\qb_{\perp}}}(t,t')$ as the phonon Green's function, according to its definition it is actually equal to the displacement-displacement correlation \cite{Giustino2017}, but we will forgo this point for the sake of brevity.

\begin{figure}[ht]
    \centering
    \includegraphics[width=\linewidth]{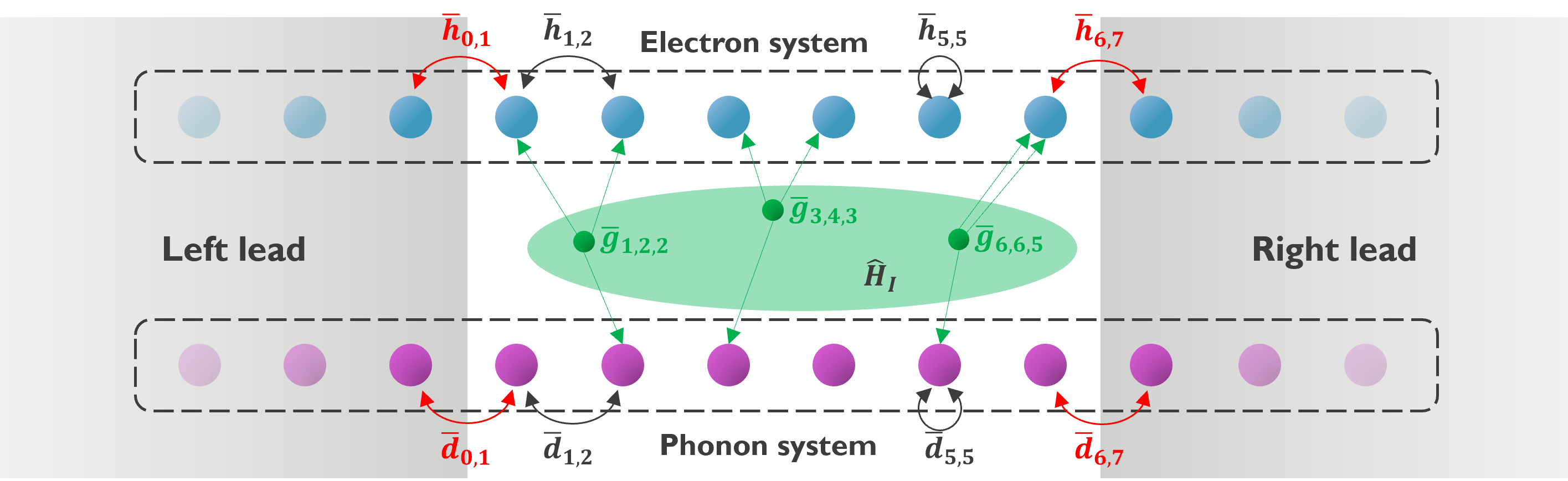}
    \caption{Schematic representation of the coupled electron-phonon system within a device with two semi-infinite leads. The $\kb_{\perp}$ and $\qb_{\perp}$ subscripts are left out for the sake of brevity. Degrees of freedom of the electron and phonon system are grouped such that the on-site energies and coupling elements form matrices. The matrices that form the perturbation terms coupling the device to the left and right leads are denoted in red. The electron-phonon interactions are denoted in green. It should be noted that the representation here is only qualitatively true for phonons. A rigorous treatment can be found in the Supplemental Material \cite{Supp}.}
    \label{fig:coupled_sys}
\end{figure}

The expressions in \eqref{eq:el_cont_G} and \eqref{eq:ph_cont_D} cannot be solved exactly due to the electron-phonon interactions in $\hat{H}_I$ and due to the unknown occupation of states for devices not in equilibrium. A solution can be found through a perturbation expansion, of which the theory is well established. Here, we follow the derivation in Ref. \cite{Maciejko2007}. The interacting system is subdivided into several non-interacting systems: the electrons and phonons in the device, a left lead and a right lead, as shown in Fig. \ref{fig:coupled_sys}. Additionally, each subsystem is divided into slabs such that every slab only interacts with its nearest neighbors \cite{Svizhenko2002}. Alternatively, one could interpret it as elements of \eqref{eq:hamiltonian_mix} being grouped into matrices such that the total device Hamiltonian for the electron system and phonon system form two block tridiagonal matrices, respectively. 

Each subsystem has a known one-particle occupation: the infinite leads are each characterized by Fermi-Dirac statistic functions for the electrons

\begin{equation}
    f_i(\omega) = \frac{1}{\exp{\frac{\hb\omega-E_{f_i}}{k_BT_i}}+1}\label{eq:Fermi-Dirac}
\end{equation}

and Bose-Einstein statistic functions for the phonons

\begin{equation}
    N_i(\omega) = \frac{1}{\exp{\frac{\hb\omega}{k_BT_i}}-1}\label{eq:Bose-Einstein}
\end{equation}

with temperatures, $T_1$ and $T_2$, and chemical potentials for the electrons, $E_{f_1}$ and $E_{f_2}$, for the left and right lead, respectively. The device itself can be kept empty before connection. The subsystems are connected by adiabatically switching on the interaction terms, $\hat{H}_I$, and the perturbation terms connecting the device to the leads. These perturbation terms can be grouped into matrices $\mathbf{U}_{\kb_{\perp}}$ ($\mathbf{V}_{\qb_{\perp}}$), related to $\{\mathbf{\bar{h}_{i,j}},\mathbf{\bar{h}_{j,i}}\}$ ($\{\mathbf{\bar{d}_{i,j}},\mathbf{\bar{d}_{j,i}}\}$) with $\mathbf{i}$, $\mathbf{j}$ equal to $\mathbf{0}$, $\mathbf{1}$ and $\mathbf{n}$, $\mathbf{n+1}$ \cite{Supp}. Note that bold indices are used to indicate slab indices instead of individual degrees of freedom.

The switching on, in combination with Wick's theorem, leads to the following Dyson equations \cite{Maciejko2007},

\begin{align}
    &\Gb_{\kb_{\perp}}(t,t') = \Gb^0_{\kb_{\perp}}(t,t')
    + \int_C dt_1\Gb^0_{\kb_{\perp}}(t,t_1)\mathbf{U}_{\kb_{\perp}}
    \Gb_{\kb_{\perp}}(t_1,t')\label{eq:exactG_as_convol_sigG}\nonumber\\
    &\qquad+ \int_C dt_1\int_C dt_2\Gb^0_{\kb_{\perp}}(t,t_1)\mathbf{\Sigma}^{s}_{\kb_{\perp}}
    (t_1,t_2)\Gb_{\kb_{\perp}}(t_2,t'),
    \\
    &\Db_{\qb_{\perp}}(t,t') = \Db^0_{\qb_{\perp}}(t,t')
    + \int_C dt_1\Db^0_{\qb_{\perp}}(t,t_1)\mathbf{V}_{\qb_{\perp}}
    \Db_{\qb_{\perp}}(t_1,t')\label{eq:exactD_as_convol_piD}\nonumber\\
    &\qquad+ \int_C dt_1\int_C dt_2\Db^0_{\qb_{\perp}}(t,t_1)\mathbf{\Pi}^{s}_{\qb_{\perp}}
    (t_1,t_2)\Db_{\qb_{\perp}}(t_2,t'),
\end{align}

where $\Gb^0_{\kb_{\perp}}$ and $\Db^0_{\qb_{\perp}}$ are the Green's function solutions for the non-interacting non-connected subsystems, $\mathbf{U}_{\kb_{\perp}}$ and $\mathbf{V}_{\qb_{\perp}}$ are defined as above, $\mathbf{\Sigma}^{s}_{\kb_{\perp}}$ and $\mathbf{\Pi}^{s}_{\qb_{\perp}}$ are the self-energies related to electron-phonon scattering and the integrals are integrals over the two-branch contours. The contour-ordered Green's function can be resolved into lesser, greater, retarded and advanced Green's functions by confinement of the time arguments to specific branches and the contour integrals in \eqref{eq:exactG_as_convol_sigG} and \eqref{eq:exactD_as_convol_piD} can be simplified to real axis integrals by using Langreth's theorem \cite{Maciejko2007,Lake2006}. Finally, Fourier transformation of the integral equations to the energy domain results in the following well-known expressions for the electron Green's function \cite{Lake2006,Dattasupp},

\begin{align}
    \Gb^{R/A}_{\kb_{\perp}}(\omega) &=\left((\hb\omega\pm i\eta)\mathbf{I}-\mathbf{H}^{\,}_{\kb_{\perp}} - \mathbf{\Sigma}_{\kb_{\perp}}^{R/A}(\omega)\right)^{-1},
\label{eq:Gr_final}\\
    \Gb_{\kb_{\perp}}^{\lessgtr}(\omega) &= 
    \Gb_{\kb_{\perp}}^R(\omega)\mathbf{\Sigma}_{\kb_{\perp}}^{\lessgtr}(\omega)\Gb_{\kb_{\perp}}^A(\omega),
\label{eq:Glsgt_final}
\end{align}

with $\Gb^{R}_{\kb_{\perp}}$, $\Gb^{A}_{\kb_{\perp}}$, $\Gb^{<}_{\kb_{\perp}}$ and $\Gb^{>}_{\kb_{\perp}}$ the retarded, advanced, lesser and greater electron Green's function of the device, respectively, and $\mathbf{\Sigma}_{\kb_{\perp}}^{R/A}$ and $\mathbf{\Sigma}_{\kb_{\perp}}^{\lessgtr}$ their corresponding self-energies. $\mathbf{H}_{\kb_{\perp}}$ is the device Hamiltonian with

\begin{equation}
\left(\mathbf{H}_{\kb_{\perp}}\right)_{i,j} = \bar{h}_{\substack{ij\\\kb_{\perp}}}.
\end{equation}

Note that the matrices defined here only contain degrees of freedom within the device and not the degrees of freedom in the leads as before. The influence of the leads is introduced by the self-energies. The self-energies thus contain contributions from both the leads and the electron-phonon interactions

\begin{equation}
\mathbf{\Sigma} = \mathbf{\Sigma}^{l} + \mathbf{\Sigma}^{s}
\label{eq:self_energy_merge}
\end{equation}

where we dropped the $\omega$ and $\kb_{\perp}$ dependency in the notation for the sake of brevity and where

\begin{align}
\mathbf{\Sigma}^{l,R/A}_{\mathbf{1,1}}&= \mathbf{\bar{h}}^{\,}_{\mathbf{1,0}}\,
\Gb^{0,R/A}_{\mathbf{0,0}}\,
\mathbf{\bar{h}}^{\,}_{\mathbf{0,1}},\\
\mathbf{\Sigma}^{l,R/A}_{\mathbf{n,n}}&= \mathbf{\bar{h}}^{\,}_{\mathbf{n,n+1}}\,
\Gb^{0,R/A}_{\mathbf{n+1,n+1}}\,
\mathbf{\bar{h}}^{\,}_{\mathbf{n+1,n}},
\end{align}

\begin{align}
\mathbf{\Sigma}^{l,<} &=  if_1\mathbf{\Gamma}_1^{l} +if_2\mathbf{\Gamma}_2^{l},\label{eq:Sigma_leads1}\\
\mathbf{\Sigma}^{l,>} &=  -i(1-f_1)\mathbf{\Gamma}^{l}_{1}-i(1-f_2)\mathbf{\Gamma}^{l}_{2},\label{eq:Sigma_leads2}
\end{align}

with

\begin{align}
\left(\mathbf{\Gamma}^{l}_1\right)_{\mathbf{1,1}} &= i\left(\mathbf{\Sigma}^{l,R}_{\mathbf{1,1}}-\mathbf{\Sigma}^{l,A}_{\mathbf{1,1}}\right),\label{eq:Gamma_1}\\
\left(\mathbf{\Gamma}^{l}_2\right)_{\mathbf{n,n}} &= i\left(\mathbf{\Sigma}^{l,R}_{\mathbf{n,n}}-\mathbf{\Sigma}^{l,A}_{\mathbf{n,n}}\right).\label{eq:Gamma_2}
\end{align}

The retarded and advanced Green's function in the non-interacting non-connected leads, $\Gb^{0,R/A}$, are readily computed using the Sancho-Rubio algorithm \cite{Sancho1985}.

Similar expressions can be found for the phonon Green's function \cite{Mingo2003,Rhyner2014},

\begin{align}
    \Db^{R/A}_{\qb_{\perp}}(\omega) 
    &= 
    \left((\hb^2\omega^2\pm i\eta)\mathbf{I}-\mathbf{K}_{\qb_{\perp}} -\mathbf{\Pi}^{R/A}_{\qb_{\perp}}(\omega)\right)^{-1},
\label{eq:Dr_final}\\
    \Db^{\lessgtr}_{\qb_{\perp}}(\omega) 
    &= 
    \Db^R_{\qb_{\perp}}(\omega)
    \mathbf{\Pi}^{\lessgtr}_{\qb_{\perp}}(\omega)
    \Db^A_{\qb_{\perp}}(\omega),
\label{eq:Dlsgt_final}
\end{align}

with $\mathbf{K}_{\qb_{\perp}}$ the Fourier transform of $\mathbf{\Phi}$, the rescaled interatomic force constants matrix,

\begin{equation}
\mathbf{\Phi}_{i,j} = \frac{\hb^2}{\sqrt{m_im_j}}\frac{\partial^2U}{\partial \tau_{i}\partial \tau_{j}}.
\label{eq:Phi_matrix}
\end{equation}

Here, $U$ denotes the internal energy, the index $i$ indicates a degree of freedom in the real space phonon system, i.e., an atom with mass $m_i$ with a polarization direction along which it is displaced over a distance $\tau_i$.

Similarly to the electron system, the effects of the leads and electron-phonon interactions are introduced through the self-energy

\begin{equation}
\mathbf{\Pi} = \mathbf{\Pi}^{l} + \mathbf{\Pi}^{s}
\label{eq:self_energy_merge2}
\end{equation}

with

\begin{align}
\mathbf{\Pi}^{l,R/A}_{\mathbf{1,1}}&= \mathbf{\bar{k}}^{\,}_{\mathbf{1,0}}\,
\Db^{0,R/A}_{\mathbf{0,0}}\,
\mathbf{\bar{k}}^{\,}_{\mathbf{0,1}},\label{eq:phonon_RA_lead_self_energy_1}\\
\mathbf{\Pi}^{l,R/A}_{\mathbf{n,n}}&= \mathbf{\bar{k}}^{\,}_{\mathbf{n,n+1}}\,
\Db^{0,R/A}_{\mathbf{n+1,n+1}}\,
\mathbf{\bar{k}}^{\,}_{\mathbf{n,n+1}},\label{eq:phonon_RA_lead_self_energy_2}
\end{align}

and

\begin{align}
\mathbf{\Pi}^{l,<} &=  -iN_1\mathbf{\Delta}_1^{l} -iN_2\mathbf{\Delta}_2^{l},\\
\mathbf{\Pi}^{l,>} &=  -i(N_1+1)\mathbf{\Delta}^{l}_{1}-i(N_2+1)\mathbf{\Delta}^{l}_{2},
\end{align}

with

\begin{align}
\left(\mathbf{\Delta}^{l}_1\right)_{\mathbf{1,1}} &= i\left(\mathbf{\Pi}^{l,R}_{\mathbf{1,1}}-\mathbf{\Pi}^{l,A}_{\mathbf{1,1}}\right),\\
\left(\mathbf{\Delta}^{l}_2\right)_{\mathbf{n,n}} &= i\left(\mathbf{\Pi}^{l,R}_{\mathbf{n,n}}-\mathbf{\Pi}^{l,A}_{\mathbf{n,n}}\right).\label{eq:Delta_2}
\end{align}

The similarity between \eqref{eq:Gr_final}-\eqref{eq:Gamma_2} and \eqref{eq:Dr_final}-\eqref{eq:Delta_2} readily allows for the adaptation of electronic transport codes to phonon transport, as was done in Ref. \cite{Duflou2023SISPAD}. However, the derivation of these expressions for the phonon system in the literature typically relies on different conventions and does not apply the same principles as used for the electron system \cite{Mingo2003,Mingo2009}. This discrepancy complicates linking the electron and phonon Green's function for the self-energy calculation. Additionally, expressions are usually obtained for full real space \cite{Rhyner2014,Mingo2003,Mingo2009} or reciprocal space \cite{Giustino2017}, neglecting mixed space, which is useful for devices. We therefore provide a derivation of the Green's function expressions provided above and their corresponding self-energies in Appendix \ref{app:Greens_functions} and Appendix \ref{app:self_energies}, respectively. The final results for the lesser and greater self-energies due to electron-phonon scattering are

\begin{widetext}
\begin{align}
\mathbf{\Sigma}^{s,\lessgtr}_{\kb_{\perp}}(\omega)
&=
  \int^{+\infty}_{0}
  \frac{2i\hb}{N_{\perp}}
  \sum_{\nu\mu\qb_{\perp}}
  \mathbf{M}^{\nu}_{\kb_{\perp}-\qb_{\perp},\qb_{\perp}}
  \bigg(
  \Gb_{\kb_{\perp}-\qb_{\perp}}^{\lessgtr}(\omega-\omega')
  D_{\substack{\nu,\mu\\\qb_{\perp}}}^\lessgtr(\omega')\label{eq:final_Sigma_expression}\\
 &\qquad\qquad\qquad\qquad\qquad\qquad\qquad\qquad+
  \Gb_{\kb_{\perp}-\qb_{\perp}}^{\lessgtr}(\omega+\omega')
  D_{\substack{\mu,\nu\\-\qb_{\perp}}}^\gtrless(\omega')
  \bigg)
  \mathbf{M}^{\mu}_{\kb_{\perp},-\qb_{\perp}}\frac{d\omega'}{2\pi},\nonumber\\
\left(\mathbf{\Pi}^{s,\lessgtr}_{\qb_{\perp}}(\omega)\right)_{\nu,\mu} &=\int^{+\infty}_{-\infty}
-\frac{2n_si\hb}{N_{\perp}}
  \sum_{\kb_{\perp}}
    \Tr\bigg(
    \mathbf{M}^{\nu}_{\kb_{\perp},-\qb_{\perp}}
    \Gb_{\kb_{\perp}}^{\lessgtr}(\omega')
    \mathbf{M}^{\mu}_{\kb_{\perp}-\qb_{\perp},\qb_{\perp}}
    \Gb_{\substack{\kb_{\perp}-\qb_{\perp}}}^{\gtrless}(\omega'-\omega)\bigg)\frac{d\omega'}{2\pi},\label{eq:final_Pi_expression}
\end{align}
\end{widetext}

where $n_s$ is a spin degeneracy factor,

\begin{equation} \bigg(\mathbf{M}^{\nu}_{\kb_{\perp},\qb_{\perp}}\bigg)_{n,n'}
= \bar{\bar{g}}_{\substack{nn'\nu\\\kb_{\perp}\qb_{\perp}}}
\label{eq:def_M_as_g}
\end{equation}

is a rescaled version of the matrix element in \eqref{eq:hamiltonian_mix}, as detailed in Appendix \ref{app:self_energies}, and $\Tr()$ denotes the trace.

The retarded and advanced self-energies are calculated from the following relation \cite{Lake2006},

\begin{align}
    \mathbf{\Sigma}^{s,R/A}_{\kb_{\perp}}(\omega)
    = \mathcal{P}\int_{-\infty}^{+\infty} \frac{\mathbf{\Gamma}^{s}_{\kb_{\perp}}(\omega')}{\omega-\omega'}\frac{d\omega'}{2\pi} 
    \mp \frac{i}{2}\mathbf{\Gamma}^{s}_{\kb_{\perp}}(\omega),
    \label{eq:sigma_freq_phoneq_ret_full}\\
    \mathbf{\Pi}^{s,R/A}_{\qb_{\perp}}(\omega)
    = \mathcal{P}\int_{-\infty}^{+\infty} \frac{\mathbf{\Delta}^{s}_{\qb_{\perp}}(\omega')}{\omega-\omega'}\frac{d\omega'}{2\pi} 
    \mp \frac{i}{2}\mathbf{\Delta}^{s}_{\qb_{\perp}}(\omega),
\label{eq:pi_freq_phoneq_ret_full}
\end{align}

with

\begin{align}
    \mathbf{\Gamma}^{s}_{\kb_{\perp}}(\omega) = i\left(\mathbf{\Sigma}^{s,>}_{\kb_{\perp}}(\omega)-\mathbf{\Sigma}^{s,<}_{\kb_{\perp}}(\omega)\right),\\
    \mathbf{\Delta}^{s}_{\qb_{\perp}}(\omega) = i\left(\mathbf{\Pi}^{s,>}_{\qb_{\perp}}(\omega)-\mathbf{\Pi}^{s,<}_{\qb_{\perp}}(\omega)\right),
\end{align}

although the first terms in \eqref{eq:sigma_freq_phoneq_ret_full} and \eqref{eq:pi_freq_phoneq_ret_full} are usually left out as the principal value integral is difficult to compute and merely results in an energy renormalization \cite{Lake2006}. 

With the lesser and greater Green's functions available, we can compute the electron density \cite{Afzalian2021}, the current and the electron and phonon heat current \cite{Rhyner2014} as

\begin{equation}
n_k=-\frac{i n_s \hb}{N_{\perp}} \sum_{\kb_{\perp}}\int_{-\infty}^{+\infty}G^<_{\substack{k,k\\\kb_{\perp}}}\frac{d\omega}{2\pi} \label{eq:carrier_conc},
\end{equation}

\begin{equation}
I_{i \rightarrow j} = -\frac{qn_s}{N_{\perp}}\sum_{\kb_{\perp}}
\int_{-\infty}^{+\infty}(\bar{h}^{\,}_{\substack{ij\\\kb_{\perp}}}G^<_{\substack{j,i\\\kb_{\perp}}}-\bar{h}^{\,}_{\substack{ji\\\kb_{\perp}}}G^<_{\substack{i,j\\\kb_{\perp}}})
\frac{d\omega}{2\pi}, \label{eq:electron_current}
\end{equation}

\begin{align}
J_{el,i \rightarrow j} &= \frac{n_s}{N_{\perp}}
\sum_{\kb_{\perp}}\int_{-\infty}^{+\infty}\hb\omega(\bar{h}^{\,}_{\substack{ij\\\kb_{\perp}}}G^<_{\substack{j,i\\\kb_{\perp}}}-\bar{h}^{\,}_{\substack{ji\\\kb_{\perp}}}G^<_{\substack{i,j\\\kb_{\perp}}})
\frac{d\omega}{2\pi}, \label{eq:electron_heat_current}\\
J_{ph,i \rightarrow j} &= \frac{1}{N_{\perp}}
\sum_{\qb_{\perp}}\int_{0}^{+\infty}\hbar\omega(\bar{k}^{\,}_{\substack{ij\\\qb_{\perp}}}D^<_{\substack{j,i\\\qb_{\perp}}}-\bar{k}^{\,}_{\substack{ji\\\qb_{\perp}}}D^<_{\substack{i,j\\\qb_{\perp}}})
\frac{d\omega}{2\pi}.\label{eq:phonon_heat_current}
\end{align}

\section{\label{sec:methods}Methods}
\subsection{\label{secsec:matpam}Material parameter extraction}
All simulations in this work use matrix elements extracted for MoS$_2$ in the 2H-phase. MoS$_2$ was selected for its experimental maturity and industrial relevance. The electronic band energies, phonon energies and electron-phonon matrix elements were extracted in reciprocal space using the QUANTUM ESPRESSO DFT code \cite{Giannozzi2009}. The structure was relaxed with an energy convergence criteria of \SI{1e-16}{\rydberg} between subsequent scf iteration steps and energy and force convergence criteria of \SI{5e-7}{\rydberg} and \SI{5e-6}{\rydberg\per\bohr}, respectively, between subsequent ionic optimization steps. The PBE exchange-correlation functional was used with ultrasoft pseudopotentials, an energy cutoff of \SI{70}{\rydberg} and a kmesh density of $16\times16\times1$. The latter two were set after a convergence test to see that this correspond to a relative energy variation of less than \SI{1e-6}{}, an absolute energy variation less than \SI{1} mRy/atom and a variation of the lattice constant of less than \SI{0.02}{\percent}. The obtained lattice constant is \SI{3.183}{\angstrom}, which differs from the experimental result of \SI{3.165}{\angstrom} \cite{Alhilli1972} due to the lack of Vander Waals corrections. Relaxation with the Grimme DFT-D3 Vander Waals correction \cite{Grimme2010} resulted in a lattice constant of \SI{3.166}{\angstrom}. However, this correction is incompatible with phonon calculations. A vacuum of \SI{15}{\angstrom} and out-of-plane screening were used to block interactions between different layers. Spin-orbit coupling was neglected in all simulations. The same kmesh was used for the phonon calculation with a convergence threshold of \SI{1e-17}{}. 

The reciprocal space parameters were converted to real space using the Wannier90 \cite{Marzari2012} and Perturbo \cite{Giustino2007} code. For the initial projections during the Wannierization process, 5 d-orbitals on Mo and 3 p-orbitals on S were used. Perturbo provides the Hamiltonian elements in the Wannier basis, the interatomic force constants and atomic masses, which are readily combined to form $\mathbf{\Phi}$ in \eqref{eq:Phi_matrix}, and the real space deformation potentials, all in the HDF5 format \cite{Giustino2007}. To retrieve the matrix elements of \eqref{eq:def_M_as_g} in real space, an additional transformation is required,

\begin{equation}
\tilde{\tilde{g}}_{\substack{mn\kappa\alpha\\\Rb_e,\Rb_p}} = \frac{\hb}{\sqrt{2m_{\kappa}}}g^{Perturbo}_{mn\kappa\alpha}(\Rb_e,\Rb_p).
\label{eq:extra_rescale_electron_phonon}
\end{equation}

The details are provided in Appendices \ref{app:transformations} and \ref{app:Greens_functions}.

\subsection{\label{secsec:device_simulation}Device simulation}
\begin{figure}[ht]
    \centering
    \includegraphics[width=\linewidth]{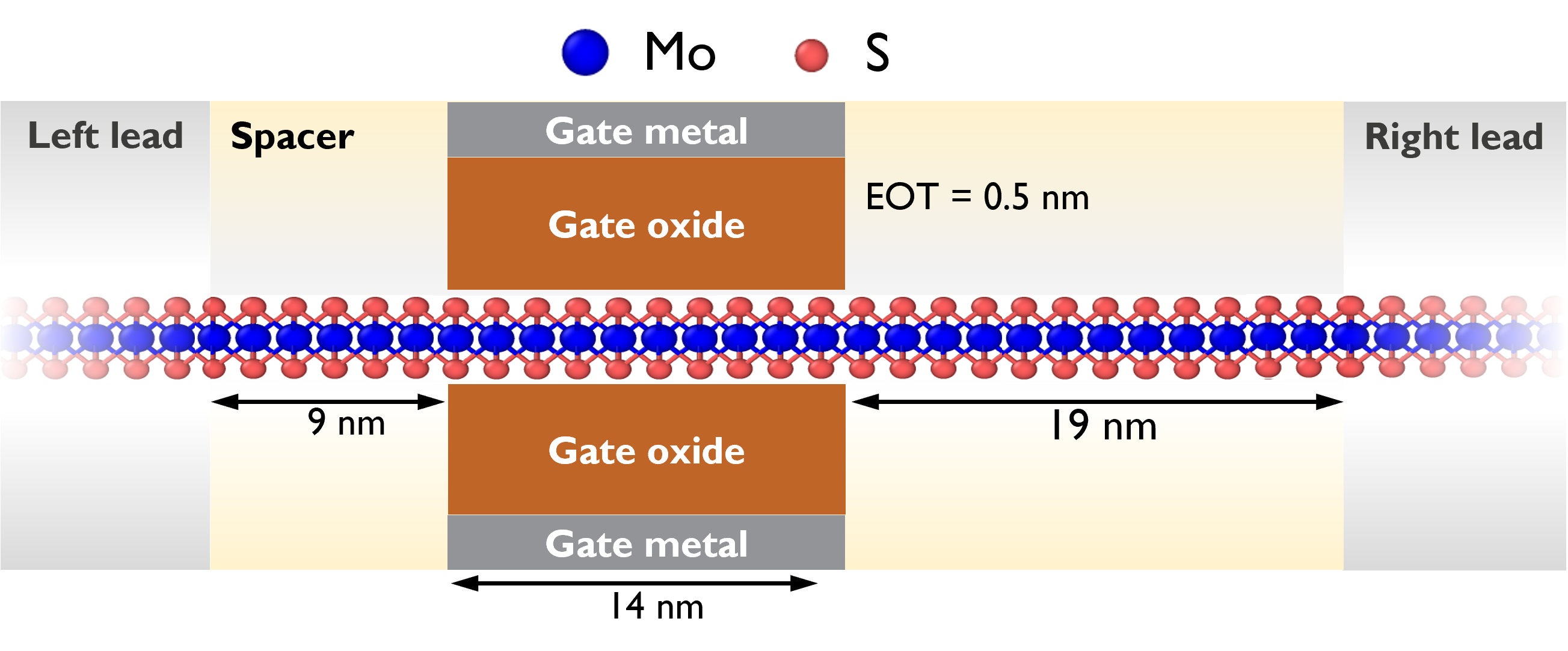}
    \caption{Schematic representation of a MoS$_2$ dual-gate transistor with its dimensions.}
    \label{fig:device}
\end{figure}

The matrix elements extracted in the previous section are grouped into device Hamiltonians for both the electrons and phonons. These Hamiltonians are then Fourier transformed to mixed space by the ATOMOS quantum transport solver. 10 k-points were used for half of the mixed space Brillouin zone. The other half can be considered equal due to symmetry. The device is a dual-gate transistor as depicted in Fig. \ref{fig:device}. The \SI{42}{\nano\meter} long MoS$_2$ sheet consists of source and drain extension regions, doped with a carrier concentration of \SI{1.8e13}{\per\centi\meter\squared}, and a \SI{14}{\nano\meter} intrinsic channel region between a top and bottom gate. The drain extension region is elongated compared to the source extension region to allow for thermalization of the carriers. Both gates have a corresponding gate oxide of \SI{2}{\nano\meter} and relative permittivity of \SI{15.6} for an effective oxide thickness of \SI{0.5}{\nano\meter}. The bias between source and drain was set to \SI{0.3}{\volt}. Unless specified otherwise, the source-gate potential was set to \SI{0.6}{\volt}, corresponding to on-state. 

\begin{figure}[ht!]
    \centering
    \includegraphics[width=0.9\linewidth]{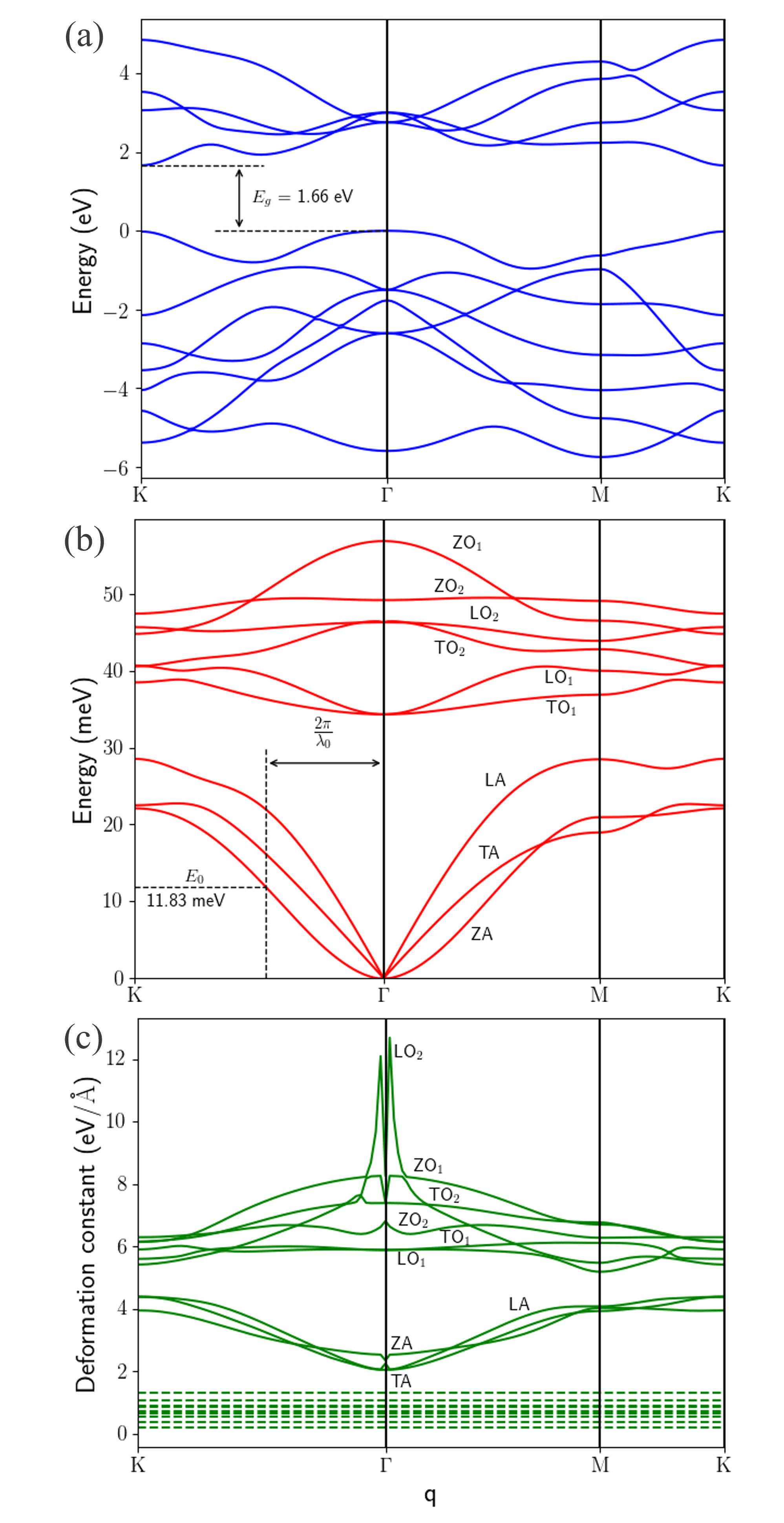}
    \caption{Wannier interpolated band structure (a) and phonon dispersion (b) of MoS$_2$ obtained by ATOMOS using the matrix elements provided by Perturbo. The band gap energy, $E_g$, threshold energy for acoustic phonon mode damping, $E_0$, and the different phonon branches are denoted. (c) The deformation constants for each phonon branch as a function of $\qb$ for $\kb$ at $\Gamma$ as provided by Perturbo (full lines) and after removing interaction parameters to reduce the computational complexity in ATOMOS (dashed lines).}
    \label{fig:bandstructures}
\end{figure}

To verify the correctness of the matrix elements, the band structure and phonon dispersion are shown in Fig. \ref{fig:bandstructures} (a) and (b), respectively. The deformation constants as a function of $\qb$ for $\kb$ at $\Gamma$ as provided by Perturbo are denoted with full lines for each phonon branch in Fig. \ref{fig:bandstructures} (c). Note that the acoustic branches show minima for $\qb$ at the $\Gamma$-point. Electronic transitions due to long wavelength acoustic phonons are prohibited by symmetry. Hence, one would expect the deformation constants to tend to zero as $D_{\nu\qb} \sim\omega_{\qb\nu}$ \cite{Fischetti2016}. A zero deformation potential is required to counter the singularity in the phonon Green's function due to both the displacement and the Bose-Einstein distribution function tending towards infinity at the $\Gamma$-point for acoustic phonons \cite{Fischetti2016}. Zero deformation constants at the $\Gamma$-point are, however, not achieved here as these rely on a cancellation of large terms which is an ill-conditioned problem.

The matter is made worse by further approximation. To reduce the significant computational cost of the self-energy calculations in \eqref{eq:final_Sigma_expression} and \eqref{eq:final_Pi_expression}, only on-site interactions are considered, i.e., within \eqref{eq:final_Sigma_expression} and \eqref{eq:final_Pi_expression}, $\mathbf{\Sigma}_{\kb_{\perp}}^{\lessgtr}$, $\mathbf{\Pi}_{\qb_{\perp}}^{\lessgtr}$, $\Gb_{\kb_{\perp}}^{\lessgtr}$, $\Db_{\qb_{\perp}}^{\lessgtr}$ and $\mathbf{M}_{\kb_{\perp},\qb_{\perp}}^{\nu}$ are assumed to be diagonal matrices and the diagonal entries of $\mathbf{M}_{\kb_{\perp},\qb_{\perp}}^{\nu}$ are only nonzero if the corresponding Wannier functions are located on the atom corresponding to $\nu$. An estimate of the influence due to this approximation is obtained by computing deformation constants with the same approximations, i.e., reciprocal space deformation constants are computed according to the principles in Ref. \cite{Giustino2007}, but matrix elements which would be neglected in our diagonal approach in \eqref{eq:final_Sigma_expression} and \eqref{eq:final_Pi_expression} are set to zero in the computation of the deformation constants as well. The resulting approximate deformation constants are denoted by the dashed curves in Fig. \ref{fig:bandstructures} (c). One can clearly see that neglecting the non-local interactions has a large influence on the deformation constants. First, the average deformation constant is significantly smaller due to neglecting non-local scattering processes. Second, the deformation constants do not show any dispersion. The neglect of non-local interactions thus also removes the minimum of the deformation constants of the acoustic modes at the $\Gamma$-point.

To compensate for the reduction of the average deformation constants, we introduce a scaling factor to increase the mixed-space matrix elements. To compensate for the non-zero deformation constants at the $\Gamma$-point, we introduce a damping of the phonon Green's function during the self-energy calculation for energies corresponding to acoustic modes. 

\begin{align}
    \mathbf{M}^{\nu}_{\kb_{\perp},\qb_{\perp}}&\leftarrow c\;\mathbf{M}^{\nu}_{\kb_{\perp},\qb_{\perp}}\\
    \Db_{\qb_{\perp}}^{\lessgtr} &\leftarrow \Db_{\qb_{\perp}}^{\lessgtr} \left(1-\left(1-\frac{E}{E_0}\right)^2\right) \;\text{ for }E < E_0
\end{align}

It was found that $c=7.409$ results in the same average deformation constants as for the case when all interactions are included. $E_0$ is calculated as the energy of the out-of-plane transverse acoustic mode \cite{Fischetti2016} at a wavelength of $\lambda_0 = \SI{1}{\nano\meter}$. This corresponds to a value of $E_0 = \SI{11.83}{\milli\electronvolt}$.

\subsection{FFT-based implementation of the self-energy calculation}
\subsubsection{Premise}
Despite the approximations made concerning the sparsity of the matrices in \eqref{eq:final_Sigma_expression} and \eqref{eq:final_Pi_expression}, the evaluation of these expressions still introduces a significant computational cost in the calculation. The energy integrals in Section \ref{secsec:Greensfunctions} are evaluated by evaluating the Green's function on an energy grid and can hence be computed in $\mathcal{O}(N_EN_{\perp})$ time, where $N_E$ denotes the number of energy points of the energy grid. Hence, \eqref{eq:final_Sigma_expression} and \eqref{eq:final_Pi_expression} also need to be evaluated $N_EN_{\perp}$ times, but a single evaluation itself scales as $\mathcal{O}(N_EN_{\perp})$ due to the fact that the self-energies depend on the Green's function at all energies and k-points. The total cost of the self-energy calculation thus scales as $\mathcal{O}(N_E^2N_{\perp}^2)$. Indeed, for the number of energy grid points and k-points required to converge to a sufficiently accurate result, a few hundreds and about 10, respectively, the computational cost of the self-energy calculation dwarfs the cost of the  Green's function evaluation. 

We propose an alternative to direct evaluation of \eqref{eq:final_Sigma_expression} and \eqref{eq:final_Pi_expression}. As $\mathbf{M}_{\kb_{\perp},\qb_{\perp}}^{\nu}$ is independent of $\omega$, \eqref{eq:final_Sigma_expression} and \eqref{eq:final_Pi_expression} can essentially be considered as a convolution of two energy dependent functions and the autocorrelation of an energy dependent function, respectively, which is expressed on a discrete energy grid as

\begin{align}
    \Sigma^{\lessgtr}[k] &= \sum_{l}
    G^{\lessgtr}[k-l]D^{\lessgtr}[l]+G^{\lessgtr}[k+l]D^{\gtrless}[l],\label{eq:Sigma_as_convolution}\\
    \Pi^{\lessgtr}[k] &= \sum_{l}
    G^{\lessgtr}[l]G^{\gtrless}[l-k],\label{eq:Pi_as_convolution}
\end{align}

where we dropped most of the subscripts for the sake of clarity and the multiplications are actually matrix multiplications involving the $\mathbf{M}_{\kb_{\perp},\qb_{\perp}}^{\nu}$ matrix. Convolutions of series with significant kernels can be evaluated efficiently by Fourier transforming both series, performing an element-wise multiplication and performing an inverse Fourier transform to obtain the final result. However, this technique requires the grids of both functions to be the same and equidistant. This is, in general, not the case. Shifts in energy are of no consequence as the integrals in \eqref{eq:final_Sigma_expression} and \eqref{eq:final_Pi_expression} do not directly depend on the energy, but the energy windows that the electron and phonon Green's functions are evaluated on are usually different in size. For instance, the energy window for the phonons of MoS$_2$ is usually no larger than \SI{65}{\milli\electronvolt}, but for the electrons the energy window can be 10 to 30 times larger depending on the potential in the device. Additionally, the Green's functions are characterized by Van Hove singularities, which require a dense energy grid to be evaluated accurately \cite{Chu2018}. ATOMOS applies an adaptive grid strategy to locally refine the energy grid near singularities \cite{Afzalian2011}. This allows the integrals in \eqref{eq:carrier_conc}-\eqref{eq:phonon_heat_current} to be evaluated efficiently without increasing the computational cost unnecessarily by also having a dense energy mesh where the Green's functions are smooth. This, however, implies that the energy meshes are usually not equidistant.

The obstacles of having non-equidistant grids over different energy windows could be resolved by refining the energy step size everywhere in the electron and phonon energy grid to its most refined part and by extending the smaller energy window, usually the phonon energy window, to the larger energy window size. This would, however, result in much higher computation times and memory requirements. The extra computation time related to evaluating additional energy points could be reduced by linearly interpolating the Green's functions between its evaluations on the nearest energy points. Indeed, interpolation is also used for direct evaluation of \eqref{eq:final_Sigma_expression} and \eqref{eq:final_Pi_expression} \cite{Afzalian2019}. The self-energy is calculated for every $\omega$ on the energy grid and uses every $\omega'$ on the energy grid. As the energy grids are not necessarily equal and equidistant, $\omega-\omega'$, $\omega+\omega'$ and $\omega'-\omega$ are not necessarily on the energy grid. However, as the energy grid is refined to capture all features of the Green's functions, it can be assumed that the Green's function can be obtained at these intermediate energies by interpolating the Green's function between its neighboring energy grid points. Likewise, extending the energy grid for the Green's function with the smaller energy window size is readily achieved by padding with zeros.

Interpolation and padding can thus reduce the cost of evaluating on a large dense grid. However, storing and Fourier transforming these interpolated and padded Green's functions still gives rise to significant increases of the memory requirements and computation times related to the Fourier transform and the evaluation of the Fourier transformed self-energies.

\subsubsection{Difference in energy window}
\begin{figure}[ht]
    \centering
    \includegraphics[width=\linewidth]{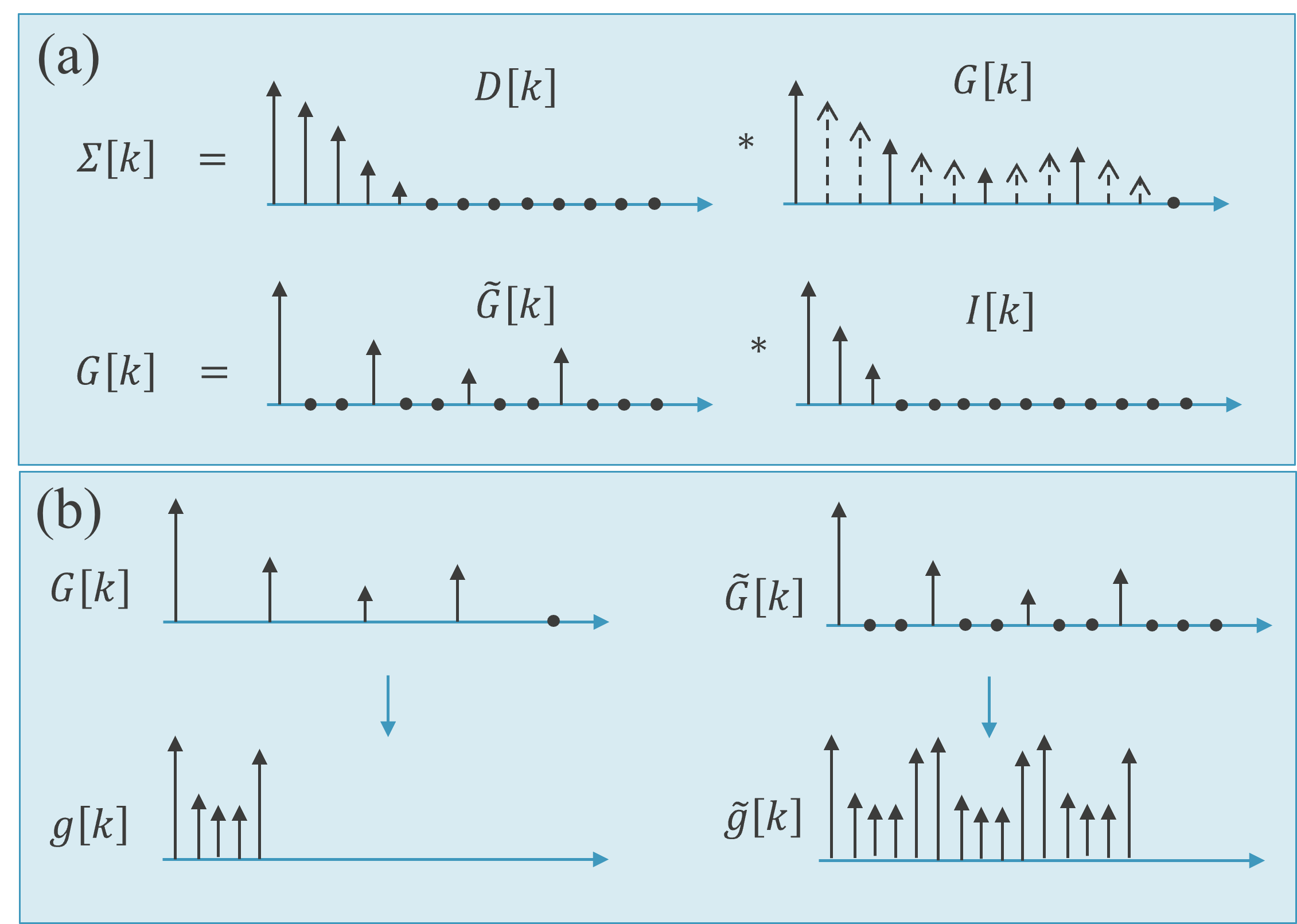}
    \caption{(a) A schematic showing how the first term in the electron self-energy is a convolution of the phonon Green's function and the interpolated Green's function and how the interpolated Green's function is itself a convolution of the original Green's function evaluations only, $\tilde{G}[k]$, and a function $I[k]$. (b) A schematic showing how $\tilde{g}[k]$, the Fourier transform of $\tilde{G}[k]$, is a repetition of the Fourier transform of $G[k]$ on the original electron energy mesh.}
    \label{fig:convolutions}
\end{figure}

The concept of interpolation can be used directly in an FFT-based evaluation of the self-energies. For this, we first assume that the electron and phonon energy meshes are equidistant and have the same number of energy points, but that their energy window sizes differ by a factor $m$. The energy grid step sizes thus also differ by a factor $m$. $m$ can be made an integer by increasing either energy window slightly if necessary. Let us focus on the first term in \eqref{eq:Sigma_as_convolution}, which, evaluated on a mesh with the step size of the phonon mesh and the energy window of the electron mesh, can be expressed as a convolution of the phonon Green's function padded with zeros, and the interpolated electron Green's function, as denoted in Fig. \ref{fig:convolutions}. Assuming the original meshes have $N_E$ mesh-points, then this refined mesh has $mN_E$ energy points. The interpolated electron Green's function can itself be seen as a convolution of the original Green's function evaluations with intermediate zeros for the interpolation points, which we'll refer to as $\tilde{G}[k]$, and the function $I[k]$, with

\begin{equation}
    I[k] = \left\{
    \begin{array}{ll}
        1-\frac{|k|}{m} & \qquad\text{if } |k| <= m \\
        0 & \qquad\text{else}
    \end{array}\right.
\end{equation}

assuming zero-based numbering.

It is readily shown how the Fourier transform of $\Sigma[k]$, is an element-wise multiplication of the Fourier transforms of $\tilde{G}[k]$, $D[k]$ and $I[k]$

\begin{align}
\begin{split}
\Sigma[k] =&
 \sum_{l,k}^{mN_E} \tilde{G}[k-l-h]D[l]I[h] \\
=&
 \sum_{l,k}^{mN_E} \sum_{k'}^{mN_E}
 \frac{1}{mN_E}
\tilde{g}[k']e^{-\frac{2\pi i(k-l-h)k'}{mN_E}}\\
 &\quad
 \sum_{l'}^{mN_E}\frac{1}{mN_E}
 d[l']e^{-\frac{2\pi ill'}{mN_E}}
\sum_{h'}^{mN_E}\frac{1}{mN_E} i[h']
e^{-\frac{2\pi ihh'}{mN_E}} \\
=& \frac{1}{mN_E}\sum_{k'}\tilde{g}[k']d[k']i[k']
e^{-\frac{2\pi ikk'}{mN_E}},
\end{split}
\end{align}

where we used the identity

\begin{equation}
    \sum_{l}^{mN_E}e^{-\frac{2\pi il(l'-k')}{mN_E}} = mN_E\delta_{l'k'}.
\end{equation}

As denoted in Fig. \ref{fig:convolutions}, $\tilde{g}[k]$ merely consists of repetitions of $g[k]$, the Fourier transform of $G[k]$ on the original electron energy mesh. Additionally, $\Sigma[k]$ is only needed on the original electron mesh, i.e., for $k$ a multiple of $m$. These considerations allow for a further simplification,

\begin{align}
\begin{split}
   \Sigma[mk]=& \frac{1}{mN_E}\sum_{k'}^{mN_E}\tilde{g}[k']d[k']i[k']
e^{\frac{-2\pi imkk'}{mN_E}} \\
=& \frac{1}{mN_E}\sum_{k'}^{N_E}\sum_{n}^{m}\tilde{g}[k'+nN_E]\\
&\qquad \quad
d[k'+nN_E]i[k'+nN_E]
e^{\frac{-2\pi ikk'}{N_E}} \\
=& \frac{1}{N_E}\sum_{k'}^{N_E}\tilde{g}[k']\tilde{d}[k']
e^{\frac{-2\pi ikk'}{N_E}},
\end{split}
\end{align}

with 

\begin{equation}
    \tilde{d}[k'] = \frac{1}{m}\sum_{n}^{m}d[k'+nN_E]i[k'+nN_E].\label{eq:D_fold}
\end{equation}

The second term of the electron self-energy can be computed in a similar way, resulting in

\begin{align}
\begin{split}
    \Sigma^{\lessgtr}[k] =& \frac{1}{N_E}\sum_{k'}^{N_E}\bigg(\tilde{g}^{\lessgtr}[k']\tilde{d}^{\lessgtr}[k']\\
    &\qquad\qquad\qquad+\tilde{g}^{\lessgtr}[k']\tilde{d}^{\gtrless}[-k']\bigg)
e^{\frac{-2\pi ikk'}{N_E}},
\label{eq:Sigma_as_convolution_2}
\end{split}
\end{align}

where $k$ is an index on the original electron energy mesh. The self-energy computation with the expensive matrix multiplications involving $\mathbf{M}_{\kb_{\perp},\qb_{\perp}}^{\nu}$ can thus be performed on a grid with size $N_E$ instead of $mN_E$ and the electron Green's function only needs to be Fourier transformed on its original energy mesh. The extra cost is an extra Fourier transformation for $I[k]$, but as this only needs to be done once and only involves a series of single values instead of Green's functions of the full device, this cost is usually negligible. The phonon Green's function still requires padding with zeros and a Fourier transform on a grid with size $mN_E$. However, the Fourier transformed phonon Green's function is not needed as is, but its entries need to be summed according to \eqref{eq:D_fold}. This Fourier transformation and summation can be done separately for every degree of freedom of the device. This implies that there is never a need to store the $mN_E$ full Green's functions at the same time, which severely reduces the memory footprint.

A similar approach is possible for the phonon self-energy. \eqref{eq:Pi_as_convolution} contains the electron Green's function twice. In the direct evaluation scheme, $G[l]$ corresponds to values on the original electron mesh and only $G[l-k]$ is interpolated,

\begin{equation}
    \Pi[k] = \sum_{l,n}^{mN_E}\tilde{G}[l]\tilde{G}[l-k-n]I[n].
\end{equation}

This, however, treats the two electron Green's functions asymmetrically. Alternatively, one can interpolate both electron Green's functions on the refined mesh,

\begin{equation}
    \Pi[k] = 
    \frac{1}{m}\sum_{l,h,n}^{mN_E}\tilde{G}[l-h]\tilde{G}[l-k-n]I[h]I[n].
\end{equation}

The resulting expressions are

\begin{equation}
    \Pi^{\lessgtr}[k] = 
    \frac{1}{mN_e}
    \sum_{k'}^{mN_E}\bigg(\tilde{g}^{\lessgtr}[k']\tilde{g}^{\gtrless}[-k']i[k']\bigg)
e^{\frac{-2\pi ikk'}{N_E}},\label{eq:Pi_as_convolution_2}
\end{equation}

and

\begin{equation}
    \Pi^{\lessgtr}[k] = 
    \frac{1}{m^2N_E} \sum_{k'}^{mN_E}\bigg(\tilde{g}^{\lessgtr}[k']\tilde{g}^{\gtrless}[-k']i[k']^2\bigg)
e^{\frac{-2\pi ikk'}{N_E}},\label{eq:Pi_as_convolution_3}
\end{equation}

respectively.

While \eqref{eq:Pi_as_convolution_2} and \eqref{eq:Pi_as_convolution_3} express the need for expensive matrix multiplications involving $\mathbf{M}_{\kb_{\perp},\qb_{\perp}}^{\nu}$ on a grid with size $mN_E$, the multiplication only needs to be performed $N_E$ times due to the periodicity of $\tilde{g}[k]$. The result can be reused multiple times, each time multiplied with a different value of $i[k]$. Similarly to the creation of $\tilde{d}[k]$, this can be done for each degree of freedom of the self-energy separately. An inverse Fourier transform on a grid with size $mN_E$ is still required, but the input and output of this Fourier transform do not ever need to be stored for the full device simultaneously, as the inverse Fourier transform can be performed separately for every degree of freedom, significantly reducing the memory that is required.

It is shown that for equidistant grids, an FFT-based calculation of the self-energies is possible, even when the electron and phonon energy mesh have significantly different ranges. Some additional considerations must be made, however. As a linear convolution is desired, both the electron and phonon Green's function must be padded to avoid a cyclic convolution. Additionally, during Fourier transformation of the energy mesh, Green's function entries of vastly different size are mixed, e.g., the electron Green's function in the band gap is mixed with the Green's function in the conduction band for an n-type transistor. After self-energy calculation and inverse Fourier transformation, the self-energy in the band gap will thus have a machine precision error relative to the self-energy in the conduction band, which can be many orders of magnitude higher than the self-energy in the band gap. We will refer to this error on the self-energy as the energy mixing error.

\subsubsection{Non-equidistant grids}

\begin{figure}
    \centering
    \includegraphics[width=\linewidth]{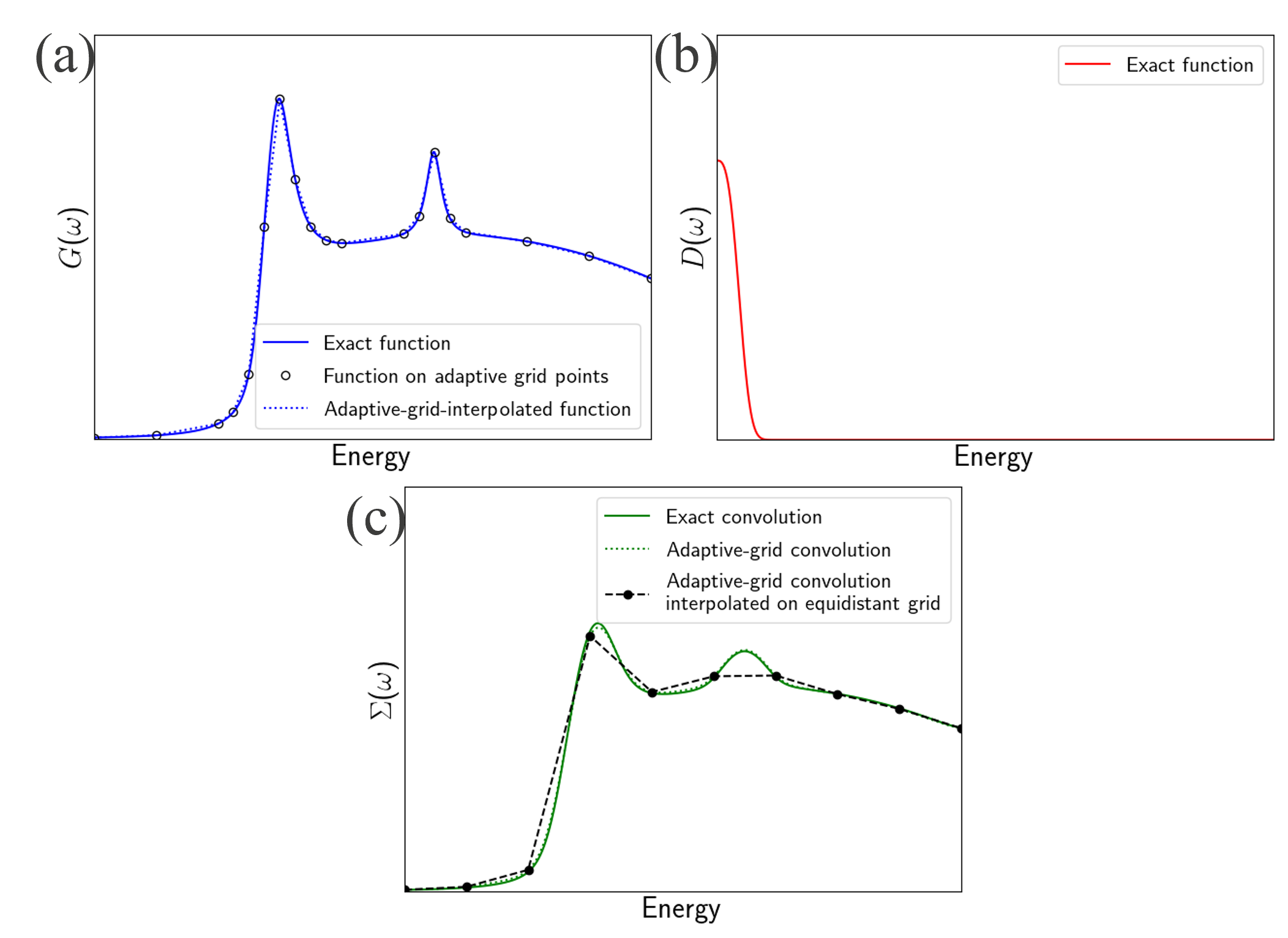}
    \caption{Calculation of the self-energy for a toy-problem set of Green's functions. (a) shows an exemplary electron Green's function, its adaptive grid evaluations and the function interpolating these adaptive grid evaluations. (b) shows the phonon Green's function and (c) shows the convolution according to \eqref{eq:Sigma_as_convolution}, both for the exact electron Green's function and the function interpolating the adaptive grid evaluations. Finally, it also shows the effect of interpolating the self-energy grid on an equidistant grid instead of direct evaluation on intermediate points.}
    \label{fig:equidistant_mesh_err1}
\end{figure}

The limitation concerning non-equidistant energy grids is not resolved as readily. Fig. \ref{fig:equidistant_mesh_err1} shows an example of a non-equidistant energy grid being used to resolve two peaks in the electron Green's function. The function interpolating the adaptive grid evaluations is nearly superimposed on the function itself. The phonon Green's function has a significantly smaller energy window of relevance and is sampled on a dense grid, which is appended with zeros. The convolution of the two according to \eqref{eq:Sigma_as_convolution}, effectively results in a smoothing of the electron Green's function. The results for the exact electron Green's function and the function interpolating the adaptive grid points are nearly identical. However, Fig. \ref{fig:equidistant_mesh_err1} shows that even if a methodology based on an equidistant grid could reproduce these results, an error will still be introduced as the results are only obtained on this equidistant grid. For intermediate points, interpolation of the self-energy is required, which does not properly capture the features of the self-energy at all energies. We will refer to this error as the self-energy interpolation error.

\begin{figure}
    \centering
    \includegraphics[width=\linewidth]{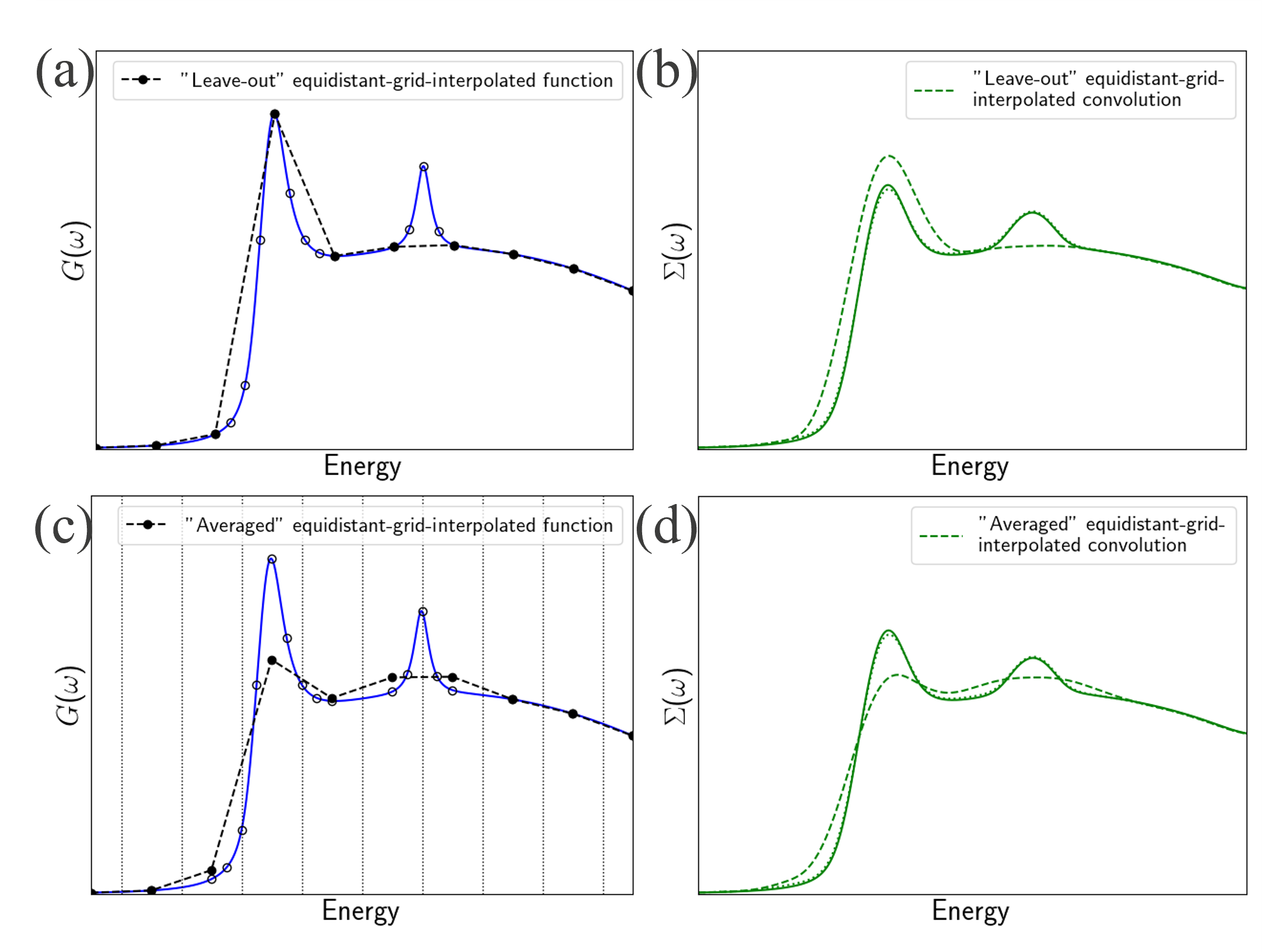}
    \caption{The effect on the self-energy of using an electron Green's function approximation interpolating an equidistant mesh. The equidistant mesh can be achieved by leaving out any intermediate adaptive grid points (a) or by averaging the nearest adaptive grid points (c). The effects on the self-energy are shown in (b) and (d), respectively.}
    \label{fig:equidistant_mesh_err2}
\end{figure}

Additionally, when an equidistant grid is used instead of an adaptive grid, the computation of the self-energy is usually susceptible to errors, even on the equidistant grid points itself. This is illustrated in Fig. \ref{fig:equidistant_mesh_err2}. Two strategies for reduction of the adaptive grid to an equidistant grid are shown. The first approach merely leaves out any grid refinements in the adaptive grid. The effect is that the total mass of certain peaks can be overestimated, underestimated, or even completely missed, with the corresponding effect on the self-energy. The second approach divides the energy window in sections, one for each equidistant grid point. The Green's function of each section is determined as the weighted average of the evaluations on the adaptive grid points in that section. This better preserves the total mass of features in the Green's function, and hence, the self-energy. However, the shapes of features in the energy profile of the self-energy are altered. We will refer to the error introduced by converting the Green's function evaluations on an adaptive grid to an equidistant grid, as the Green's function conversion error.

The influence of these three types of errors, the energy mixing error, the self-energy interpolation error and the Green's function conversion error, on macroscopic properties such as the current, heat current and charge, are the focus of Section \ref{secsec:error_testing}.

\section{\label{sec:model_testing}Model testing}
\subsection{\label{secsec:error_testing}Error estimate of FFT-based self-energy calculation}

In Section \ref{sec:methods}, we introduced definitions for different types of errors made by the approximations in the FFT-based self-energy calculation. They are summarized here as

\begin{itemize}
\item{The energy mixing error}, arising during Fourier transformation because the Green's function at all energies in the energy window are mixed. Some of these entries are orders of magnitude larger than others. Machine precision errors on a large-value Green's function, e.g., at the bottom of the conduction band, thus result in large relative errors on small-value Green's functions, e.g., in the middle of the band gap. The same is true for the self-energies during the inverse Fourier transformation. In the conventional implementation of \eqref{eq:final_Sigma_expression} and \eqref{eq:final_Pi_expression}, the Green's functions are only mixed with entries relatively close in energy due to the limited energy window of the phonon Green's function. In the FFT-based implementation, all energies are mixed, which can amplify this error. 
\item{The self-energy interpolation error}, arising because the self-energy computed by the FFT-based implementation is only provided on an equidistant grid. The adaptive grid also requires the self-energy on intermediate points, which are obtained by interpolation. 
\item{The Green's function conversion error}, arising because the Green's function evaluations on a non-equidistant adaptive grid need to be converted to an equidistant grid for Fourier transformation. Fig. \ref{fig:equidistant_mesh_err2} showed two ways of achieving this: either by leaving out any evaluations on adaptively added grid points or by averaging the adaptive grid evaluations.
\end{itemize}

These errors have to be compared with the integration error, introduced by performing the integral in \eqref{eq:carrier_conc}-\eqref{eq:phonon_heat_current} using a finite set of integration points. The integration error is thus not related to the FFT-based implementation of the self-energy calculation, but it determines how the adaptive grid is refined. ATOMOS starts from an initial equidistant grid, which is then further refined in order to reduce the error below a certain threshold. Here, we impose a relative error threshold of at most 1\% on certain macroscopic parameters such as the electron current, phonon heat current and electronic charge density. However, a minimal initial grid density is required for the adaptive grid refinement to work. We show in Appendix \ref{app:error_testing} that this minimal initial grid density corresponds to $\sim$100 equidistant initial grid points. We also show that a significantly denser equidistant grid is required if we desire the errors introduced by the FFT-based computation to be of the same magnitude as the imposed threshold.

The energy mixing error is several orders of magnitude smaller than the integration error for all energy grid densities. The self-energy interpolation error and the Green's function conversion error require an equidistant grid of $\sim$1000 and $\sim$500 grid points, respectively, to achieve a relative error of 1\% on the macroscopic electronic parameters. Additionally, the "averaged" approach should be used to convert the Green's functions on the non-equidistant grid to an equidistant grid. Finally, even for the rather dense initial equidistant grid of $\sim$1000 grid points, an error of several percent can be present on the phonon heat current. This error is due to the Green's function conversion error creating phonons at slightly shifted energies, which has a direct effect on the magnitude of the heat current. However, it should be noted that the shift in phonon energy is small compared to the electron energy grid distance. Additionally, it was verified that the number of phonons at 1000 initial equidistant grid points was not altered more than 1\% by the "averaged" approach. This implies that the effect on the electron scattering will be small, which is the main interest in our research on the self-heating.

\subsection{\label{secsec:timing_testing}Computation time reduction by FFT-based self-energy calculation}

\begin{figure}[ht]
    \centering
    \includegraphics[width=\linewidth]{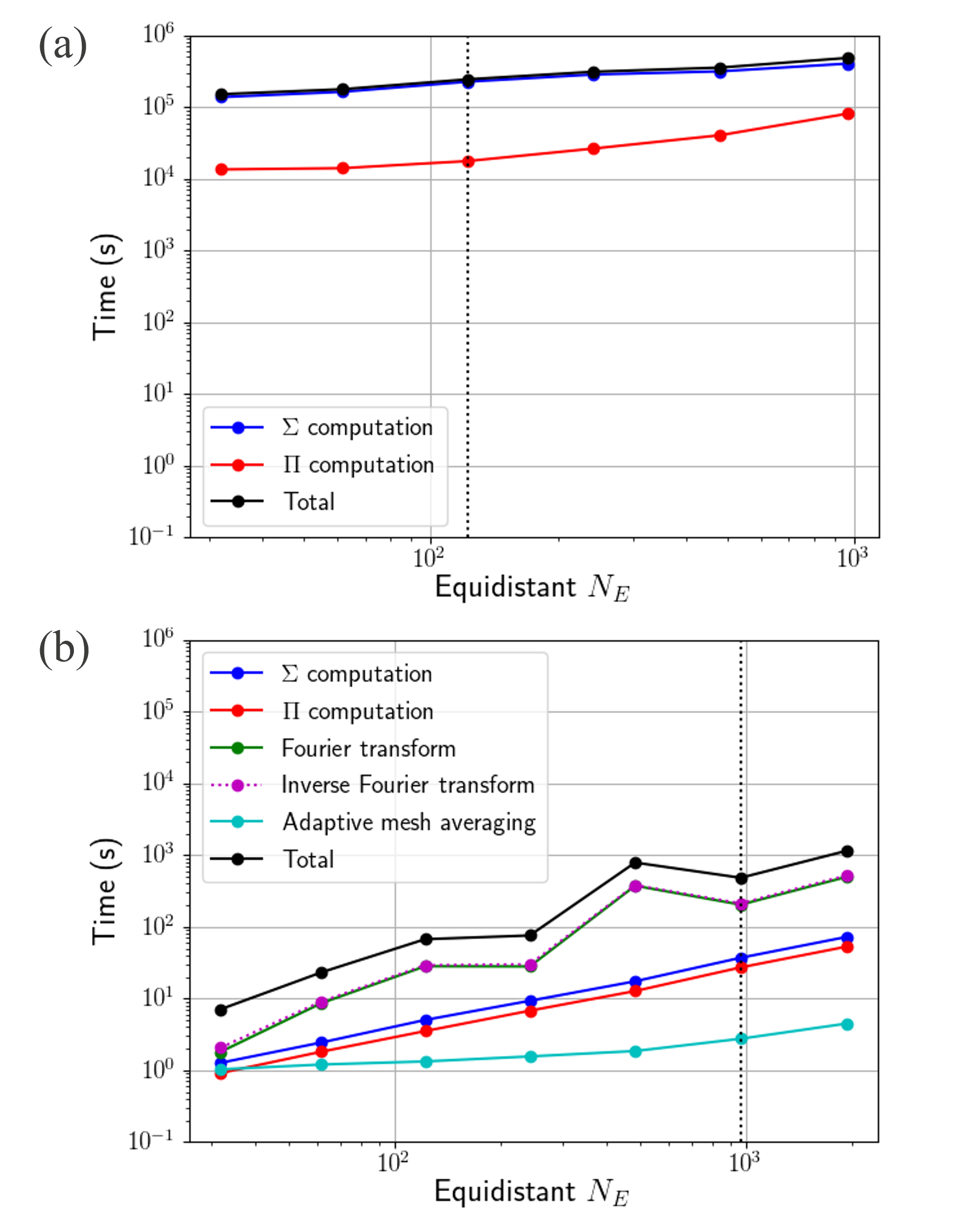}
    \caption{The computation time of calculating the self-energies for a single non-self-consistent iteration of the system in Section \ref{secsec:device_simulation} as a function of the number of initial equidistant grid energy points. (a) and (b) show the computation times for the conventional convolution-based implementation and the FFT-based implementation, respectively. The dotted lines indicate the required number of initial energy grid points for an estimated relative error of 1\% on the electronic properties.}
    \label{fig:timings}
\end{figure}

In Section \ref{secsec:error_testing}, it was shown that the self-energy can be computed using an FFT-based implementation at the cost of increasing the number of initial grid points, from $\sim$100 to $\sim$1000, and introducing a 1\% error on the electronic output parameters of the device and a few percent error on the heat current. Note that this tenfold increase of the number of initial grid points does not give rise to a tenfold increase of the Green's function computation time. As shown in Appendix \ref{app:error_testing}, the number of adaptively added points does not increase accordingly or even decreases. Timing tests on a single Intel Xeon Gold 6132 processor showed an increase of the computation time by only 60\% by this tenfold increase of the initial of the initial grid. The reduction in the computation time of the self-energy is, however, significant, as shown in Fig. \ref{fig:timings}. 

Fig. \ref{fig:timings} (a) shows the computation time to compute the self-energies using the conventional convolution-based implementation for a single non-self-consistent iteration on a single Intel Xeon Gold 6132 processor. Note that no quadratic dependency can be observed as the computation time depends on the total number of grid points, which is not linearly dependent on the number of initial grid points. The computation time is dominated by the electron self-energy as this depends on the phonon Green's function, which is characterized by more grid points than the electron Green's function. 

Fig. \ref{fig:timings} (b) shows the computation time for the self-energies using the FFT-based implementation. The computation of the Fourier transformed self-energies scales linearly with the number of equidistant initial grid points, as expected. The averaging approach to convert the adaptive grid to an equidistant grid scales slower than linearly as it scales with the number of adaptive grid points. The majority of the computation time is related to the Fourier transform and the inverse Fourier transform, which scales as $\mathcal{O}(N_E\log N_E)$. Note that, despite the higher required initial grid size of $\sim$1000 energy points, the total computation time is still significantly lower than for the conventional implementation with $\sim$100 initial grid points, differing by a factor of $\sim$500. 

\begin{figure}[ht]
    \centering
    \includegraphics[width=\linewidth]{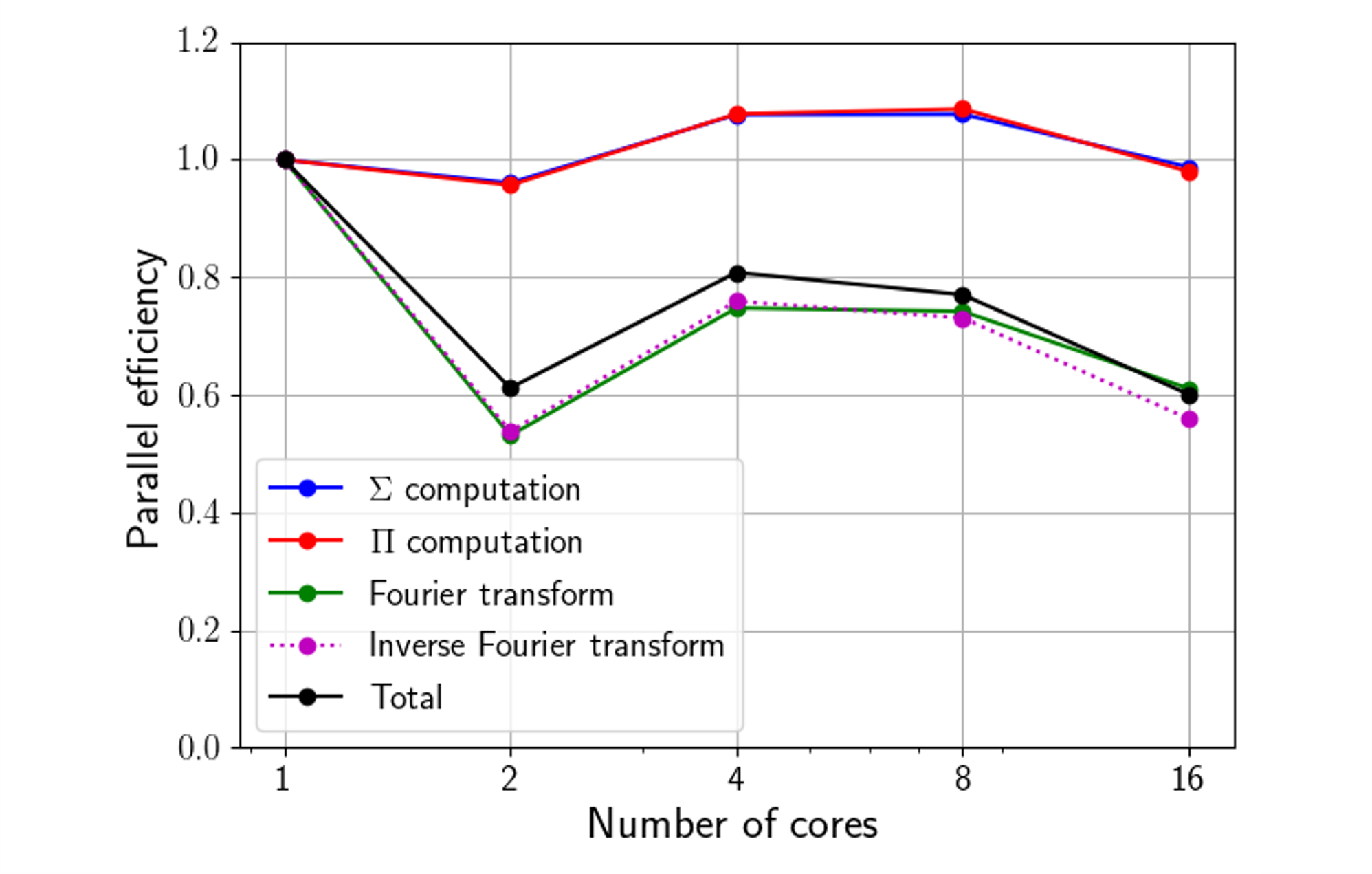}
    \caption{The parallel efficiency of calculating the self-energies as a function of the number of cores for a single non-self-consistent iteration of the system in Section \ref{secsec:device_simulation} with a parallelized FFT-based implementation with 1000 equidistant grid points.}
    \label{fig:timings_parallel}
\end{figure}

With the conventional implementation, the self-energy computation takes 98\% of the total computation time for a single Born iteration, when performed on a single processor. With the FFT-based implementation, this reduces to 9\%. Additionally, the Fourier transform computation time scales linearly with the number of k-points and the computation can be done in parallel for every degree of freedom of the system, i.e., for the Green's function on every orbital or atom. The FFT-based implementation thus readily allows for an increase of the number of k-points or parallelization of the self-energy calculation. Also the computation of the Fourier transformed self-energies can be parallelized more easily than the conventional convolution-based self-energy calculation. As can be seen from \eqref{eq:Sigma_as_convolution_2} and \eqref{eq:Pi_as_convolution_2}, the Fourier transformed self-energy at grid point $k'$ only depends on the Fourier transformed Green's functions on grid points $k'$ and $-k'$, in contrast to the conventional convolution-based computation, which requires the Green's functions on every grid point. Parallelization of the FFT-based self-energy computation thus requires at most a doubling of the memory requirements, whereas for the conventional convolution-based implementation, the memory requirements scale with the number of cores. Fig. \ref{fig:timings_parallel} shows the parallel efficiency of our parallel implementation for a single non-self-consistent self-energy calculation as a function of the number of cores for the FFT-based implementation, where parallel efficiency is defined as

\begin{equation}
\textrm{Efficiency}(N) = \frac{\textrm{Time on 1 core}}{N \cdot\textrm{Time on $N$ cores}}.
\end{equation}

\subsection{\label{secsec:double_G_vs_single_G}Choice of phonon self-energy interpolation scheme}

\begin{figure}[ht]
    \centering
    \includegraphics[width=\linewidth]{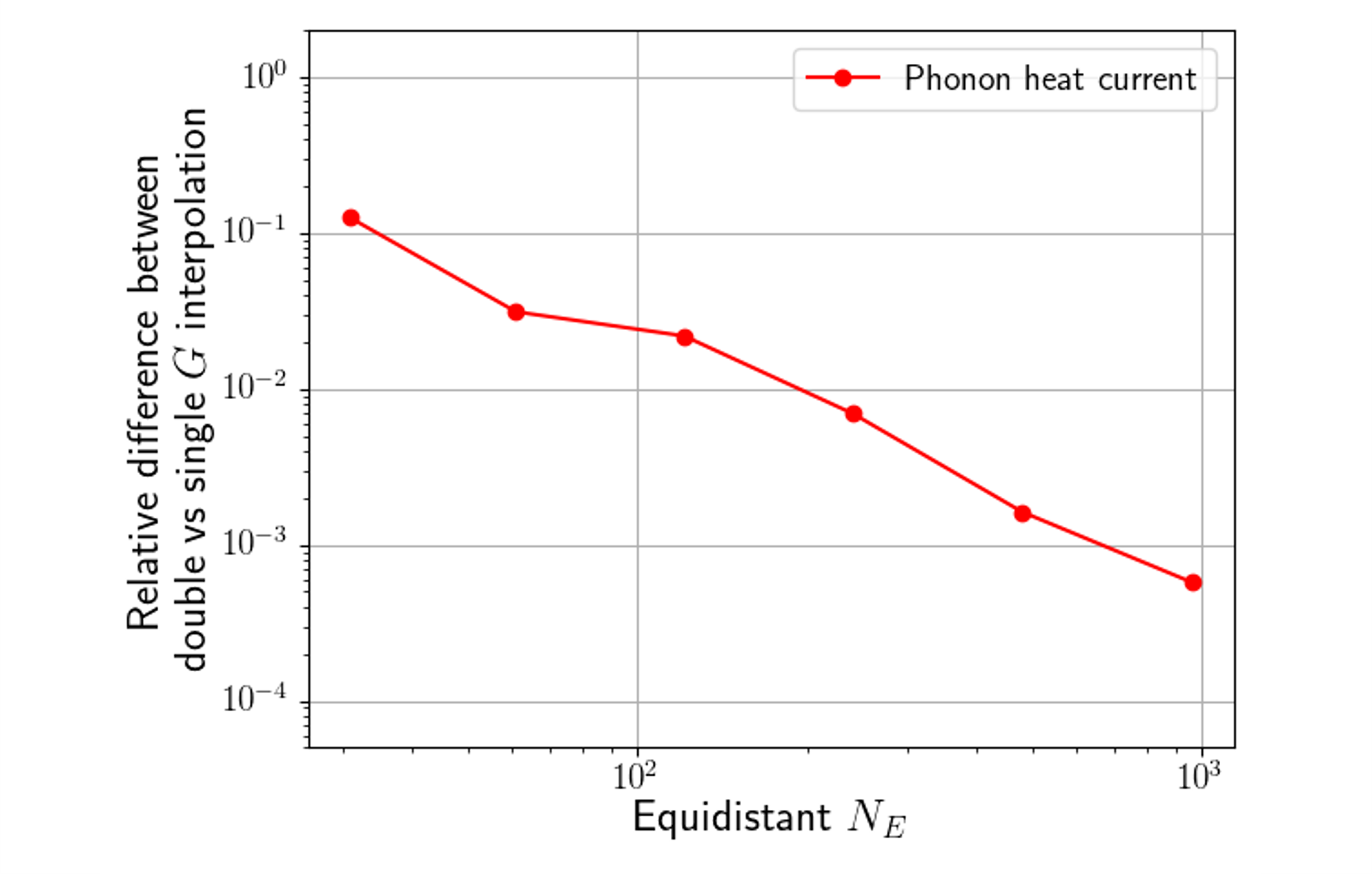}
    \caption{The relative difference between the phonon heat currents computed with \eqref{eq:Pi_as_convolution_2} and \eqref{eq:Pi_as_convolution_3} for the system in Section \ref{secsec:device_simulation} as a function of the number of initial equidistant grid energy points.}
    \label{fig:double_G_vs_single_G}
\end{figure}

In the FFT-based self-energy computation, there are two choices for the interpolation scheme of the electron Green's function during the calculation of the phonon self-energy, corresponding to \eqref{eq:Pi_as_convolution_2} and \eqref{eq:Pi_as_convolution_3}, respectively. In the discussion above, \eqref{eq:Pi_as_convolution_2} was used in all simulations with the FFT-based implementation. This choice was not well-founded. However, Fig. \ref{fig:double_G_vs_single_G} demonstrates that the difference between these two choices is significantly lower than the equidistant-grid errors discussed above.

\section{\label{sec:self_consist_device}Self-consistent coupled electron phonon transport in a device}
\begin{figure}[ht]
    \centering
    \includegraphics[width=\linewidth]{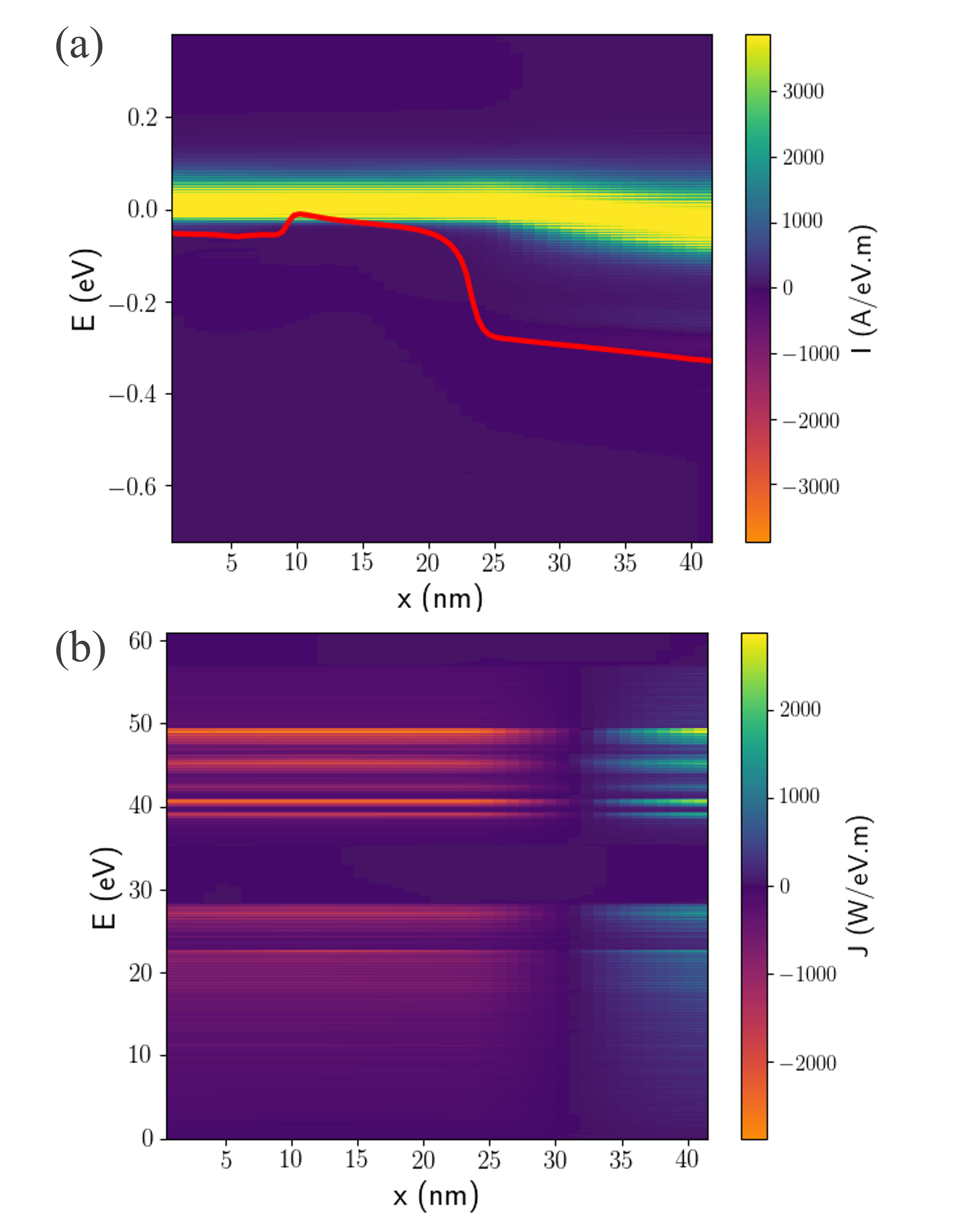}
    \caption{The electron current spectrum (a) and phonon heat current spectrum (b) for a converged simulation of the system in Section \ref{secsec:device_simulation}. The spectra are given as a function of energy and the position along the transport direction. The red line in the electron current spectrum denote the bottom of the conduction band.}
    \label{fig:scattering_spectra}
\end{figure}

\begin{figure}[ht]
    \centering
    \includegraphics[width=\linewidth]{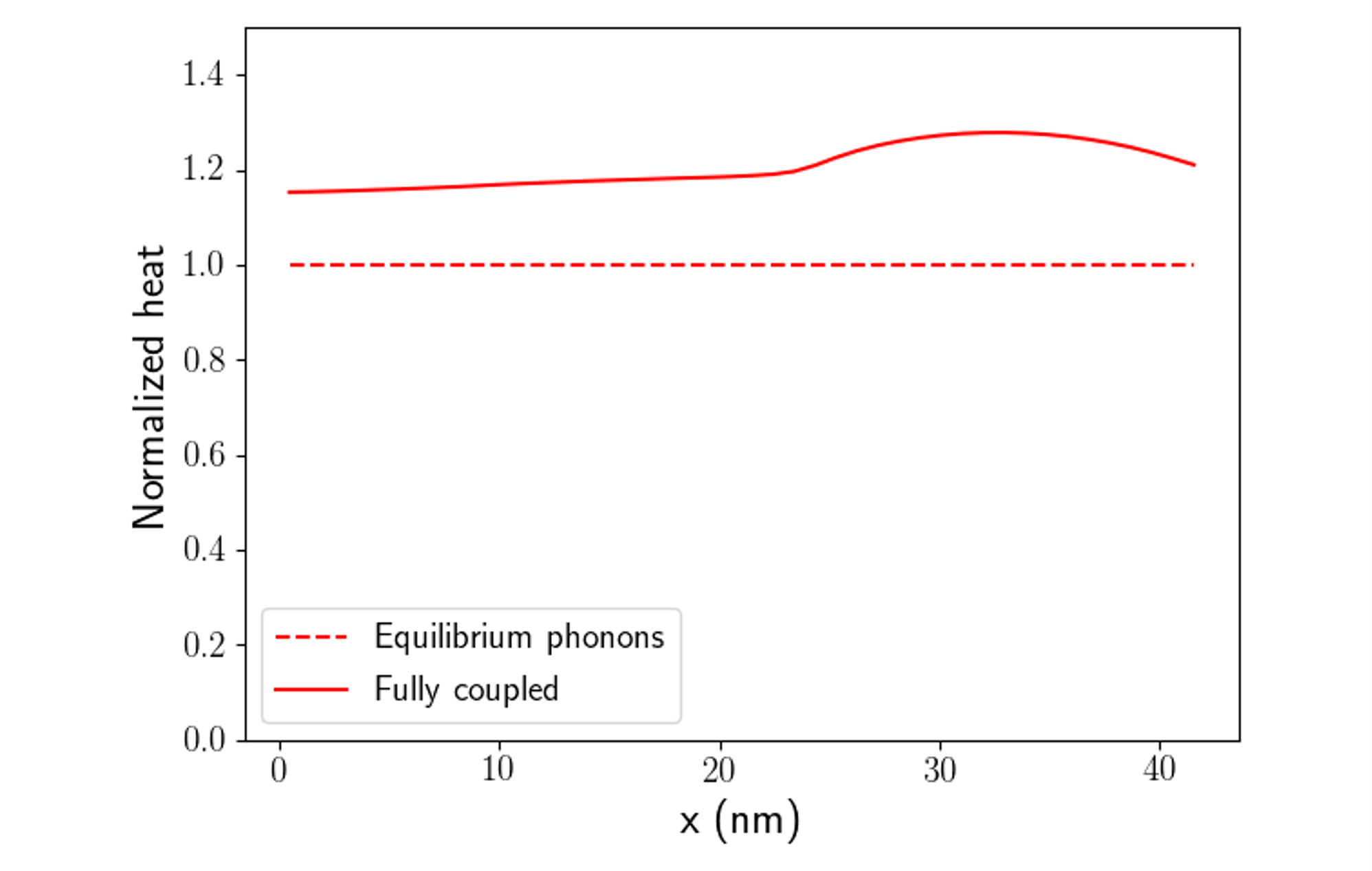}
    \caption{The heat related to the phonon Green's function for a converged fully coupled simulation of the system in Section \ref{secsec:device_simulation}, and a converged dissipative simulation with phonons kept in equilibrium. Both curves are given as a function of the position along the transport direction and are normalized by the equilibrium phonon heat in the middle of the device.}
    \label{fig:heat_generation}
\end{figure}

The discussion in Section \ref{sec:model_testing} showed that the FFT-based self-energy computation can provide an efficient way to perform fully coupled transport simulations. We now extend that discussion to more than a single Born iteration. The results for a converged simulation are given in Fig. \ref{fig:scattering_spectra}. Fig. \ref{fig:scattering_spectra} (a) shows the electron current spectrum as a function of the position in the device. It can be seen that the current behaves quasi-ballistic until it reaches the drain extension region. The availability of lower-energy states then allows for heat dissipation as the electrons lose energy to generate additional phonons. This is confirmed by the phonon heat current spectrum in Fig. \ref{fig:scattering_spectra} (b). Phonons are generated in the drain extension region and travel away in both directions. Fig. \ref{fig:heat_generation} compares this heat generation with a simulation with equilibrium phonons. The heat of the equilibrium phonons is, as expected, independent of position. Compared to the equilibrium phonon case, the heat of the fully coupled simulation is higher everywhere in the device. It reaches a maximum in the drain extension region, where the phonons are generated, and decreases further away. 

\clearpage

\begin{figure}[ht]
    \centering
    \includegraphics[width=\linewidth]{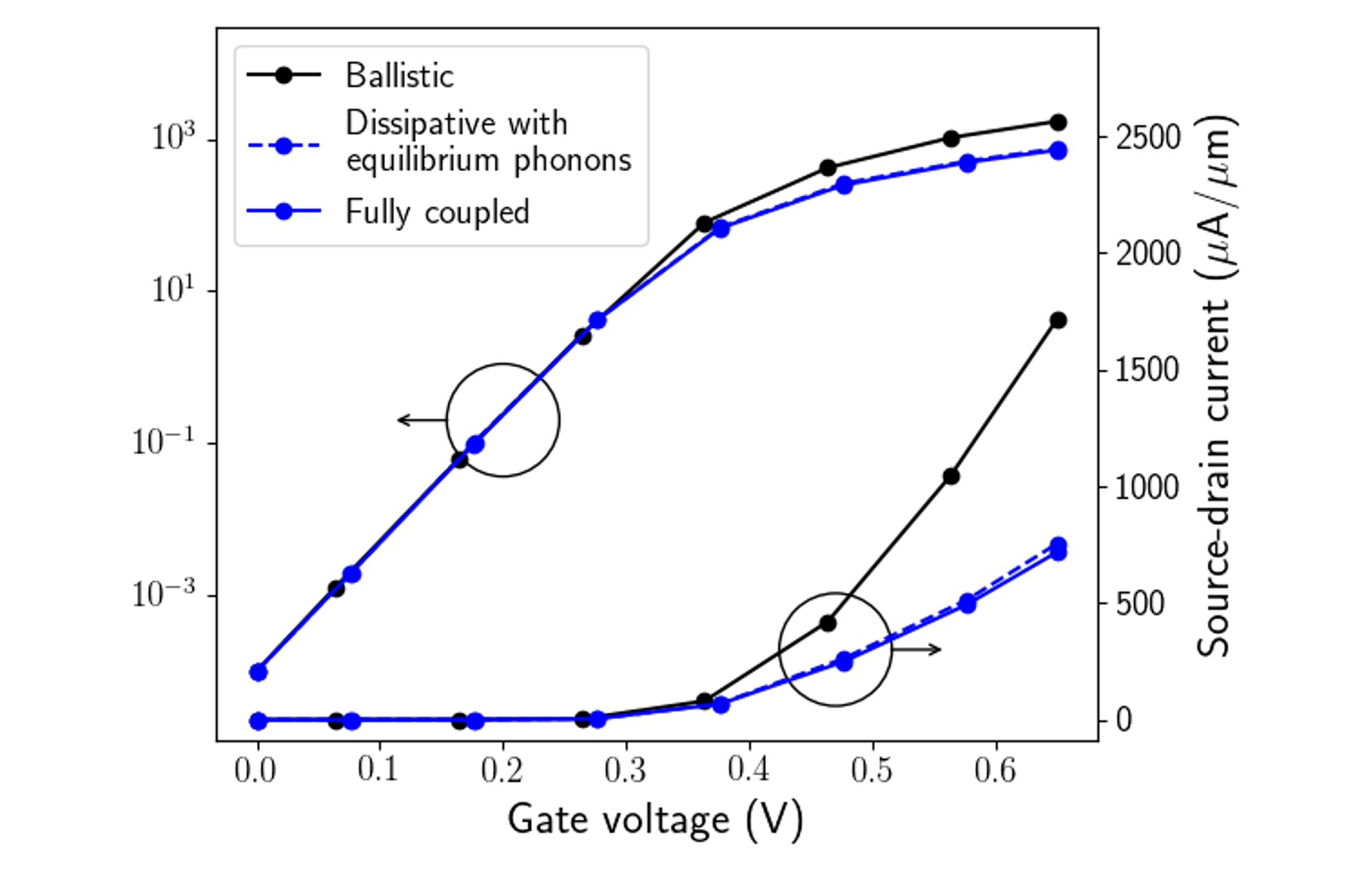}
    \caption{The source-drain current for converged simulations of the system in Section \ref{secsec:device_simulation} as a function of the gate potential. The current is shown for the ballistic case, the case with scattering by phonons in equilibrium and the fully coupled case.}
    \label{fig:IV}
\end{figure}

The influence on the source-drain current is shown in Fig. \ref{fig:IV}. The influence of scattering on the subthreshold regime is limited. The ON-state current, however, is reduced by 57\%. The results for a fully coupled simulation and a simulation with equilibrium phonons are nearly identical, differing by a mere 5\% in the ON-state current. For the system discussed here, self-heating effects thus have negligible influence on device performance. It should be noted that Fig. \ref{fig:scattering_spectra} (a) shows that the electrons are not fully thermalized, despite the elongated drain extension region. This implies that the remaining heat is lost in the metal leads and would thus introduce a temperature increase there. A full incorporation of the metal contacts is, however, outside the scope of this work.

\section{\label{sec:conclusion}Conclusion and Future Work}

In this work, we showed how real space matrix elements provided by readily available codes such as QUANTUM ESPRESSO and Perturbo can be used to perform fully coupled electron-phonon transport simulations using NEGF. Additionally, we showed how the computationally demanding self-energy calculation step can be made significantly more efficient by using an FFT-based implementation. For a serial computation, a $\sim$500 times speedup for the self-energy calculation step was achieved by using this FFT-based implementation, going from 98\% of the total computation time to a mere 9\%. Additionally, we showed that the FFT-based implementation is readily parallelized, without the significant additional memory requirements coming with parallelization of the conventional convolution-based implementation. 

For equidistant energy grids, the error introduced by this FFT-based implementation is limited to an energy mixing error, which is predicted to be multiple orders smaller than the integration error. For adaptive non-equidistant energy grids, the relative error on macroscopic electronic properties, such as the charge density and current, can be made less than 1\% while the error on the heat current in the device can be limited to a few percent.

Additionally, our simulations predict that, while scattering by electron-phonon interactions is important, self-heating effects can be neglected. However, this conclusion is based on the fact that further thermalization of electrons in the metallic leads does not introduce significant heating effects as the metal leads were not included in our simulations.

Finally, the simulations in this work rely on approximations concerning the self-energy computation. Certain electron-phonon matrix elements and Green's function elements were neglected to reduce the significant computational cost of the full self-energy computation. To compensate for this, the remaining electron-phonon matrix elements were rescaled. An exact self-energy calculation, using all electron-phonon matrix elements and Green's function elements, is prohibitively expensive with the conventional convolution-based implementation. The FFT-based implementation could, however, provide a means to estimate the influence of these currently neglected elements.

\begin{acknowledgments}
This research was funded by the FWO as part of the PhD fellowship 1100321N.
\end{acknowledgments}

\clearpage

\appendix

\section{Reciprocal-real space transformations}\label{app:transformations}

We elaborate on the different reciprocal-real space and real-mixed space transformations of Section \ref{secsec:Hamiltonian}. The reciprocal-real space transformation of the electron creation operator in \eqref{eq:el_rec2rel_cre}, repeated here for the sake of clarity,

\begin{equation}
\ctc_{\Rb_e m} = \frac{1}{\sqrt{N_e}}\sum_{n\kb}e^{-i\kb\cdot\Rb_e}U_{nm,\kb}\cc_{\kb n},
\label{eq:el_rec2rel_cre2}
\end{equation}

can be derived from the fact that

\begin{align}
\ket{n\kb} &= \cc_{\kb n} \ket{0},\\
\ket{m\Rb_e} &= \ctc_{\Rb_e m} \ket{0},
\end{align}

and that

\begin{equation}
\ket{m\Rb_e} = \frac{1}{\sqrt{N_e}}\sum_{n\kb}e^{-i\kb\cdot\Rb_e}U_{nm,\kb}\ket{n\kb}.
\end{equation}

Note that the normalization convention is different than the ones in Ref. \cite{Giustino2007,Marzari2012}. The states $\ket{n\kb}$ and $\ket{m\Rb_e}$ are normalized over all of space. This is required to let both $\cc_{\kb n}$ and $\cc_{\Rb_e m}$ be proper creation operators. Taking the Hermitian conjugate of \eqref{eq:el_rec2rel_cre}, we namely find the transformation expression of the electron annihilation operators,

\begin{equation}
\ct_{\Rb_e m} = \frac{1}{\sqrt{N_e}}\sum_{n\kb}e^{i\kb\cdot\Rb_e}U^{\dagger}_{mn,\kb}\ch_{\kb n}.
    \label{eq:el_rec2rel_ann}
\end{equation}

For these operators to be proper creation and annihilation operators, they need to obey the anticommutation relations,

\begin{align}
\begin{split}
    \{\ch_{\kb n},\cc_{\kb'n'}\} &= \delta_{n,n'}\delta_{\kb,\kb'},\\
    \{\cc_{\kb n},\cc_{\kb'n'}\} &= \{\ch_{\kb n},\ch_{\kb'n'}\} =  0,\\
    \{\ct_{\Rb_e m},\ctc_{\Rb_e'm'}\} &= \delta_{m,m'}\delta_{\Rb_e,\Rb_e'},\\
    \{\ctc_{\Rb_e m},\ctc_{\Rb_e'm'}\} &= \{\ct_{\Rb_e m},\ct_{\Rb_e'm'}\} =  0.
\end{split}
\label{eq:anticommutation}
\end{align}

This is only true for both the reciprocal and real space operators for the normalization convention of \eqref{eq:el_rec2rel_cre2}.

In a similar fashion, the reciprocal-real space transformation for the phonon operators can be found as

\begin{align}
    \atc_{\Rb_p\ka\al} &= \frac{1}{\sqrt{N_p}}\sum_{\nu\qb}e^{-i\qb\cdot\Rb_p}e^*_{\ka\al\nu,\qb}\ac_{\qb \nu},
    \label{eq:ph_rec2rel_cre2}\\
    \at_{\Rb_p\ka\al} &= \frac{1}{\sqrt{N_p}}\sum_{\nu\qb}e^{i\qb\cdot\Rb_p}e_{\ka\al\nu,\qb}\ah_{\qb \nu}.
    \label{eq:ph_rec2rel_ann}
\end{align}

Again, a different convention from Ref. \cite{Giustino2007} is used. Similarly to the electron case, the normalization with the number of k-points is symmetric over the two transformations. Additionally, no mass rescaling is performed. For these operators to be proper bosonic operators, they need to obey the commutation relations,

\begin{align}
\begin{split}
    [\ah_{\qb \nu},\ac_{\qb'\nu'}] &= \delta_{\nu,\nu'}\delta_{\qb,\qb'},\\
    [\ac_{\qb \nu},\ac_{\qb'\nu'}] &= [\ah_{\qb \nu},\ah_{\qb'\nu'}] =  0,\\
    [\at_{\Rb_p \nu},\atc_{\Rb_p'\nu'}] &= \delta_{\nu,\nu'}\delta_{\Rb_p,\Rb_p'}\\
    [\atc_{\Rb_p \nu},\atc_{\Rb_p'\nu'}] &= [\at_{\Rb_p \nu},\at_{\Rb_p'\nu'}] = 0.
\end{split}
\label{eq:commutation}
\end{align}

This, and the fact that the creation and annihilation operator are each other's Hermitian conjugate, can only satisfied for the normalization conventions of \eqref{eq:ph_rec2rel_cre2} and \eqref{eq:ph_rec2rel_ann}.

The inverse transformations of \eqref{eq:el_rec2rel_cre2}, \eqref{eq:el_rec2rel_ann}, \eqref{eq:ph_rec2rel_cre2}  and \eqref{eq:ph_rec2rel_ann} are given by

\begin{align}
    \cc_{\kb n} &= \frac{1}{\sqrt{N_e}}\sum_{m\Rb_e}e^{i\kb\cdot\Rb_e}U^{\dagger}_{mn,\kb}\ctc_{\Rb_e m},
    \label{eq:el_rel2rec_cre}\\
    \ch_{\kb n} &= \frac{1}{\sqrt{N_e}}\sum_{m\Rb_e}e^{-i\kb\cdot\Rb_e}U_{nm,\kb}\ct_{\Rb_e m},
    \label{eq:el_rel2rec_ann}\\
    \ac_{\qb \nu} &= \frac{1}{\sqrt{N_p}}\sum_{\ka\al \Rb_p}e^{i\qb\cdot\Rb_p}e_{\ka\al\nu,\qb}\atc_{\Rb_p\ka\al},
    \label{eq:ph_rel2rec_cre}\\
    \ah_{\qb \nu} &= \frac{1}{\sqrt{N_p}}\sum_{\ka\al \Rb_p}e^{-i\qb\cdot\Rb_p}e_{\ka\al\nu,\qb}^*\at_{\Rb_p\ka\al}.
    \label{eq:ph_rel2rec_ann}
\end{align}

Inserting these relations into \eqref{eq:hamiltonian_rec}, we obtain

\begin{align}
\begin{split}
\hat{H} =&
\sum_{\substack{m m'\\\Rb_e\Rb_e'}} 
    \tilde{h}_{\substack{m m'\\\Rb_e'-\Rb_e}} \ctc_{\Rb_e m}\ct_{\Rb_e' m'}\\
&+ \sum_{\substack{\ka\al\ka'\al'\\\Rb_p\Rb_p'}}   
    \tilde{d}_{\substack{\ka\al\ka'\al'\\\Rb_p'-\Rb_p}} \atc_{\Rb_p\ka\al}\at_{\Rb_p'\ka'\al'} + H_{zero}\\
&+\sum_{\substack{m' n'\ka\al\\\Rb_e\Rb_e'\Rb_p}}
    \tilde{g}_{\substack{m'n'\ka\al\\\Rb_e'-\Rb_e\\\Rb_p-\Rb_e}}
    \ctc_{\Rb_e m'}\ct_{\Rb_e' n'}
    (\at_{\Rb_p\ka\al}+\atc_{\Rb_p\ka\al})
\end{split}
\label{eq:hamiltonian_rel}
\end{align}

with $H_{zero}$ the zero-point energy of the phonon system, which will be neglected from here forth. 

\begin{equation}
    H_{zero} = \sum_{\nu\qb}\frac{\hb\omega_{\qb\nu}}{2}
\end{equation}

The matrix elements in \eqref{eq:hamiltonian_rel} are defined as

\begin{align}
\tilde{h}_{\substack{m m'\\\Rb_e'-\Rb_e}}
     &= \frac{1}{N_e} \sum_{n\kb} e^{-i\kb\cdot(\Rb_e'-\Rb_e)} U^{\dagger}_{mn,\kb}e_{\kb n}U_{nm',\kb},
    \label{eq:el_rel2rec_h}\\
\tilde{d}_{\substack{\ka\al\ka'\al'\\\Rb_p'-\Rb_p}}
    &= \frac{1}{N_p} \sum_{\nu\qb} e^{-i\qb\cdot(\Rb_p'-\Rb_p)} e_{\ka\al\nu,\qb}\hb\omega_{\qb \nu}e_{\ka'\al'\nu,\qb}^*,
    \label{eq:ph_rel2rec_d}\\
\tilde{g}_{\substack{m'n'\ka\al\\\Rb_e'-\Rb_e\\\Rb_p-\Rb_e}}
    &= \frac{1}{N_pN_e}\sum_{\substack{mn\kb\\\qb\nu}}e^{-i\kb\cdot(\Rb_e'-\Rb_e)-i\qb\cdot(\Rb_p-\Rb_e)} \nonumber\\
    &\qquad\qquad U^{\dagger}_{m'm,\kb+\qb}g_{mn\nu}(\kb,\qb)U_{nn',\kb}e_{\ka\al\nu,\qb}^*.
\label{eq:rel2rec_M}
\end{align}

The electron Hamiltonian elements of \eqref{eq:el_rel2rec_h} are equal to the ones computed by the Wannier90 code \cite{Marzari2012} and presented in the flexible HDF5 format by Perturbo \cite{Giustino2007}. Equation 26 in Ref. \cite{Giustino2007} has a different normalization convention. However, attempts to match the Wannier interpolated electronic bands as provided by Perturbo only succeeded assuming Perturbo instead uses the normalization convention of \eqref{eq:el_rel2rec_h}.

ATOMOS does not use the phonon Hamiltonian elements, but the interatomic force constants. Grouping the indices ($\kappa$, $\alpha$, $\Rb_p$) into a single row or column index, $\mathbf{\bar{d}}$ can be considered a matrix. The square of this matrix can be linked to the required interatomic force constants,

\begin{align}
\begin{split}
\sum_{\ka'\al'\Rb_p'}&\tilde{d}_{\substack{\ka\al\ka'\al'\\\Rb_p'-\Rb_p}}\tilde{d}_{\substack{\ka'\al'\ka''\al''\\\Rb_p''-\Rb_p'}} \\
&= \frac{\hb^2}{N_p}\sum_{\nu\qb} e^{-i\qb\cdot(\Rb_p''-\Rb_p)} e_{\ka\al\nu,\qb}\omega_{\qb \nu}^2e_{\ka''\al''\nu,\qb}^*,
\end{split}
\label{eq:ph_square_d}
\end{align}

where we used that 

\begin{equation}
\sum_{\Rb_p'}e^{-i(\qb-\qb')\cdot \Rb_p'} = N_p \delta_{\qb,\qb'},
\end{equation}

and used the orthonormality of $e_{\ka\al\nu,\qb}$ \cite{Giustino2007},

\begin{equation}
\sum_{\ka'\al'} e_{\ka'\al'\nu,\qb}^* e_{\ka'\al'\nu',\qb} = \delta_{\nu,\nu'}.
\end{equation}

The right-hand side of \eqref{eq:ph_square_d} is a rescaled version of interatomic force constant defined in \eqref{eq:Phi_matrix} \cite{Giustino2007,Giustino2017},

\begin{align}
\begin{split}
\sum_{\ka'\al'\Rb_p'}&\tilde{d}_{\substack{\ka\al\ka'\al'\\\Rb_p'-\Rb_p}}\tilde{d}_{\substack{\ka'\al'\ka''\al''\\\Rb_p''-\Rb_p'}} \\
&= \frac{\hb^2}{\sqrt{m_{\ka}m_{\ka''}}}\frac{\partial^2U}{\partial \tau_{\ka\al p}\partial \tau_{\ka''\al'' p''}}.
\end{split}
\label{eq:ph_square_d2}
\end{align}

The interatomic force constants are also provided in the HDF5 format by Perturbo \cite{Giustino2007}.

As will be detailed in Appendix \ref{app:Greens_functions} and \ref{app:self_energies}, ATOMOS does not use the matrix elements in \eqref{eq:rel2rec_M}, but requires a modification,

\begin{align}
\begin{split}
    \tilde{\tilde{g}}_{\substack{m'n'\ka\al\\\Rb_e'-\Rb_e\\\Rb_p-\Rb_e}}
    &= \frac{1}{N_pN_e}\sum_{\substack{mn\kb\\\qb\nu}}e^{-i\kb\cdot(\Rb_e'-\Rb_e)-i\qb\cdot(\Rb_p-\Rb_e)}\\
    &\qquad\qquad U^{\dagger}_{m'm,\kb+\qb}g^{def}_{mn\nu}(\kb,\qb)U_{nn',\kb}e_{\ka\al\nu,\qb}^*
\end{split}
\label{eq:rel2rec_M2}
\end{align}

with

\begin{equation}
    g^{def}_{mn\nu}(\kb,\qb) = \sqrt{\hb\omega_{\qb\nu}}g_{mn\nu}(\kb,\qb).
\end{equation}

Comparing the definition of $g_{mn\nu}(\kb,\qb)$ (Equations 21 and 31-38 in Ref. \cite{Giustino2017}) and the definition of the deformation potentials (Equations 6 and 17 in Ref. \cite{Giustino2007}) we find that

\begin{equation}
    g^{def}_{mn\nu}(\kb,\qb) = \frac{\hb}{\sqrt{2m_0}}g^{Perturbo}_{mn\nu}(\kb,\qb).
\end{equation}

Next, comparing the Perturbo reciprocal-real space transformation (Equation 24 in Ref. \cite{Giustino2007}) with \eqref{eq:rel2rec_M2}, we find a difference in the normalization convention related to $N_e$ and an additional mass rescaling. Concerning the normalization convention, attempts to match the Wannier interpolated deformation potentials as provided by Perturbo again only succeeded if the normalization convention of \eqref{eq:rel2rec_M2} were used instead. The mass rescaling can be incorporated, resulting in 

\begin{align}
\begin{split}
\tilde{\tilde{g}}_{\substack{mn\kappa\alpha\\\Rb_e,\Rb_p}} = \frac{\hb}{\sqrt{2m_{\kappa}}}g^{Perturbo}_{mn\kappa\alpha}(\Rb_e,\Rb_p).
\end{split}
\end{align}

The full real space Hamiltonian can be further transformed to mixed space by Fourier transforming the periodic directions using the following transformations,

\begin{align}
    \cbc_{\kb_{\perp}\Rb_{e,\parallel} n} &= \frac{1}{\sqrt{N_{\perp}}}\sum_{\Rb_{e,\perp}}e^{i\kb_{\perp}\cdot\Rb_{e,\perp}}\ctc_{\Rb_{e,\perp}\Rb_{e,\parallel} n},
    \label{eq:el_rel2mix_cre}\\
    \cb_{\kb_{\perp}\Rb_{e,\parallel} n} &= \frac{1}{\sqrt{N_{\perp}}}\sum_{\Rb_{e,\perp}}e^{-i\kb_{\perp}\cdot\Rb_{e,\perp}}\ct_{\Rb_{e,\perp}\Rb_{e,\parallel} n},
    \label{eq:el_rel2mix_ann}\\
    \abc_{\qb_{\perp}\Rb_{p,\parallel} \ka\al} &= \frac{1}{\sqrt{N_{\perp}}}\sum_{\Rb_{p,\perp}}e^{i\qb_{\perp}\cdot\Rb_{p,\perp}}\atc_{\Rb_{p,\perp}\Rb_{p,\parallel}\ka\al},
    \label{eq:ph_rel2mix_cre}\\
    \ab_{\qb_{\perp}\Rb_{p,\parallel} \ka\al} &= \frac{1}{\sqrt{N_{\perp}}}\sum_{\Rb_{p,\perp}}e^{-i\qb_{\perp}\cdot\Rb_{p,\perp}}\at_{\Rb_{p,\perp}\Rb_{p,\parallel}\ka\al}.
    \label{eq:ph_rel2mix_ann}
\end{align}

Currently no assumptions are made about the number of periodic directions and the number of periodic k-points, $N_{\perp}$, except that this number is equal for the electron and phonon system. The number of periodic directions can range from 0 for fully real space nanowires with $N_{\perp} = 1$ and the systems being unchanged, to 1 for planar or 2 for diode-like or resistor-like systems. The inverse transformations are given by

\begin{align}
    \ctc_{\Rb_{e,\perp}\Rb_{e,\parallel} n} &= \frac{1}{\sqrt{N_{\perp}}}\sum_{\kb_{\perp}}e^{-i\kb_{\perp}\cdot\Rb_{e,\perp}}\cbc_{\kb_{\perp}\Rb_{e,\parallel} n},
    \label{eq:el_mix2rel_cre}\\
    \ct_{\Rb_{e,\perp}\Rb_{e,\parallel} n} &= \frac{1}{\sqrt{N_{\perp}}}\sum_{\kb_{\perp}}e^{i\kb_{\perp}\cdot\Rb_{e,\perp}}\cb_{\kb_{\perp}\Rb_{e,\parallel} n},
    \label{eq:el_mix2rel_ann}
\end{align}
    
\begin{align}    
    \atc_{\Rb_{p,\perp}\Rb_{p,\parallel} \ka\al} &= \frac{1}{\sqrt{N_{\perp}}}\sum_{\qb_{\perp}}e^{-i\qb_{\perp}\cdot\Rb_{p,\perp}}\abc_{\qb_{\perp}\Rb_{p,\parallel} \ka\al},
    \label{eq:ph_mix2rel_cre}\\
    \at_{\Rb_{\perp}\Rb_{p,\parallel} \ka\al} &= \frac{1}{\sqrt{N_{\perp}}}\sum_{\qb_{\perp}}e^{i\qb_{\perp}\cdot\Rb_{p,\perp}}\ab_{\qb_{\perp}\Rb_{p,\parallel} \ka\al}.
    \label{eq:ph_mix2rel_ann}
\end{align}

It is readily verified that the mixed space operators obey the required commutation and anticommutations relations. Insertion in \eqref{eq:hamiltonian_rel} results in

\begin{widetext}
\begin{align}
\begin{split}
    \hat{H} =
    &\sum_{\substack{m n\kb_{\perp}\\\Rb_{e,\parallel}\Rb_{e,\parallel}'}}
\bar{h}_{\substack{m n\kb_{\perp}\\\Rb_{e,\parallel}\Rb_{e,\parallel}'}}
 \cbc_{\kb_{\perp}\Rb_{e,\parallel} m}\cb_{\kb_{\perp}\Rb_{e,\parallel}' n} + \sum_{\substack{\ka\al\ka'\al'\qb_{\perp}\\\Rb_{p,\parallel}\Rb_{p,\parallel}'}}
    \bar{d}_{\substack{\ka\al\ka'\al'\qb_{\perp}\\\Rb_{p,\parallel}\Rb_{p,\parallel}'}}
     \abc_{\qb_{\perp}\Rb_{p,\parallel} \ka\al}\ab_{\qb_{\perp}\Rb_{p,\parallel}' \ka'\al'}\\ 
 &+N_{\perp}^{-\frac{1}{2}}\sum_{\substack{m n\ka\al\\\Rb_{e,\parallel}\Rb_{e,\parallel}'\\\kb_{\perp}\qb_{\perp}\Rb_{p,\parallel}}}
    \bar{g}_{\substack{mn\ka\al\\\Rb_{e,\parallel}\Rb_{e,\parallel}'\\\kb_{\perp}\qb_{\perp}\Rb_{p,\parallel}}}
    \cbc_{\kb_{\perp}+\qb_{\perp}\Rb_{e,\parallel} m}\cb_{\kb_{\perp}\Rb_{e,\parallel}' n}(\ab_{\qb_{\perp}\Rb_{p,\parallel} \ka\al}+\abc_{-\qb_{\perp}\Rb_{p,\parallel} \ka\al}).
\end{split}
\end{align}
\end{widetext}

Grouping the real space indices into a single index results in \eqref{eq:hamiltonian_mix}. The matrix element transformations are performed by ATOMOS and correspond to

\begin{align}   \bar{h}_{\substack{mn\kb_{\perp}\\\Rb_{e,\parallel}\Rb_{e,\parallel}'}}
    =& \sum_{\Rb_{e,\perp}}
    \tilde{h}_{\substack{m n\\\Rb_{e,\perp}\\\Rb_{e,\parallel}'-\Rb_{e,\parallel}}}
     e^{i\kb_{\perp}\cdot\Rb_{e,\perp}},\\
\bar{d}_{\substack{\ka\al\ka'\al'\qb_{\perp}\\\Rb_{p,\parallel}\Rb_{p,\parallel}'}}
    =& \sum_{\Rb_{p,\perp}}
    \tilde{d}_{\substack{\ka\al\ka'\al'\\\Rb_{p,\perp}\\\Rb_{p,\parallel}'-\Rb_{p,\parallel}}}
    e^{i\qb_{\perp}\cdot\Rb_{p,\perp}},\label{eq:ph_mix2rel_d}\\
\bar{g}_{\substack{mn\ka\al\\\Rb_{e,\parallel}\Rb_{e,\parallel}'\\\kb_{\perp}\qb_{\perp}\Rb_{p,\parallel}}} 
=&
    \sum_{\Rb_{e,\perp}\Rb_{p,\perp}}
    \tilde{g}_{\substack{mn\ka\al\\\Rb_{e,\perp}\\\Rb_{e,\parallel}'-\Rb_{e,\parallel}\\\Rb_{p,\perp}\\\Rb_{p,\parallel}-\Rb_{e,\parallel}}}
    e^{i\kb_{\perp}\cdot\Rb_{e,\perp}+i\qb_{\perp}\cdot\Rb_{p,\perp}}.\label{eq:mix2rel_M}
\end{align}

\section{Green's function expressions}\label{app:Greens_functions}

We provide a full derivation of the electron and phonon Green's functions expression introduced in Section \ref{secsec:Greensfunctions}. We start with the electron Green's function, despite derivations for these expressions being readily available in the literature \cite{Lake2006,Dattasupp}, for the sake of completeness and to provide a reference for the phonon case. Consider the definition of the retarded, advanced, greater and lesser Green's function

\begin{align}
    G_{\substack{n,m\\\kb_{\perp}}}^R(t,t') &= \frac{-i}{\hb}\theta(t-t')\label{eq:retarded_def}\\
    &\qquad\quad\braket{\cb_{\kb_{\perp}n}(t)\cbc_{\kb_{\perp}m}(t')+\cbc_{\kb_{\perp}m}(t')\cb_{\kb_{\perp}n}(t)},\nonumber\\
    G_{\substack{n,m\\\kb_{\perp}}}^A(t,t') &= \frac{i}{\hb}\theta(t'-t)\label{eq:advanced_def}\\
    &\qquad\quad\braket{\cb_{\kb_{\perp}n}(t)\cbc_{\kb_{\perp}m}(t')+\cbc_{\kb_{\perp}m}(t')\cb_{\kb_{\perp}n}(t)},\nonumber\\
    G_{\substack{n,m\\\kb_{\perp}}}^>(t,t') &= \frac{-i}{\hb}\braket{\cb_{\kb_{\perp}n}(t)\cbc_{\kb_{\perp}m}(t')},\label{eq:greater_def}\\
    G_{\substack{n,m\\\kb_{\perp}}}^<(t,t') &= \frac{i}{\hb}\braket{\cbc_{\kb_{\perp}m}(t')\cb_{\kb_{\perp}n}(t)}.\label{eq:lesser_def}
\end{align}

When an expression for the contour-ordered Green's function is known involving convolution integrals on a two-branch contour, e.g., \eqref{eq:exactG_as_convol_sigG}, then Langreth's theorem can be used to find equivalent expressions for the retarded, advanced, greater and lesser Green's functions involving only real-time convolutions \cite{Maciejko2007}. Convolutions in the time domain become multiplications in the frequency domain. Hence, using the Fourier transform

\begin{align}
    \Gb_{\kb_{\perp}}(\omega) &= \int^{+\infty}_{-\infty} \Gb_{\kb_{\perp}}(t,t')e^{i\omega(t-t')} d(t-t'),\label{eq:fourier_forward_def}\\
    \Gb_{\kb_{\perp}}(t,t') &= \int^{+\infty}_{-\infty} \Gb_{\kb_{\perp}}(\omega)e^{-i\omega(t-t')} \frac{d\omega}{2\pi},\label{eq:fourier_inverse_def}
\end{align}

with the Green's function denoting any of the Green's functions in \eqref{eq:retarded_def}-\eqref{eq:lesser_def}, results in

\begin{align}
\begin{split}
\Gb_{\kb_{\perp}}^{R/A}(\omega) 
    =& \Gb_{\kb_{\perp}}^{0,R/A}(\omega)+
    \Gb_{\kb_{\perp}}^{0,R/A}(\omega)\mathbf{U}_{\kb_{\perp}}\Gb_{\kb_{\perp}}^{R/A}(\omega)\\
    &\qquad+
    \Gb_{\kb_{\perp}}^{0,R/A}(\omega)\mathbf{\Sigma}_{\kb_{\perp}}^{s,R/A}
    (\omega)\Gb_{\kb_{\perp}}^{R/A}(\omega),
    \end{split}
\label{eq:Langreth1}
\end{align}

\begin{align}
\begin{split}
\Gb_{\kb_{\perp}}^{\gtrless}(\omega) 
    =& \Gb_{\kb_{\perp}}^{0,\gtrless}(\omega)+
    \Gb_{\kb_{\perp}}^{0,\gtrless}(\omega)\mathbf{U}_{\kb_{\perp}}
    \Gb_{\kb_{\perp}}^{A}(\omega)\\
     &\qquad\quad\ \ +
     \Gb_{\kb_{\perp}}^{0,R}(\omega)\mathbf{U}_{\kb_{\perp}}\Gb_{\kb_{\perp}}^{\gtrless}(\omega)\\
     &\qquad\quad\ \ + 
    \Gb_{\kb_{\perp}}^{0,R}(\omega)\mathbf{\Sigma}_{\kb_{\perp}}^{s,R}
    (\omega)\Gb_{\kb_{\perp}}^{\gtrless}(\omega)\\
    &\qquad\quad\ \ + 
    \Gb_{\kb_{\perp}}^{0,R}(\omega)\mathbf{\Sigma}_{\kb_{\perp}}^{s,\gtrless}
    (\omega)\Gb_{\kb_{\perp}}^{A}(\omega)\\
    &\qquad\quad\ \ + 
    \Gb_{\kb_{\perp}}^{0,\gtrless}(\omega)\mathbf{\Sigma}_{\kb_{\perp}}^{s,A}
    (\omega)\Gb_{\kb_{\perp}}^{A}(\omega).
\end{split}
\label{eq:Langreth2}
\end{align}

For the electrons, the matrix $\mathbf{U}_{\kb_{\perp}}$ contains the elements $\bar{h}_{\substack{mn}\\\kb{\perp}}$ connecting the device with the leads, as denoted in Fig. \ref{fig:coupled_sys}. The remaining elements form

\begin{equation}
	\hat{H}^{el}_0 = \sum_{\substack{n m\\\kb_{\perp}}}
	\bar{h}_{\substack{n m\\\kb_{\perp}}}
	\cbc_{\kb_{\perp} n}\cb_{\kb_{\perp} m}.
\end{equation}

with the summation leaving out interaction elements connecting the device and the leads. Note that all leftmost factors in \eqref{eq:Langreth1} and \eqref{eq:Langreth2} are non-interacting, which implies that for the sake of these Green's functions, the Heisenberg picture operators in their definitions are equivalent to interaction picture operators. The time-dependency of interaction picture operators is readily found as \cite{Fetterch3}

\begin{align}
\begin{split}
    i\hb \frac{\partial}{\partial t}\cb_{\kb_{\perp}n}(t) &= \left[\cb_{\kb_{\perp}n}(t),\hat{H}^{el}_0\right] \\
     &= \sum_{m}\bar{h}_{\substack{n m\\\kb_{\perp}}}\cb_{\kb_{\perp} m}(t).
\end{split}
\end{align}

The time-derivative of the non-interacting retarded Green's functions is then given by

\begin{equation}
        i\hb\frac{\partial}{\partial t}G_{\substack{n,m\\\kb_{\perp}}}^{0,R}(t,t') 
= \delta(t-t')\delta_{n,m}+\sum_{n'}\bar{h}_{\substack{n n'\\\kb_{\perp}}}G_{\substack{n',m\\\kb_{\perp}}}^{0,R}(t,t') 
\label{eq:tempnotimportant2}
\end{equation}

or expressed as a matrix equation,

\begin{equation}
        i\hb\frac{\partial}{\partial t}\Gb_{\kb_{\perp}}^{0,R}(t,t') 
= \delta(t-t')\mathbf{I}+\mathbf{H}^{\,}_{\kb_{\perp}}\Gb_{\kb_{\perp}}^{0,R}(t,t').
\label{eq:tempnotimportant3}
\end{equation}

Inserting \eqref{eq:fourier_inverse_def} and using $\delta(t-t') = \int^{+\infty}_{-\infty} e^{-i\omega(t-t')}\frac{d\omega}{2\pi}$, we find

\begin{equation}
        \left(\hb\omega\mathbf{I}-\mathbf{H}_{\kb_{\perp}}\right)
        \Gb_{\kb_{\perp}}^{0,R}(\omega)
= \mathbf{I}.
\label{eq:tempnotimportant4}
\end{equation}

The time-derivative of the lesser and greater Green's function is found in a similar fashion, 

\begin{equation}
    i\hb\frac{\partial}{\partial t}G_{\substack{n,m\\\kb_{\perp}}}^{0,\lessgtr}(t,t')
    = \sum_{n'}\bar{h}_{\substack{n n'\\\kb_{\perp}}}G_{\substack{n',m\\\kb_{\perp}}}^{0,\lessgtr}(t,t')
\end{equation}

or expressed as a matrix expression

\begin{equation}
    i\hb\frac{\partial}{\partial t}\Gb_{\kb_{\perp}}^{0,\lessgtr}(t,t')
    = \mathbf{H}_{\kb_{\perp}}\Gb_{\kb_{\perp}}^{0,\lessgtr}(t,t')
\end{equation}

which, after Fourier transformation, becomes
    
\begin{equation}
    \left(\hb\omega\mathbf{I}- \mathbf{H}_{\kb_{\perp}}\right)
    \Gb_{\\\kb_{\perp}}^{0,\lessgtr}(\omega) = \mathbf{0}.\label{eq:Glsgt_noninteract_2}
\end{equation}

Leaving out the $\omega$ and $\kb_{\perp}$ dependency in the notation for the sake of brevity and left-multiplying \eqref{eq:Langreth1} and \eqref{eq:Langreth2} with $\left(\hb\omega\mathbf{I}- \mathbf{H}\right)$, results in

\begin{align}
    \left(\hb\omega\mathbf{I}-\mathbf{H} \right)\Gb^{R} &= \mathbf{I}+\mathbf{U}\Gb^R+ \mathbf{\Sigma}^{s,R}\Gb^R,\label{eq:tempnotimportant5}\\
 \left(\hb\omega\mathbf{I}-\mathbf{H} \right)\Gb^{\lessgtr} &= \mathbf{U}\Gb^{\lessgtr}+\mathbf{\Sigma}^{s,R}\Gb^{\lessgtr} +\mathbf{\Sigma}^{s,\lessgtr}\Gb^A.\label{eq:tempnotimportant6}
\end{align}

Inserting \eqref{eq:Langreth1} into the second term on the right-hand side of \eqref{eq:tempnotimportant5}, we find

\begin{align}
\begin{split}
    \left(\hb\omega\mathbf{I}-\mathbf{H} \right)\Gb^{R} &= \mathbf{I}+\mathbf{U}\Gb^{0,R}+\mathbf{U}\Gb^{0,R}\mathbf{U}\Gb^{R}\\
    &\quad\ \ +\mathbf{U}\Gb^{0,R}\mathbf{\Sigma}^{s,R}\Gb^R
    +\mathbf{\Sigma}^{s,R}\Gb^R.
\end{split}
\end{align}

Note that the matrices so far contain row and column indices corresponding to all degrees of freedom in both the device itself and the left and right leads. As the two leads are infinite, this prohibits building and solving for these matrices in practice. However, to find the currents and charges in the device, only the Green's function elements with both row and column indices within the device are required. We therefore confine the matrices to their subblocks pertaining to the device itself. Note that this does not introduce complications for the multiplication with $\mathbf{H}$ and $\mathbf{\Sigma}^{s,R}$. $\mathbf{H}$ has a disconnected subblock corresponding to the device as the connections with the leads are contained in $\mathbf{U}$. $\mathbf{\Sigma}^{s,R}$ is non-zero only within the device as can be seen from its definition in \eqref{eq:final_Sigma_expression} and \eqref{eq:sigma_freq_phoneq_ret_full} and the fact that the electron-phonon subsystems are only connected within the device as demonstrated in Fig. \ref{fig:coupled_sys}. This is not the case for the term $\mathbf{U}\Gb^0$. $\Gb^0$ corresponds to the disconnected device and thus has disconnected blocks for the device and the leads. $\mathbf{U}$, on the other hand, only contains cross terms, connecting the device and the leads. Their matrix product will hence, also only contain cross terms connecting device to leads. When both row and column indices of the expressions are thus confined to the device, this cross term disappears. The same is true for the product $\mathbf{U}\Gb^0\mathbf{\Sigma}^{s,R}\Gb^R$ as $\mathbf{\Sigma}^{s,R}$ is only non-zero within the device. Leaving out these cross terms, we find

\begin{equation}    
    \left(\hb\omega\mathbf{I}-\mathbf{H} -\mathbf{\Sigma}^{s,R} -\mathbf{U}\Gb^{0,R}\mathbf{U}\right)\Gb^{R} = \mathbf{I}.
\end{equation}

The term $\mathbf{U}\Gb^{0,R}\mathbf{U}$ connects the device to the leads and is only non-zero in the subblocks corresponding to the leftmost and rightmost slab, i.e., slab $\mathbf{1}$ and $\mathbf{n}$,

\begin{align}
    \left(\mathbf{U}\Gb^{0,R}\mathbf{U}\right)_{\mathbf{1},\mathbf{1}} &= \mathbf{\bar{h}}_{\mathbf{1,0}}\,
\Gb^{0,R/A}_{\mathbf{0,0}}\,
\mathbf{\bar{h}}_{\mathbf{0,1}},\\
\left(\mathbf{U}\Gb^{0,R}\mathbf{U}\right)_{\mathbf{n},\mathbf{n}} &= \mathbf{\bar{h}}_{\mathbf{n,n+1}}\,
\Gb^{0,R/A}_{\mathbf{n+1,n+1}}\,
\mathbf{\bar{h}}_{\mathbf{n+1,n}}.
\end{align}

Defining

\begin{equation}
    \mathbf{\Sigma}^{l} = \mathbf{U}\Gb^{0}\mathbf{U},\label{eq:def_sigma_leads}
\end{equation}

we can merge these two self-energies with \eqref{eq:self_energy_merge} and obtain

\begin{equation}
\Gb^{R} = \left(\hb\omega\mathbf{I}-\mathbf{H} -\mathbf{\Sigma}^R\right)^{-1}.
\label{eq:Gr_final2}
\end{equation}

The only difference between \eqref{eq:Gr_final} and \eqref{eq:Gr_final2} is the presence of the convergence term $i\eta$, usually introduced to ensure the convergence of the Fourier transform \cite{Dattach8}. It can be readily achieved by adding a factor $e^{-\eta(t-t')}$ to the definition of the retarded Green's function in \eqref{eq:retarded_def} to make it absolutely integrable. A similar procedure is possible for the advanced Green's function using a factor $e^{\eta(t-t')}$, which will result in a convergence term $-i\eta$. We forgo this point and just add it retroactively.

For the lesser and greater Green's function, we insert \eqref{eq:Langreth2} into the first term on the right-hand side of \eqref{eq:tempnotimportant6},

\begin{align}
\begin{split}
    \mathbf{U}\Gb^{\lessgtr}  
    &= \mathbf{U}\Gb^{0,\gtrless}
    +
    \mathbf{U}\Gb^{0,\gtrless}\mathbf{U}
    \Gb^{A}
    +
     \mathbf{U}\Gb^{0,R}\mathbf{U}\Gb^{\gtrless}\\
     &\qquad\quad\ \ + 
    \mathbf{U}\Gb^{0,R}\mathbf{\Sigma}^{s,R}
    \Gb^{\gtrless}
    + 
    \mathbf{U}\Gb^{0,R}\mathbf{\Sigma}^{s,\gtrless}
    \Gb^{A}\\
    &\qquad\quad\ \ + 
    \mathbf{U}\Gb^{0,\gtrless}\mathbf{\Sigma}^{s,A}
    \Gb^{A}.
\end{split}
\label{eq:tempnotimportant7}
\end{align}

The first term and last three terms in \eqref{eq:tempnotimportant7} are cross terms, which become zero when the matrices are limited to their device subblocks. We thus have from \eqref{eq:tempnotimportant6},

\begin{align}
\begin{split}
&\left(\hb\omega\mathbf{I}-\mathbf{H} -\mathbf{\Sigma}^{s,R}-\mathbf{U}\Gb^{0,R}\mathbf{U}\right)\Gb^{\lessgtr} \\
&\qquad\qquad\qquad\qquad
= \left(\mathbf{\Sigma}^{s,\lessgtr} + \mathbf{U}\Gb^{0,\lessgtr}\mathbf{U}\right) \Gb^{A}.
\end{split}
\end{align}

Using \eqref{eq:def_sigma_leads} and \eqref{eq:self_energy_merge}, we obtain

\begin{align}
\left(\hb\omega\mathbf{I}-\mathbf{H} -\mathbf{\Sigma}^R\right)\Gb^{\lessgtr}
= \mathbf{\Sigma}^{\lessgtr} \Gb^{A}.
\end{align}

Finally, left-multiplication with $\Gb^R$ results in

\begin{equation}
    \Gb^{\lessgtr} = \Gb^R\mathbf{\Sigma}^{\lessgtr} \Gb^{A}\label{eq:Glsgt_final2},
\end{equation}

which is identical to \eqref{eq:Glsgt_final}. The expressions for the lead self-energies, \eqref{eq:Sigma_leads1}-\eqref{eq:Gamma_2}, are obtained by noting that the density of states in the non-connected leads is given by \cite{Dattasupp}

\begin{equation}
\bm{A}^0 = i \left(\Gb^{0,R}-\Gb^{0,A}\right),
\end{equation}

and the fact that the lesser (greater) Green's function is linked to the density of electrons (holes). In the non-connected leads, the electrons are in equilibrium and fill the density of states according to the Fermi-Dirac statistics function \eqref{eq:Fermi-Dirac}

\begin{align}
\Gb^{0,<}_{\mathbf{0},\mathbf{0}} &=  if_1\mathbf{A}^{0}_{\mathbf{0},\mathbf{0}},\\
\Gb^{0,<}_{\mathbf{n+1},\mathbf{n+1}} &=  if_2\mathbf{A}^{0}_{\mathbf{n+1},\mathbf{n+1}},\\
\Gb^{0,>}_{\mathbf{0},\mathbf{0}} &=  -i(1-f_1)\mathbf{A}^{0}_{\mathbf{0},\mathbf{0}},\\
\Gb^{0,>}_{\mathbf{n+1},\mathbf{n+1}} &=  -i(1-f_2)\mathbf{A}^{0}_{\mathbf{n+1},\mathbf{n+1}}.
\end{align}

We now perform the same procedure for the phonon Green's function. Making use of the commuting behavior of the phonon operators for the contour-ordering operator \cite{Fetterch3} and the definition of the retarded, advanced, lesser and greater Green's function \cite{Maciejko2007}, we obtain


\begin{widetext}    
\begin{align}
    D_{\substack{\nu,\mu\\\qb_{\perp}}}^{R}(t,t') &= -\frac{i}{\hb}\theta(t-t')\left<(\ab_{\qb_{\perp}\nu}(t)+\abc_{-\qb_{\perp}\nu}(t))(\abc_{\qb_{\perp}\mu}(t')+\ab_{-\qb_{\perp}\mu}(t'))
    \right.\nonumber\\&\qquad\qquad\qquad\qquad\left.
    -(\abc_{\qb_{\perp}\mu}(t')+\ab_{-\qb_{\perp}\mu}(t'))(\ab_{\qb_{\perp}\nu}(t)+\abc_{-\qb_{\perp}\nu}(t))\right>,\\
    D_{\substack{\nu,\mu\\\qb_{\perp}}}^{A}(t,t') &= \frac{i}{\hb}\theta(t'-t)\left<(\ab_{\qb_{\perp}\nu}(t)+\abc_{-\qb_{\perp}\nu}(t))(\abc_{\qb_{\perp}\mu}(t')+\ab_{-\qb_{\perp}\mu}(t'))
    \right.\nonumber\\&\qquad\qquad\qquad\qquad\left.
    -(\abc_{\qb_{\perp}\mu}(t')+\ab_{-\qb_{\perp}\mu}(t'))(\ab_{\qb_{\perp}\nu}(t)+\abc_{-\qb_{\perp}\nu}(t))\right>,\\
    D_{\substack{\nu,\mu\\\qb_{\perp}}}^{>}(t,t') &= -\frac{i}{\hb}\left<(\ab_{\qb_{\perp}\nu}(t)+\abc_{-\qb_{\perp}\nu}(t))(\abc_{\qb_{\perp}\mu}(t')+\ab_{-\qb_{\perp}\mu}(t'))\right>,\\
    D_{\substack{\nu,\mu\\\qb_{\perp}}}^{<}(t,t') &= - \frac{i}{\hb}\left<(\abc_{\qb_{\perp}\mu}(t')+\ab_{-\qb_{\perp}\mu}(t'))(\ab_{\qb_{\perp}\nu}(t)+\abc_{-\qb_{\perp}\nu}(t))\right>.
\end{align}

\clearpage
\end{widetext}

Using Langreth's theorem on \eqref{eq:exactD_as_convol_piD} and Fourier transforming results in expressions very similar to \eqref{eq:Langreth1} and \eqref{eq:Langreth2}, with all leftmost factors being non-interacting,

\begin{align}
\begin{split}
\Db_{\qb_{\perp}}^{R/A}(\omega) 
    =& \Db_{\qb_{\perp}}^{0,R/A}(\omega)+
    \Db_{\qb_{\perp}}^{0,R/A}(\omega)\mathbf{V}_{\qb_{\perp}}\Db_{\qb_{\perp}}^{R/A}(\omega)\\
    &\qquad+
    \Db_{\qb_{\perp}}^{0,R/A}(\omega)\mathbf{\Pi}_{\qb_{\perp}}^{s,R/A}
    (\omega)\Db_{\qb_{\perp}}^{R/A}(\omega),
    \end{split}
\label{eq:Langreth3}
\end{align}

\begin{align}
\begin{split}
\Db_{\qb_{\perp}}^{\gtrless}(\omega) 
    =& \Db_{\qb_{\perp}}^{0,\gtrless}(\omega)+
    \Db_{\qb_{\perp}}^{0,\gtrless}(\omega)\mathbf{V}_{\qb_{\perp}}
    \Db_{\qb_{\perp}}^{A}(\omega)\\
     &\qquad\quad\ \ +
     \Db_{\qb_{\perp}}^{0,R}(\omega)\mathbf{V}_{\qb_{\perp}}\Db^{\gtrless}_{\qb_{\perp}}(\omega)\\
     &\qquad\quad\ \ + 
    \Db_{\qb_{\perp}}^{0,R}(\omega)\mathbf{\Pi}_{\qb_{\perp}}^{s,R}
    (\omega)\Db_{\qb_{\perp}}^{\gtrless}(\omega)\\
    &\qquad\quad\ \ + 
    \Db_{\qb_{\perp}}^{0,R}(\omega)\mathbf{\Pi}_{\qb_{\perp}}^{s,\gtrless}
    (\omega)\Db_{\qb_{\perp}}^{A}(\omega)\\
    &\qquad\quad\ \ + 
    \Db^{0,\gtrless}_{\qb_{\perp}}(\omega)\mathbf{\Pi}_{\qb_{\perp}}^{s,A}
    (\omega)\Db_{\qb_{\perp}}^{A}(\omega).
\end{split}
\label{eq:Langreth4}
\end{align}

For the phonons, Fig. \ref{fig:coupled_sys} is only qualitatively correct. To obtain a Dyson equation of the form \eqref{eq:Langreth2}, and hence \eqref{eq:Langreth3} and \eqref{eq:Langreth4}, a slightly different subdivision in device and connecting terms is required. The full derivation is, however, cumbersome and has been moved to the Supporting Material \cite{Supp}. The final result is that the elements,

\begin{equation}
	\left(\mathbf{V}_{\qb_{\perp}}\right)_{\nu,\mu} = v_{\substack{\nu\mu\\\qb_{\perp}}},
\end{equation}

are the result of a perturbation Hamiltonian,

\begin{align}
\begin{split}
	\hat{H}^{ph}_P = \sum_{\substack{\nu\mu\\\qb_{\perp}}} &v_{\substack{\nu\mu\\\qb_{\perp}}} \big(\abc_{\qb_{\perp} \nu}\ab_{\qb_{\perp} \mu}\\
	&\quad+\frac{1}{2}\big(
	\abc_{\qb_{\perp} \nu}\abc_{-\qb_{\perp} \mu}
	+\ab_{-\qb_{\perp} \nu}\ab_{\qb_{\perp} \mu}
	\big)\big),
\end{split}
\end{align}

and that the non-interacting Hamiltonian for the phonons is given by

\begin{align}
	\hat{H}^{ph}_0 = 
	\sum_{\substack{\nu\mu\\\qb_{\perp}}}&
	(\bar{d}_{\substack{\nu\mu\\\qb_{\perp}}}-v_{\substack{\nu\mu\\\qb_{\perp}}})
	\abc_{\qb_{\perp}\nu}\ab_{\qb_{\perp}\mu}\\
	 &- \frac{v_{\qb_{\perp}\nu\mu}}{2}\big(
	\abc_{\qb_{\perp} \nu}\abc_{-\qb_{\perp} \mu}
	+\ab_{-\qb_{\perp} \nu}\ab_{\qb_{\perp} \mu}\big).\nonumber
\end{align}

The summation here extends over all matrix elements, also elements connecting the leads with the device. Note that despite the rather unintuitive form of $\hat{H}^{ph}_P$ and $\hat{H}^{ph}_0$, their sum is still equal to the ballistic phonon Hamiltonian in \eqref{eq:hamiltonian_mix}, implying that the material properties are unchanged.

The time-dependency of the non-interacting Green's functions in \eqref{eq:Langreth3} and \eqref{eq:Langreth4} can be evaluated explicitly. However, this time the time-dependency of both the annihilation and creation operator is required,

\begin{align}
\begin{split}
    i\hb \frac{\partial}{\partial t}\ab_{\qb_{\perp}\nu}(t) &= \left[\ab_{\qb_{\perp}\nu}(t),\hat{H}^{ph}_0\right] \\
     &= \sum_{\mu}(\bar{d}_{\substack{\nu \mu\\\qb_{\perp}}}-v_{\substack{\nu \mu\\\qb_{\perp}}})\ab_{\qb_{\perp} \mu}(t)\\
     &\qquad\quad-\frac{v_{\substack{\nu \mu\\\qb_{\perp}}}+v_{\substack{\mu \nu\\-\qb_{\perp}}}}{2}\abc_{-\qb_{\perp}\mu}(t)
\end{split}
\end{align}

and 

\begin{align}
\begin{split}
    i\hb \frac{\partial}{\partial t}\abc_{\qb_{\perp}\nu}(t) &= \left[\abc_{\qb_{\perp}\nu}(t),\hat{H}^{ph}_0\right] \\
     &= \sum_{\mu}-(\bar{d}_{\substack{\mu \nu\\\qb_{\perp}}}-v_{\substack{\mu \nu\\\qb_{\perp}}})\abc_{\qb_{\perp} \mu}(t)\\
     &\qquad\quad+\frac{v_{\substack{\mu \nu\\\qb_{\perp}}}+v_{\substack{\nu \mu\\-\qb_{\perp}}}}{2}\ab_{-\qb_{\perp}\mu}(t).
\end{split}
\end{align}

As $e_{\ka\al\nu,\qb} = e_{\ka\al\nu,-\qb}^*$ \cite{Giustino2007}, it follows from \eqref{eq:ph_rel2rec_d} that $\tilde{d}_{\substack{\ka\al\ka'\al'\\\Rb_p'-\Rb_p}} = \tilde{d}_{\substack{\ka'\al'\ka\al\\\Rb_p-\Rb_p'}}$. This, in combination with \eqref{eq:ph_mix2rel_d}, in turn implies that $\bar{d}_{\substack{\mu \nu\\\qb_{\perp}}}=\bar{d}_{\substack{\nu \mu\\-\qb_{\perp}}}$. We can choose to impose the same symmetry on $v_{\substack{\mu \nu\\\qb_{\perp}}}$. We thus have

\begin{align}
\begin{split}
	i\hb \frac{\partial}{\partial t}\ab_{\qb_{\perp}\nu}(t) &= \sum_{\mu}(\bar{d}_{\substack{\nu \mu\\\qb_{\perp}}}-v_{\substack{\nu \mu\\\qb_{\perp}}})\ab_{\qb_{\perp} \mu}(t)\\
	&\qquad\quad-v_{\substack{\nu \mu\\\qb_{\perp}}}\abc_{-\qb_{\perp}\mu}(t)
\end{split}
\end{align}

and 

\begin{align}
\begin{split}
	i\hb \frac{\partial}{\partial t}\abc_{-\qb_{\perp}\nu}(t) &= \sum_{\mu}-(\bar{d}_{\substack{\nu \mu\\\qb_{\perp}}}-v_{\substack{\nu \mu\\\qb_{\perp}}})\abc_{-\qb_{\perp} \mu}(t)\\
	&\qquad\quad+v_{\substack{\nu \mu\\\qb_{\perp}}}\ab_{\qb_{\perp}\mu}(t).
\end{split}
\end{align}

The sign change by the time-derivative of the creation operator has the effect that the first time-derivative of the retarded phonon Green's function is not readily expressed as a function of the Green's function itself. We therefore take the second time-derivative,

\begin{widetext}
\begin{align}
    \begin{split}
        -\hb^2\frac{\partial^2}{\partial t^2}D_{\substack{\nu,\mu\\\qb_{\perp}}}^{0,R}(t,t') =&
    i\hb\delta(t-t')'\left<(\ab_{\qb_{\perp}\nu}(t)+\abc_{-\qb_{\perp}\nu}(t))(\abc_{\qb_{\perp}\mu}(t')+\ab_{-\qb_{\perp}\mu}(t'))\right.\\
&\qquad\qquad\qquad\qquad
    \left.-(\abc_{\qb_{\perp}\mu}(t')+\ab_{-\qb_{\perp}\mu}(t'))(\ab_{\qb_{\perp}\nu}(t)+\abc_{-\qb_{\perp}\nu}(t))\right>\\
&
    +\sum_{\nu'}\bar{d}_{\substack{\nu\nu'\\\qb_{\perp}}}2\delta(t-t')\left<(\ab_{\qb_{\perp}\nu'}(t)-\abc_{-\qb_{\perp}\nu'}(t))(\abc_{\qb_{\perp}\mu}(t')+\ab_{-\qb_{\perp}\mu}(t'))\right.\\
&\qquad\qquad\qquad\qquad
    \left.-(\abc_{\qb_{\perp}\mu}(t')+\ab_{-\qb_{\perp}\mu}(t'))(\ab_{\qb_{\perp}\nu'}(t)-\abc_{-\qb_{\perp}\nu'}(t))\right>\\
&
    +\sum_{\nu'\nu''}
    (\bar{d}_{\substack{\nu\nu'\\\qb_{\perp}}}\bar{d}_{\substack{\nu'\nu''\\\qb_{\perp}}}
    -2\bar{d}_{\substack{\nu\nu'\\\qb_{\perp}}}v_{\substack{\nu'\nu''\\\qb_{\perp}}})
    D_{\substack{\nu'',\mu\\\qb_{\perp}}}^{0,R}(t,t').
    \end{split}
    \label{eq:second_deriv_D}
\end{align}

The time-derivative of the delta function $\delta(t-t')'$ is ill-defined except inside an integral. We therefore perform a Fourier transform by multiplying with $e^{i\omega(t-t')}$ and integrating over time. Integration by parts of the first term on the right-hand side of \eqref{eq:second_deriv_D} results in 

\begin{align}
\begin{split}
    &\int^{+\infty}_{-\infty}i\hb\delta(t-t')'\left<\ab_{\qb_{\perp}\nu}(t)+\abc_{-\qb_{\perp}\nu}(t))(\abc_{\qb_{\perp}\mu}(t')+\ab_{-\qb_{\perp}\mu}(t'))\right.\\
&\qquad\quad
    \left.-(\abc_{\qb_{\perp}\mu}(t')+\ab_{-\qb_{\perp}\mu}(t'))(\ab_{\qb_{\perp}\nu}(t)+\abc_{-\qb_{\perp}\nu}(t))\right> e^{i\omega(t-t')} d(t-t')\\
&\quad= -\sum_{\nu'}\bar{d}_{\substack{\nu\nu'\\\qb_{\perp}}}\left(
\left[\ab_{\qb_{\perp}\nu'}(t),\abc_{\qb_{\perp}\mu}(t)\right]
+\left[\ab_{-\qb_{\perp}\mu}(t),\abc_{-\qb_{\perp}\nu'}(t)\right]
\right)\\
&\quad\quad\qquad\quad +\hb\omega\left(
\left[\ab_{\qb_{\perp}\nu'}(t),\abc_{\qb_{\perp}\mu}(t)\right]
-\left[\ab_{-\qb_{\perp}\mu}(t),\abc_{-\qb_{\perp}\nu'}(t)\right]\right)\\
&\quad= -2\sum_{\nu'}\bar{d}_{\substack{\nu\nu'\\\qb_{\perp}}}\delta_{\nu'\mu} = -2\bar{d}_{\substack{\nu\mu\\\qb_{\perp}}}.
\end{split}
\end{align}
\end{widetext}

Similarly, it can be shown that the second term on the right-hand side of \eqref{eq:second_deriv_D} is equal to $4\bar{d}_{\substack{\nu\mu\\\qb_{\perp}}}$. In total we thus have

\begin{align}
\begin{split}
    \hb^2\omega^2 D_{\substack{\nu,\mu\\\qb_{\perp}}}^{0,R}(\omega) =
    2&\bar{d}_{\substack{\nu\mu\\\qb_{\perp}}}+\sum_{\nu'\nu''}\bar{d}_{\substack{\nu\nu'\\\qb_{\perp}}}\bar{d}_{\substack{\nu'\nu''\\\qb_{\perp}}}D_{\substack{\nu'',\mu\\\qb_{\perp}}}^{0,R}(\omega)\\
    &\quad-\sum_{\nu'\nu''}
    2\bar{d}_{\substack{\nu\nu'\\\qb_{\perp}}}v_{\substack{\nu'\nu''\\\qb_{\perp}}}D_{\substack{\nu'',\mu\\\qb_{\perp}}}^{0,R}(\omega).
\end{split}
\label{eq:Dr_noninteract_1}
\end{align}

The sum $\sum_{\nu'}\bar{d}_{\substack{\nu\nu'\\\qb_{\perp}}}\bar{d}_{\substack{\nu'\nu''\\\qb_{\perp}}}$ corresponds to a matrix squaring. It is readily verified from the definition in \eqref{eq:ph_mix2rel_d}, that the square of a Fourier transformed matrix is equal to the Fourier transform of the square. Additionally, \eqref{eq:ph_square_d2} showed that the square of $\mathbf{\tilde{d}}$ is equal to the interatomic force constants matrix. We can thus state that $\mathbf{\bar{d}}_{\qb_{\perp}}^2=\mathbf{K}^{tot}_{\qb_{\perp}}$ or, alternatively,

\begin{equation}
    \left(\hb^2\omega^2
    \mathbf{I}-\mathbf{K}^{tot}_{\qb_{\perp}}+2\mathbf{\bar{d}}_{\qb_{\perp}}\mathbf{V}_{\qb_{\perp}}\right) \Db_{\qb_{\perp}}^{0,R}(\omega) =
    2\mathbf{\bar{d}}_{\qb_{\perp}}.\label{eq:Dr_noninteract_2}
\end{equation}

In a similar fashion, it can be shown that

\begin{equation}
	\hb^2\frac{\partial^2}{\partial t^2}D_{\substack{\nu,\mu\\\qb_{\perp}}}^{0,\lessgtr}(t,t') =\sum_{\nu'\nu''}
	(\bar{d}_{\substack{\nu\nu'\\\qb_{\perp}}}\bar{d}_{\substack{\nu'\nu''\\\qb_{\perp}}}
	-2\bar{d}_{\substack{\nu\nu'\\\qb_{\perp}}}v_{\substack{\nu'\nu''\\\qb_{\perp}}})
	D_{\substack{\nu'',\mu\\\qb_{\perp}}}^{0,\lessgtr}(t,t'),\label{eq:Dlsgt_noninteract}
\end{equation}

and hence,

\begin{equation}
	\left(\hb^2\omega^2-\mathbf{K}^{tot}_{\qb_{\perp}}+2\mathbf{\bar{d}}_{\qb_{\perp}}\mathbf{V}_{\qb}\right)\Db_{\qb_{\perp}}^{0,\lessgtr}(\omega) =\mathbf{0}.\label{eq:Dlsgt_noninteract_2}
\end{equation}

The forms of \eqref{eq:Dr_noninteract_2} and \eqref{eq:Dlsgt_noninteract_2} differ significantly from \eqref{eq:tempnotimportant4} and \eqref{eq:Glsgt_noninteract_2}. First, as the superscript denotes, $\mathbf{K}^{tot}_{\qb_{\perp}}$ contains all matrix elements, including elements connecting the leads with the device, whereas $\mathbf{H}_{\kb_{\perp}}$ did not. Second, there is an extra term $2\mathbf{\bar{d}}_{\qb_{\perp}}\mathbf{V}_{\qb_{\perp}}$ and the right-hand side is not unity for \eqref{eq:Dr_noninteract_2}. These differences will prevent us from eliminating cross terms when we limit the degrees of freedom to the device. To resolve this, we propose the following transformations,

\begin{align}
	\Dbb_{\qb_{\perp}}=& \left(2\mathbf{\bar{d}}_{\qb_{\perp}}\right)^{-\frac{1}{2}}
	\Db_{\qb_{\perp}}
	\left(2\mathbf{\bar{d}}_{\qb_{\perp}}\right)^{-\frac{1}{2}},
	\label{eq:third_conv_D}\\
	\mathbf{\bar{V}}_{\qb_{\perp}}=& \left(2\mathbf{\bar{d}}_{\qb_{\perp}}\right)^{\frac{1}{2}}
	\mathbf{V}_{\qb_{\perp}}
	\left(2\mathbf{\bar{d}}_{\qb_{\perp}}\right)^{\frac{1}{2}},
	\label{eq:third_conv_V}\\
	\mathbf{\bar{\Pi}}_{\qb_{\perp}} =& \left(2\mathbf{\bar{d}}_{\qb_{\perp}}\right)^{\frac{1}{2}}
	\mathbf{\Pi}_{\qb_{\perp}}
	\left(2\mathbf{\bar{d}}_{\qb_{\perp}}\right)^{\frac{1}{2}}.
	\label{eq:third_conv_Pi}
\end{align}

Applying \eqref{eq:third_conv_D}-\eqref{eq:third_conv_V} to \eqref{eq:Dr_noninteract_2} and \eqref{eq:Dlsgt_noninteract_2}, we obtain

\begin{align}
	\left(\hb^2\omega^2
	\mathbf{I}-\mathbf{K}^{tot}_{\qb_{\perp}}+\mathbf{\bar{V}}_{\qb_{\perp}}\right) \Db_{\qb_{\perp}}^{0,R}(\omega) &=
	\mathbf{I},\label{eq:Dr_noninteract_3}\\
	\left(\hb^2\omega^2
	\mathbf{I}-\mathbf{K}^{tot}_{\qb_{\perp}}+\mathbf{\bar{V}}_{\qb_{\perp}}\right) \Db_{\qb_{\perp}}^{0,\lessgtr}(\omega) &=
	\mathbf{0}.\label{eq:Dlsgt_noninteract_3}
\end{align}

We can now choose the matrix elements $v_{\substack{\nu\mu\\\qb_{\perp}}}$ such that $\mathbf{\bar{V}}_{\qb_{\perp}}$ contains the elements of $\mathbf{K}^{tot}_{\qb_{\perp}}$ connecting the leads with the device. We can then define $\mathbf{K}_{\qb_{\perp}}=\mathbf{K}^{tot}_{\qb_{\perp}}-\mathbf{\bar{V}}_{\qb_{\perp}}$, which has the same block matrix structure as $\mathbf{H}_{\kb_{\perp}}$. \eqref{eq:Dr_noninteract_3} and \eqref{eq:Dlsgt_noninteract_3} then have a formally equivalent structure to \eqref{eq:tempnotimportant4} and \eqref{eq:Glsgt_noninteract_2}. Additionally, the transformations \eqref{eq:third_conv_D}-\eqref{eq:third_conv_Pi} leave the Dyson equations \eqref{eq:Langreth3} and \eqref{eq:Langreth4} unchanged. The remainder of the derivation is thus identical to the electron case. Leaving out the $\omega$ and $\kb_{\perp}$ dependency in the notation, defining

\begin{equation}
	\mathbf{\bar{\Pi}}^{l} = \mathbf{\bar{V}}\Dbb^{0}\mathbf{\bar{V}}\label{eq:def_sigma_leads2}
\end{equation}

and merging the self-energies with \eqref{eq:self_energy_merge2}, we obtain

\begin{align} 
	\Dbb^{R} &=    
	\left(\hb^2\omega^2
	\mathbf{I}-\mathbf{K}-\mathbf{\bar{\Pi}}\right)^{-1},
	\label{eq:Dr_final2}\\
    \Dbb^{\lessgtr} &= \Dbb^{R}\mathbf{\bar{\Pi}}^{\lessgtr}\Dbb^{A}.
\label{eq:Dlsgt_final2} 
\end{align}

The expressions in \eqref{eq:Dr_final} and \eqref{eq:Dr_final2} and \eqref{eq:Dlsgt_final} and \eqref{eq:Dlsgt_final2} differ only by the presence of the convergence term $i\eta$, which we will again add retroactively, and by the bar notation. It is hereby shown that the expressions for the phonon Green's function usually found in literature and derived from semi-classical principles do not pertain to the Green's functions defined in \eqref{eq:ph_cont_D}, but that additional transformations, \eqref{eq:third_conv_D} and \eqref{eq:third_conv_Pi}, are required. As will be shown in Appendix \ref{app:self_energies}, the transformations \eqref{eq:third_conv_D} and \eqref{eq:third_conv_Pi} will result in a modification of the electron-phonon matrix elements in \eqref{eq:mix2rel_M}.

\section{Self-energy expressions}\label{app:self_energies}

We provide a derivation of the self-energy expressions related to electron-phonon scattering in \eqref{eq:final_Sigma_expression} and \eqref{eq:final_Pi_expression}. The self-energy expressions due to the leads were obtained in Appendix \ref{app:Greens_functions} so from here on out the influence of the leads is neglected. It can be shown that when interactions are introduced, the electron Green's function becomes \cite{Fetterch3,Maciejko2007}

\begin{equation}
    iG_{n,m}(t,t') = \frac{1}{\hb}\braket{T_c\left[e^{\frac{-i}{\hb}\int_C \hat{H}_{I}(t_1) dt_1}\cb_{n}(t)\cbc_{m}(t')\right]},
    \label{eq:Greens_func_int}
\end{equation}

where $\hat{H}_{I}$ is defined in Section \ref{secsec:Hamiltonian} and the integral is taken over the two-branch contour, defined in Section \ref{secsec:Greensfunctions}. Averaging here is done according to the occupation of the non-interacting and non-contacted and, hence, one-particle states described in Section \ref{secsec:Greensfunctions} and the operators are described in the interaction picture. Note that an exact treatment actually requires a three-branch contour and that $\hat{H}_{I}$ requires a separate definition on this third branch \cite{Wagner1991}. The influence of this third branch is the incorporation of correlation effects after switching on the interactions and contacts. However, these correlations can usually be neglected in steady state \cite{Maciejko2007}. Limiting \eqref{eq:Greens_func_int} to a second order expansion results in 
\begin{align}
    &iG_{n,m}(t,t') = \frac{1}{\hb}\braket{T_c\left[\cb_{n}(t)\cbc_{m}(t')\right]} \nonumber\\
    &\quad+ \braket{T_c\left[\frac{-i}{\hb^2}\int_C \hat{H}_{I}(t_1) dt_1\cb_{n}(t)\cbc_{m}(t')\right]} \\
    &\quad+ \braket{T_c\left[\frac{-1}{2\hb^3}\int_C\int_C \hat{H}_{I}(t_1)\hat{H}_{I}(t_2) dt_1dt_2\cb_{n}(t)\cbc_{m}(t')\right]}.\nonumber
\label{eq:el_G_perturb}
\end{align}

The first term is just the non-interacting Green's function $iG_{n,m}^0$ defined in \eqref{eq:exactG_as_convol_sigG}. The second term can be neglected due to an odd number of phonon creation or annihilation operators. Since averaging is done over non-interacting one-particle states and the operators are non-interacting operators, Wick's theorem can be used to write the third term as \cite{Wagner1991} 

\begin{widetext}
\begin{align}
  \frac{-1}{2\hb^3N_{\perp}}
  \sum_{\substack{m_1 n_1\nu_1\\\kb_{\perp,1}\qb_{\perp,1}}}
    \bar{g}_{\substack{m_1n_1\nu_1\\\kb_{\perp,1}\qb_{\perp,1}}}&
    \sum_{\substack{m_2 n_2\nu_2\\\kb_{\perp,2}\qb_{\perp,2}}}
    \bar{g}_{\substack{m_2n_2\nu_2\\\kb_{\perp,2}\qb_{\perp,2}}}\int_C dt_1\int_C dt_2\braket{(\ab_{\qb_{\perp,1}\nu_1}(t_1)+\abc_{-\qb_{\perp,1}\nu_1}(t_1))(\ab_{\qb_{\perp,2}\nu_2}(t_2)+\abc_{-\qb_{\perp,2}\nu_2}(t_2))}\nonumber\\
\big(&\braket{
    \cb_{\kb_{\perp,1} n_1}(t_1)
    \cbc_{\substack{\kb_{\perp,1}+\qb_{\perp,1}\\m_1}}(t_1)}
    \braket{\cb_{\kb_{\perp,2} n_2}(t_2)
    \cbc_{\substack{\kb_{\perp,2}+\qb_{\perp,2}\\m_2}}(t_2)}
    \braket{\cb_{\kb_{\perp}n}(t)\cbc_{\kb_{\perp}m}(t')}
    \nonumber\\
-&\braket{
    \cb_{\kb_{\perp,2} n_2}(t_2)
    \cbc_{\substack{\kb_{\perp,1}+\qb_{\perp,1}\\m_1}}(t_1)}
    \braket{\cb_{\kb_{\perp,1} n_1}(t_1)}
    \cbc_{\substack{\kb_{\perp,2}+\qb_{\perp,2}\\m_2}}(t_2)
    \braket{\cb_{\kb_{\perp}n}(t)\cbc_{\kb_{\perp}m}(t')}
    \nonumber\\
-&\braket{
    \cb_{\kb_{\perp}n}(t)
    \cbc_{\substack{\kb_{\perp,1}+\qb_{\perp,1}\\m_1}}(t_1)}
    \braket{\cb_{\kb_{\perp,2} n_2}(t_2)
    \cbc_{\substack{\kb_{\perp,2}+\qb_{\perp,2}\\m_2}}(t_2)}
    \braket{\cb_{\kb_{\perp,1} n_1}(t_1)\cbc_{\kb_{\perp}m}(t')}
    \\
-&\braket{
    \cb_{\kb_{\perp,1} n_1}(t_1)
    \cbc_{\substack{\kb_{\perp,1}+\qb_{\perp,1}\\m_1}}(t_1)}
    \braket{\cb_{\kb_{\perp}n}(t)
    \cbc_{\substack{\kb_{\perp,2}+\qb_{\perp,2}\\m_2}}(t_2)}
    \braket{\cb_{\kb_{\perp,2} n_2}(t_2)\cbc_{\kb_{\perp}m}(t')}
    \nonumber\\
+&\braket{
    \cb_{\kb_{\perp}n}(t)
    \cbc_{\substack{\kb_{\perp,1}+\qb_{\perp,1}\\m_1}}(t_1)}
    \braket{\cb_{\kb_{\perp,1} n_1}(t_1)
    \cbc_{\substack{\kb_{\perp,2}+\qb_{\perp,2}\\m_2}}(t_2)}
    \braket{\cb_{\kb_{\perp,2} n_2}(t_2)
    \cbc_{\kb_{\perp}m}(t')}
    \nonumber\\
+&\braket{
    \cb_{\kb_{\perp}n}(t)
    \cbc_{\substack{\kb_{\perp,2}+\qb_{\perp,2}\\m_2}}(t_2)}
    \braket{\cb_{\kb_{\perp,2} n_2}(t_2)
    \cbc_{\substack{\kb_{\perp,1}+\qb_{\perp,1}\\m_1}}(t_1)}
    \braket{\cb_{\kb_{\perp,1} n_1}(t_1)
    \cbc_{\kb_{\perp}m}(t')}\big).\nonumber
\end{align}

The first two terms correspond to disconnected diagrams and can be neglected. The third and fourth term result in an exchange of zero energy and momentum and can thus also be neglected. The last two terms are identical except for an exchange of the indices and thus cancel the factor of 2 in the denominator. Applying momentum conservation and substituting in \eqref{eq:el_G_perturb} results in

\begin{align}
\begin{split}
    G_{\substack{n,m\\\kb_{\perp}}}(t,t') &= G^0_{\substack{n,m\\\kb_{\perp}}}(t,t')+ \sum_{\substack{m_1n_1m_2n_2\\\nu_1\nu_2\qb_{\perp}}} \bar{g}_{\substack{m_1n_1\nu_1\\\kb_{\perp}-\qb_{\perp},\qb_{\perp}}}
    \bar{g}_{\substack{m_2n_2\nu_2\\\kb_{\perp},-\qb_{\perp}}}\\
    &\qquad\qquad\qquad\int_C dt_1\int_C dt_2 \frac{i\hb}{N_{\perp}}G^0_{\substack{n,m_1\\\kb_{\perp}}}(t,t_1)
    G^0_{\substack{n_1,m_2\\\kb_{\perp}-\qb_{\perp}}}(t_1,t_2)
    D^0_{\substack{\nu_1,\nu_2\\\qb_{\perp}}}(t_1,t_2)
    G^0_{\substack{n_2,m\\\kb_{\perp}}}(t_2,t').
    \end{split}
\end{align}

Note that the summation over indices corresponds to a matrix multiplication. Higher-order terms of the perturbation expansion in \eqref{eq:el_G_perturb} can be obtained by turning this into a Dyson equation \cite{Fetterch3}. Replacing all but the leftmost non-interacting Green's functions with interacting Green's functions, we obtain \eqref{eq:exactG_as_convol_sigG} with

\begin{align}
\left(\mathbf{\Sigma}^{s}_{\kb_{\perp}}(t_1,t_2)\right)_{n,m}
&=
  \frac{i\hb}{N_{\perp}}
  \sum_{\substack{n_1m_2\\\nu_1\nu_2\qb_{\perp}}}
  \bar{g}_{\substack{m_1n_1\nu_1\\\kb_{\perp}-\qb_{\perp},\qb_{\perp}}}
  G_{\substack{n_1,m_2\\\kb_{\perp}-\qb_{\perp}}}(t_1,t_2)
  D_{\substack{\nu_1,\nu_2\\\qb_{\perp}}}(t_1,t_2)
  \bar{g}_{\substack{m_2n_2\nu_2\\\kb_{\perp},-\qb_{\perp}}}.
\end{align}

We are interested in an expression in terms of $\tilde{\Db}$, which requires an additional transformation according to \eqref{eq:third_conv_D}. This introduces an extra factor of 2. Additionally, it can be shown from \eqref{eq:ph_rel2rec_d}, \eqref{eq:rel2rec_M}, \eqref{eq:ph_mix2rel_d}, \eqref{eq:mix2rel_M} and \eqref{eq:rel2rec_M2} that

\begin{align}
    \sum_\nu \bar{g}_{\substack{nm\nu\\\kb_{\perp}\qb_{\perp}}}(\mathbf{\bar{d}_{\qb_{\perp}}}^{\frac{1}{2}})_{\nu\nu'} &= \bar{\bar{g}}_{\substack{nm\nu'\\\kb_{\perp}\qb_{\perp}}},\label{eq:third_conv_M1}\\
    \sum_\nu \bar{g}_{\substack{nm\nu\\\kb_{\perp},-\qb_{\perp}}}(\mathbf{\bar{d}_{\qb_{\perp}}}^{\frac{1}{2}})_{\nu'\nu} &= \bar{\bar{g}}_{\substack{nm\nu'\\\kb_{\perp},-\qb_{\perp}}}.\label{eq:third_conv_M2}
\end{align}

This, together with \eqref{eq:def_M_as_g}, results in

\begin{equation}
    \mathbf{\Sigma}^{s}_{\kb_{\perp}}(t_1,t_2)
=
  \frac{2i\hb}{N_{\perp}}
  \sum_{\nu\mu\qb_{\perp}}
  \mathbf{M}^{\nu}_{\kb_{\perp}-\qb_{\perp},\qb_{\perp}}
  \Gb_{\kb_{\perp}-\qb_{\perp}}(t_1,t_2)
  \bar{D}_{\substack{\nu,\mu\\\qb_{\perp}}}(t_1,t_2)
  \mathbf{M}^{\mu}_{\kb_{\perp},-\qb_{\perp}}.
\end{equation}

The lesser and greater Green's function can be extracted from the contour-ordered Green's function by confining the time arguments to specific branches. The same is true for the self-energy. Fourier transformation according to \eqref{eq:fourier_inverse_def} then gives

\begin{equation}
\mathbf{\Sigma}^{s,\lessgtr}_{\kb_{\perp}}(\omega)
= \int_{-\infty}^{\infty}
  \frac{2i\hb}{N_{\perp}}
  \sum_{\nu\mu\qb_{\perp}}
  \mathbf{M}^{\nu}_{\kb_{\perp}-\qb_{\perp},\qb_{\perp}}
  \Gb^{\lessgtr}_{\kb_{\perp}-\qb_{\perp}}(\omega-\omega')
  \bar{D}^{\lessgtr}_{\substack{\nu,\mu\\\qb_{\perp}}}(\omega')
  \mathbf{M}^{\mu}_{\kb_{\perp},-\qb_{\perp}}
  \frac{d\omega'}{2\pi}.
\end{equation}

Finally, we drop the bar notation on $\bar{D}_{\nu,\mu}$ and use the fact that

\begin{equation}
    D^{\lessgtr}_{\substack{\nu,\mu\\\qb_{\perp}}}(\omega) = 
    D^{\gtrless}_{\substack{\mu,\nu\\-\qb_{\perp}}}(-\omega)
\end{equation}

to confine the time integral to the positive axis, obtaining \eqref{eq:final_Sigma_expression}.

Similarly, we can find a perturbation expansion of the phonon Green's function,

\begin{align}
\begin{split}
    iD_{\substack{\nu,\mu\\\qb_{\perp}}}&(t,t') = \frac{1}{\hb}\braket{T_c\left[(\ab_{\qb_{\perp}\nu}(t)+\abc_{-\qb_{\perp}\nu}(t))(\abc_{\qb_{\perp}\mu}(t')+\ab_{-\qb_{\perp}\mu}(t'))\right]} \\
&+ 
    \braket{T_c\left[\frac{-i}{\hb^2}\int_C \hat{H}_{I}(t_1) dt_1(\ab_{\qb_{\perp}\nu}(t)+\abc_{-\qb_{\perp}\nu}(t))(\abc_{\qb_{\perp}\mu}(t')+\ab_{-\qb_{\perp}\mu}(t'))\right]} \\
&+ 
    \big<T_c\left[\frac{-1}{2\hb^3}\int_C\int_C \hat{H}_{I}(t_1)\hat{H}_{I}(t_2) dt_1dt_2
    (\ab_{\qb_{\perp}\nu}(t)+\abc_{-\qb_{\perp}\nu}(t))(\abc_{\qb_{\perp}\mu}(t')+\ab_{-\qb_{\perp}\mu}(t'))\right]\big>.
\end{split}
\label{eq:ph_D_perturb}
\end{align}

The first term is just the non-interacting phonon Green's function and the second term is zero due to having an odd number of creation or annihilation operators. The third term can be written as the following Wick decomposition, 

\begin{align}
    \frac{-1}{2\hb^3N_{\perp}}&
    \sum_{\substack{m_1 n_1 \nu_1\\\kb_{\perp,1}\qb_{\perp,1}}}
    \bar{g}_{\substack{m_1n_1\nu_1\\\kb_{\perp,1}\qb_{\perp,1}}}\sum_{\substack{m_2 n_2 \nu_2\\\kb_{\perp,2}\qb_{\perp,2}}}
    \bar{g}_{\substack{m_2n_2\nu_2\\\kb_{\perp,2}\qb_{\perp,2}}}\int_C dt_1\int_C dt_2\nonumber\\
&\big(\braket{
    \cb_{\kb_{\perp,1} n_1}(t_1)
    \cbc_{\substack{\kb_{\perp,1}+\qb_{\perp,1}\\m_1}}(t_1)}
    \braket{\cb_{\kb_{\perp,2} n_2}(t_2)
    \cbc_{\substack{\kb_{\perp,2}+\qb_{\perp,2}\\m_2}}(t_2)}\\
    &\qquad\qquad\qquad\qquad\qquad\qquad\qquad\qquad-\braket{
    \cb_{\kb_{\perp,1} n_1}(t_1)
    \cbc_{\substack{\kb_{\perp,2}+\qb_{\perp,2}\\m_2}}(t_2)}
    \braket{\cb_{\kb_{\perp,2} n_2}(t_2)
    \cbc_{\substack{\kb_{\perp,1}+\qb_{\perp,1}\\m_1}}(t_1)}\big)\nonumber\\
\big(&\braket{
    (\ah_{\qb_{\perp,1}\nu_1}(t_1)+\ac_{-\qb_{\perp,1}\nu_1}(t_1))
    (\ah_{\qb_{\perp,2}\nu_2}(t_2)+\ac_{-\qb_{\perp,2}\nu_2}(t_2))}\braket{
    (\ab_{\qb_{\perp}\nu}(t)+\abc_{-\qb_{\perp}\nu}(t))
    (\abc_{\qb_{\perp}\mu}(t')+\ab_{-\qb_{\perp}\mu}(t'))}\nonumber\\
+&\braket{
    (\ab_{\qb_{\perp}\nu}(t)+\abc_{-\qb_{\perp}\nu}(t))
    (\ah_{\qb_{\perp,1}\nu_1}(t_1)+\ac_{-\qb_{\perp,1}\nu_1}(t_1))}\braket{
    (\ah_{\qb_{\perp,2}\nu_2}(t_2)+\ac_{-\qb_{\perp,2}\nu_2}(t_2))
    (\abc_{\qb_{\perp}\mu}(t')+\ab_{-\qb_{\perp}\mu}(t'))}\nonumber\\
+&\braket{
    (\ab_{\qb_{\perp}\nu}(t)+\abc_{-\qb_{\perp}\nu}(t))
    (\ah_{\qb_{\perp,2}\nu_2}(t_2)+\ac_{-\qb_{\perp,2}\nu_2}(t_2))}\braket{
    (\ah_{\qb_{\perp,1}\nu_1}(t_1)+\ac_{-\qb_{\perp,1}\nu_1}(t_1))
    (\abc_{\qb_{\perp}\mu}(t')+\ab_{-\qb_{\perp}\mu}(t'))}\big)\nonumber.
\end{align}

The first term in the electron part corresponds to an exchange of zero energy and momentum and will be neglected. The first term of the phonon part corresponds to a disconnected diagram and will be neglected as well. The last two terms of the phonon part are equal except for an exchange of indices and can thus be summed to compensate for the factor of 2 in the denominator. Applying momentum conservation, adding a factor due to spin degeneracy and substituting in \eqref{eq:ph_D_perturb} results in

\begin{align}
\begin{split}
    D_{\substack{\nu,\mu\\\qb_{\perp}}}(t,t') &= D^0_{\substack{\nu,\mu\\\qb_{\perp}}}(t,t')
-\frac{n_si\hb}{N_{\perp}}
    \sum_{\substack{m_1n_1\nu_1\\m_2n_2\nu_2\\\kb_{\perp}}}
    \bar{g}_{\substack{m_1n_1\nu_1\\\kb_{\perp},-\qb_{\perp}}}
    \bar{g}_{\substack{m_2n_2\nu_2\\\kb_{\perp}-\qb_{\perp},\qb_{\perp}}}\\
    &\qquad\qquad\qquad\int_C dt_1\int_C dt_2
    G^0_{\substack{n_1,m_2\\\kb_{\perp}}}(t_1,t_2)
    G^0_{\substack{n_2,m_1\\\kb_{\perp}-\qb_{\perp}}}(t_2,t_1)
    D^0_{\substack{\nu,\nu_1\\\qb_{\perp}}}(t,t_1)
    D^0_{\substack{\nu_2,\mu\\\qb_{\perp}}}(t_2,t').
    \end{split}
\end{align}

Converting to a Dyson equation to include higher-order perturbation terms, we obtain \eqref{eq:exactD_as_convol_piD} with

\begin{equation}
    \left(\mathbf{\Pi}^{s}_{\qb_{\perp}}(t_1,t_2)\right)_{\nu,\mu} =
-\frac{n_si\hb}{N_{\perp}}
  \sum_{\substack{m_1n_1\\m_2n_2\kb_{\perp}}}
    \bar{g}_{\substack{m_1n_1\nu\\\kb_{\perp},-\qb_{\perp}}}
     G_{\substack{n_1,m_2\\\kb_{\perp}}}(t_1,t_2)
    \bar{g}_{\substack{m_2n_2\mu\\\kb_{\perp}-\qb_{\perp},\qb_{\perp}}}
    G_{\substack{n_2,m_1\\\kb_{\perp}-\qb_{\perp}}}(t_2,t_1).
\end{equation}

We want an expression for $\mathbf{\bar{\Pi}}$ however. Using \eqref{eq:third_conv_Pi}, \eqref{eq:third_conv_M1}, \eqref{eq:third_conv_M2} and \eqref{eq:def_M_as_g}, it can be shown that

\begin{equation}
        \left(\mathbf{\bar{\Pi}}^{s}_{\qb_{\perp}}(t_1,t_2)\right)_{\nu,\mu} =
-\frac{2n_si\hb}{N_{\perp}}
  \sum_{\kb_{\perp}}
  \Tr
  \Big(
    \mathbf{M}^{\nu}_{\kb_{\perp},-\qb_{\perp}}
     \Gb_{\kb_{\perp}}(t_1,t_2)
    \mathbf{M}^{\mu}_{\kb_{\perp}-\qb_{\perp},\qb_{\perp}}
    \Gb_{\kb_{\perp}-\qb_{\perp}}(t_2,t_1)\Big).
\end{equation}

Confining the time arguments to specific branches and Fourier transforming according to \eqref{eq:fourier_inverse_def} then gives

\begin{equation}
\left(\mathbf{\bar{\Pi}}^{s,\lessgtr}_{\qb_{\perp}}(\omega)\right)_{\nu,\mu} =\int^{+\infty}_{-\infty}
-\frac{2n_si\hb}{N_{\perp}}
  \sum_{\kb_{\perp}}
    \Tr\bigg(
    \mathbf{M}^{\nu}_{\kb_{\perp},-\qb_{\perp}}
    \Gb_{\kb_{\perp}}^{\lessgtr}(\omega')
    \mathbf{M}^{\mu}_{\kb_{\perp}-\qb_{\perp},\qb_{\perp}}
    \Gb_{\substack{\kb_{\perp}-\qb_{\perp}}}^{\gtrless}(\omega'-\omega)\bigg)\frac{d\omega'}{2\pi}.
\end{equation}

Dropping the bar notation on $\left(\mathbf{\bar{\Pi}}_{\qb_{\perp}}(t_1,t_2)\right)_{\nu,\mu}$ results in \eqref{eq:final_Pi_expression}.

\clearpage
\end{widetext}

\section{Error estimates for the FFT-based self-energy computation}\label{app:error_testing}

\subsection{The integration error}
\begin{figure}[ht]
    \centering
    \includegraphics[width=\linewidth]{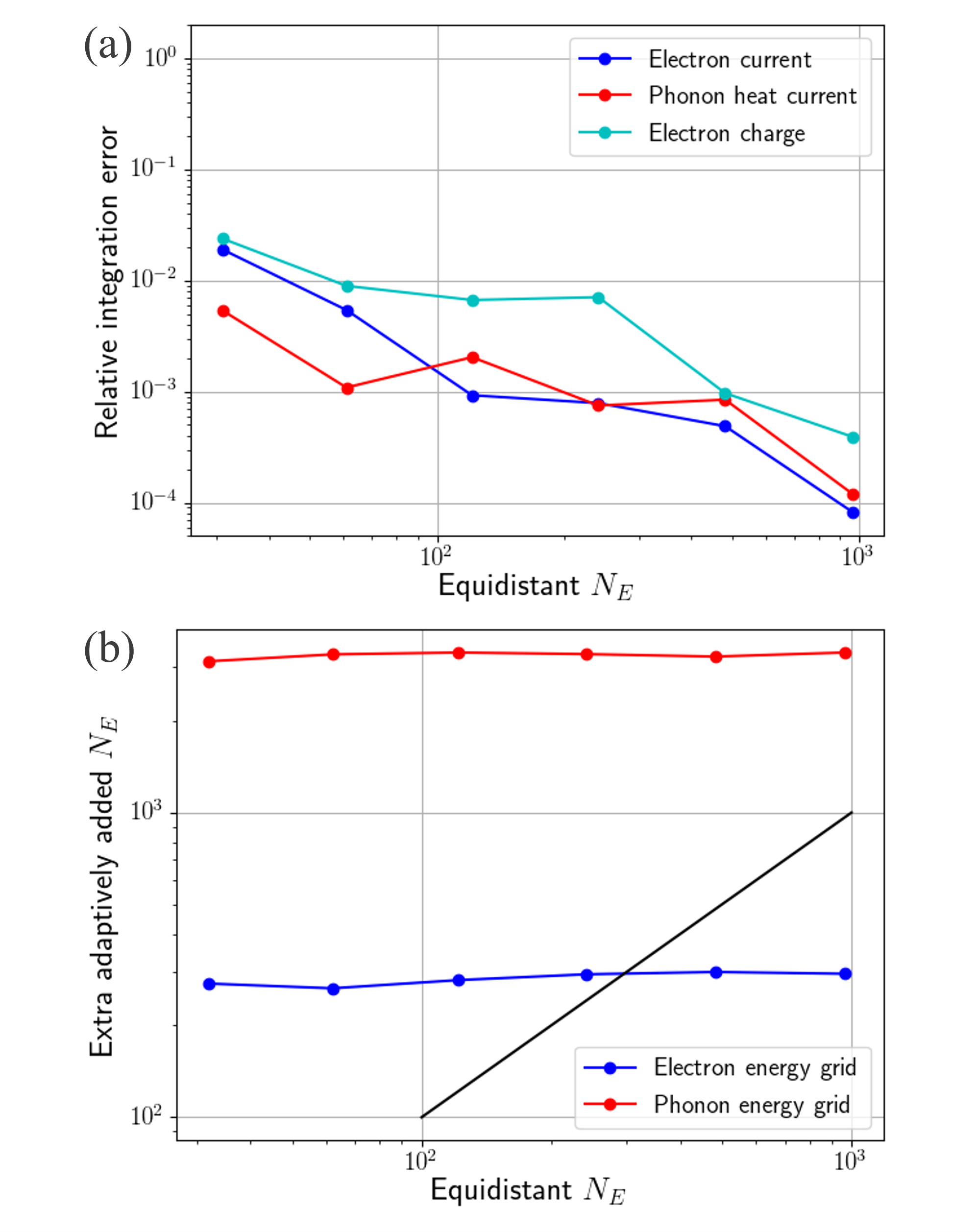}
    \caption{(a) The relative integration error on the electron current, phonon heat current and charge density for a ballistic simulation of the system in Section \ref{secsec:device_simulation} as a function of the number of equidistant grid energy points. The error is obtained as the difference with a fine-initial-grid simulation with 2000 initial grid points. A temperature difference of \SI{0.1}{\kelvin} was applied to provide a net heat current through the device. (b) The number of adaptively added points as a function of the number of equidistant initial grid points, for both electrons and phonons. The black line denotes the cross-over line where the adaptive grid points dominate the computational cost.}
    \label{fig:integration_err}
\end{figure}

The grid refinement strategy and corresponding integration error are illustrated by the results shown in Fig. \ref{fig:integration_err}. First, we obtained a potential profile within the device described in Section \ref{secsec:device_simulation} with a fine initial energy grid. Then, the macroscopic parameters of interest were obtained with a single ballistic non-self-consistent iteration for several initial grid sizes. Fig. \ref{fig:integration_err} (a) shows the relative difference compared to a single ballistic iteration with a fine initial grid of 2000 energy points for both the electrons and phonons as an estimate of the integration error. Fig. \ref{fig:integration_err} (b) shows the number of adaptively added grid points as a function of the initial grid size. Note that significantly more points are added to the phonon energy grid as the full phonon dispersion needs to be integrated. For the electrons only the bottom of the conduction band is of interest. Additionally, it can be seen that a minimum of initial grid points is required to achieve accurate integration despite the adaptive grid. This can be understood from Fig. \ref{fig:equidistant_mesh_err1} (a) where the second peak is nearly missed by the initial grid. The adaptive grid will not refine itself around features that are too fine to be captured by the initial grid. From Fig. \ref{fig:integration_err} (b) we can see that the number of adaptively added grid points increases slightly up to initial grid sizes of 100-250 points. For higher number of initial grid sizes, the number of adaptively added points does not increase or even decreases as some of the adaptively added points are now already introduced by the initial grid. Fig. \ref{fig:integration_err} (a) shows that an initial grid size of $\sim$100 points is sufficient to reach the 1\% error threshold on the macroscopic properties of interest. Further refinement of the initial grid does reduce the error further, but not by introducing more adaptive grid points as the error threshold is already reached.

\subsection{The energy mixing error}

\begin{figure}[ht]
    \centering
    \includegraphics[width=\linewidth]{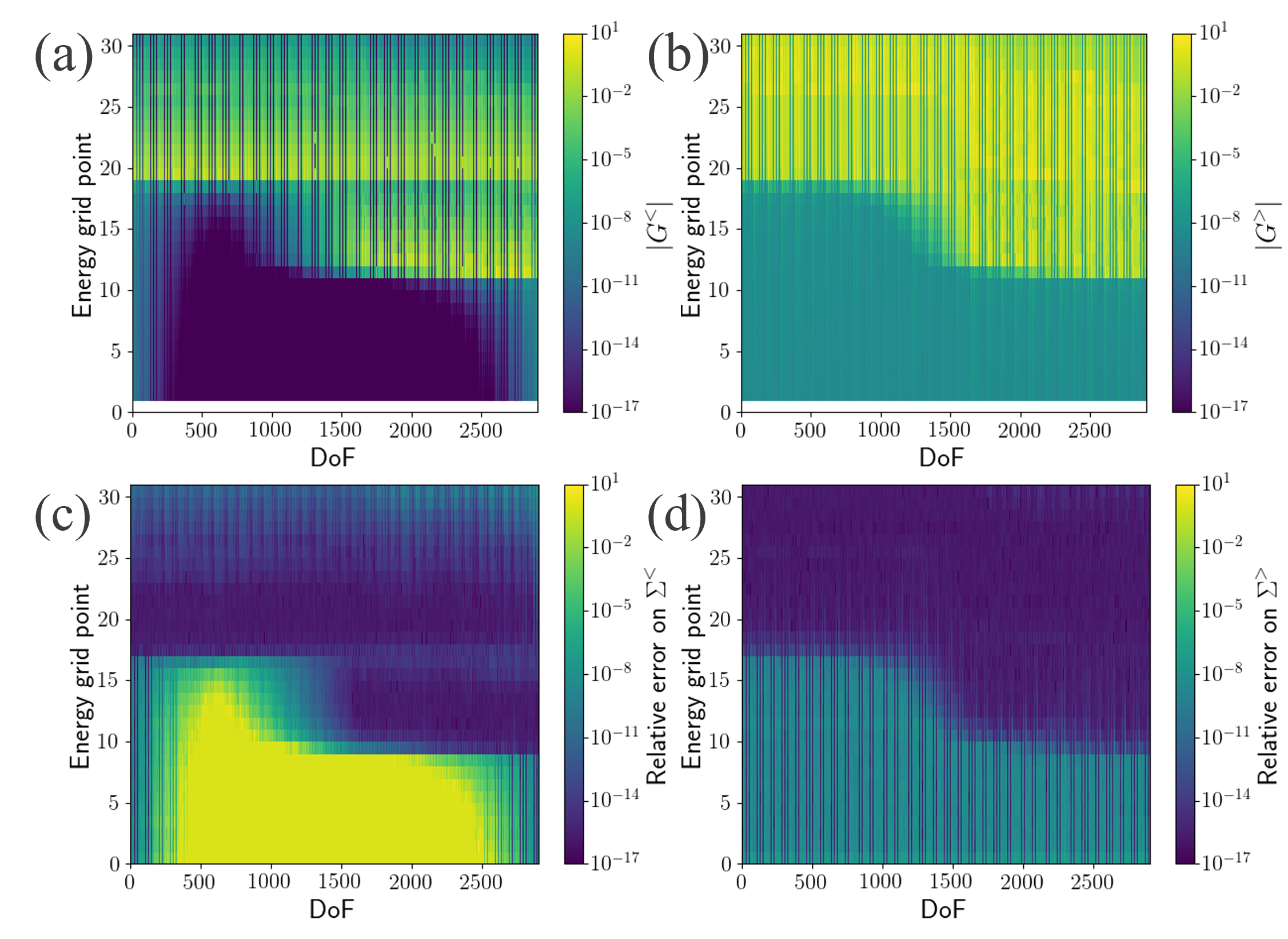}
    \caption{Green's function and self-energy results after a single self-consistent Born iteration for the system in Section \ref{secsec:device_simulation}. Only a single k-point and 32 energy grid points were used without further energy grid refinements. (a) and (b) show the lesser and greater Green's function as a function of energy and the degree of freedom, i.e., atom orbital index, of the device. (c) and (d) show the relative error on the lesser and greater self-energy, calculated as the difference between the self-energy calculated with the conventional convolution-based implementation and the FFT-based implementation. }
    \label{fig:energy_mix_err1}
\end{figure}

An estimate of the energy mixing error is shown in Fig. \ref{fig:energy_mix_err1}, which shows the self-energies of the system described in Section \ref{secsec:device_simulation} after one non-self-consistent iteration. The potential was obtained with a fine initial energy grid, after which the energy grids were limited to an equidistant grid with 32 points with no further adaptive grid refinements and a single k-point for the calculation of the self-energy. The removal of adaptive grid refinements was to eliminate the self-energy interpolation error and the Green's function conversion error. Fig. \ref{fig:energy_mix_err1} (a) and (b) clearly shown the band gap and the bottom of the conduction band. Fig. \ref{fig:energy_mix_err1} (c) and (d) show that the lesser self-energy in the middle of the band gap does not have a single correct digit. 

\begin{figure}[ht]
    \centering
    \includegraphics[width=\linewidth]{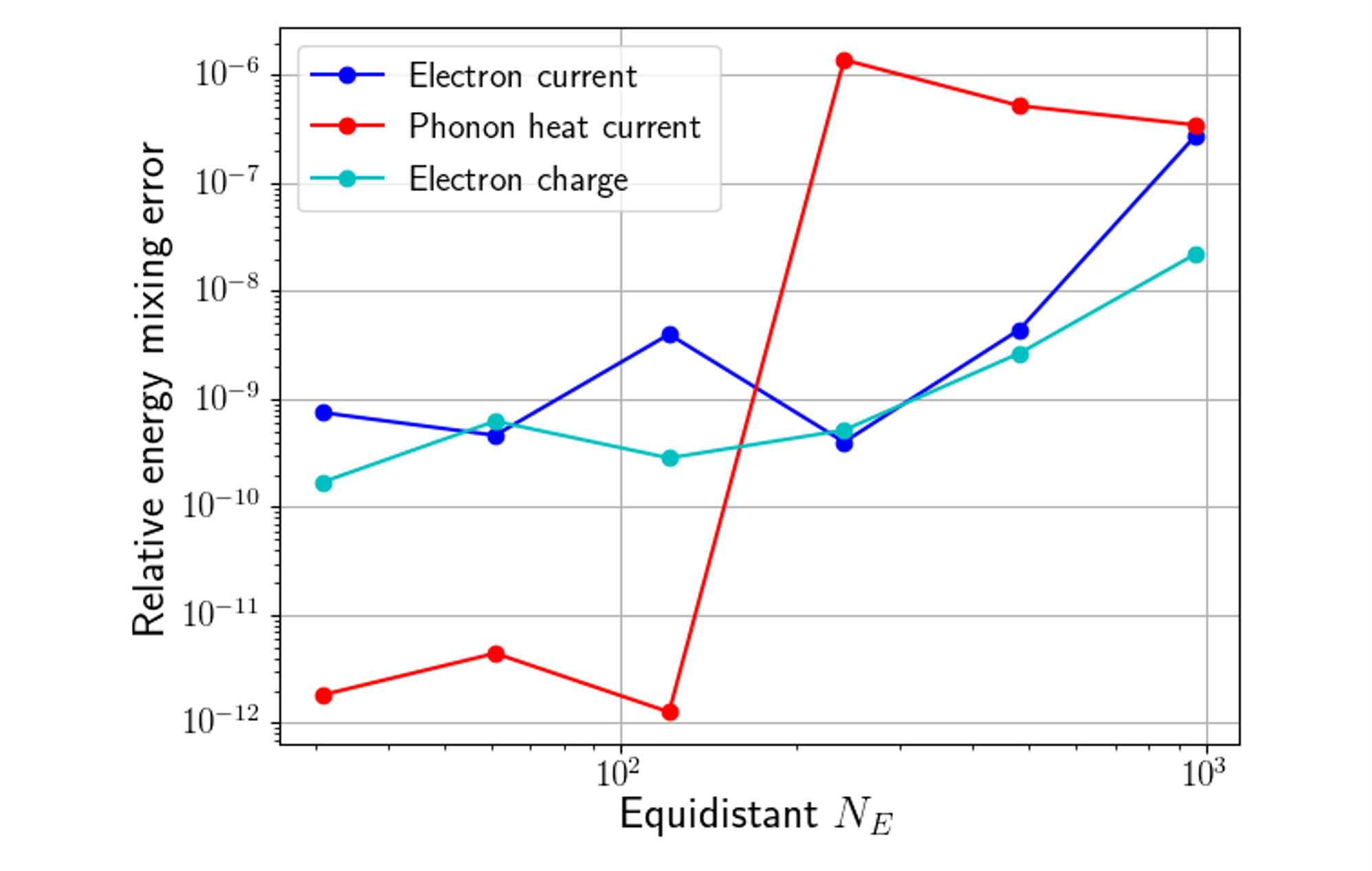}
    \caption{The relative energy mixing error on the electron current, phonon heat current and charge density for the system in Section \ref{secsec:device_simulation} as a function of the number of equidistant grid energy points without further refinements. The error is calculated as the difference between the calculation using the conventional convolution-based implementation and the FFT-based implementation.}
    \label{fig:energy_mix_err2}
\end{figure}

An estimate of the errors on the macroscopic device properties for a simulation with 10 k-points are shown in Fig. \ref{fig:energy_mix_err2} as a function of the number of equidistant energy grid points. It can be seen that the energy mixing error increases with the number of energy points. However, for the number of energy points considered here, the energy mixing error is still multiple orders of magnitude smaller than the integration error in Fig. \ref{fig:integration_err} (a). The large relative error on the self-energy thus does not pose problems for obtaining correct macroscopic parameters. This can be attributed to the fact that the large relative error is present at energy ranges where the density of states is low, and hence, at energies which do not have a large effect on the macroscopic properties.

\subsection{The self-energy interpolation error}

\begin{figure}[ht]
    \centering
    \includegraphics[width=\linewidth]{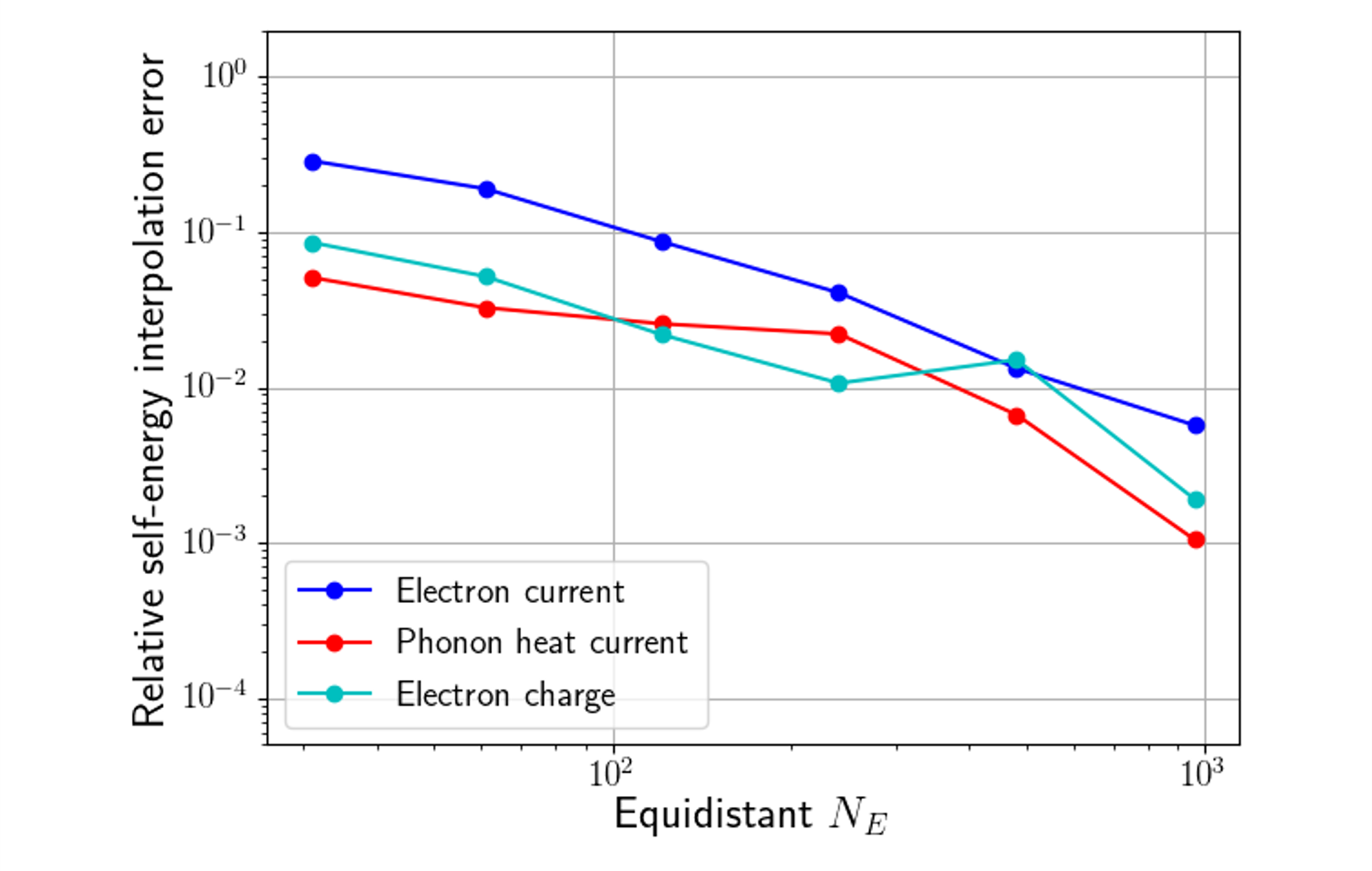}
    \caption{The relative self-energy interpolation error on the electron current, phonon heat current and charge density for the system in Section \ref{secsec:device_simulation} as a function of the number of initial equidistant grid energy points. The error is calculated as the difference between the results when the self-energy is calculated with the conventional convolution-based implementation for every energy grid point and when it is done only for the initial grid with the rest obtained through interpolation.}
    \label{fig:self_energy_interp_err}
\end{figure}

An estimate of the self-energy interpolation error is shown in Fig. \ref{fig:self_energy_interp_err} as a function of the number of initial equidistant energy grid points. To reach a relative error below 1\% on all macroscopic parameters, $\sim$1000 initial equidistant grid energy points are required. The self-energy interpolation energy thus requires more equidistant initial grid points than the integration error.

\subsection{The Green's function conversion error}

\begin{figure}[ht]
    \centering
    \includegraphics[width=\linewidth]{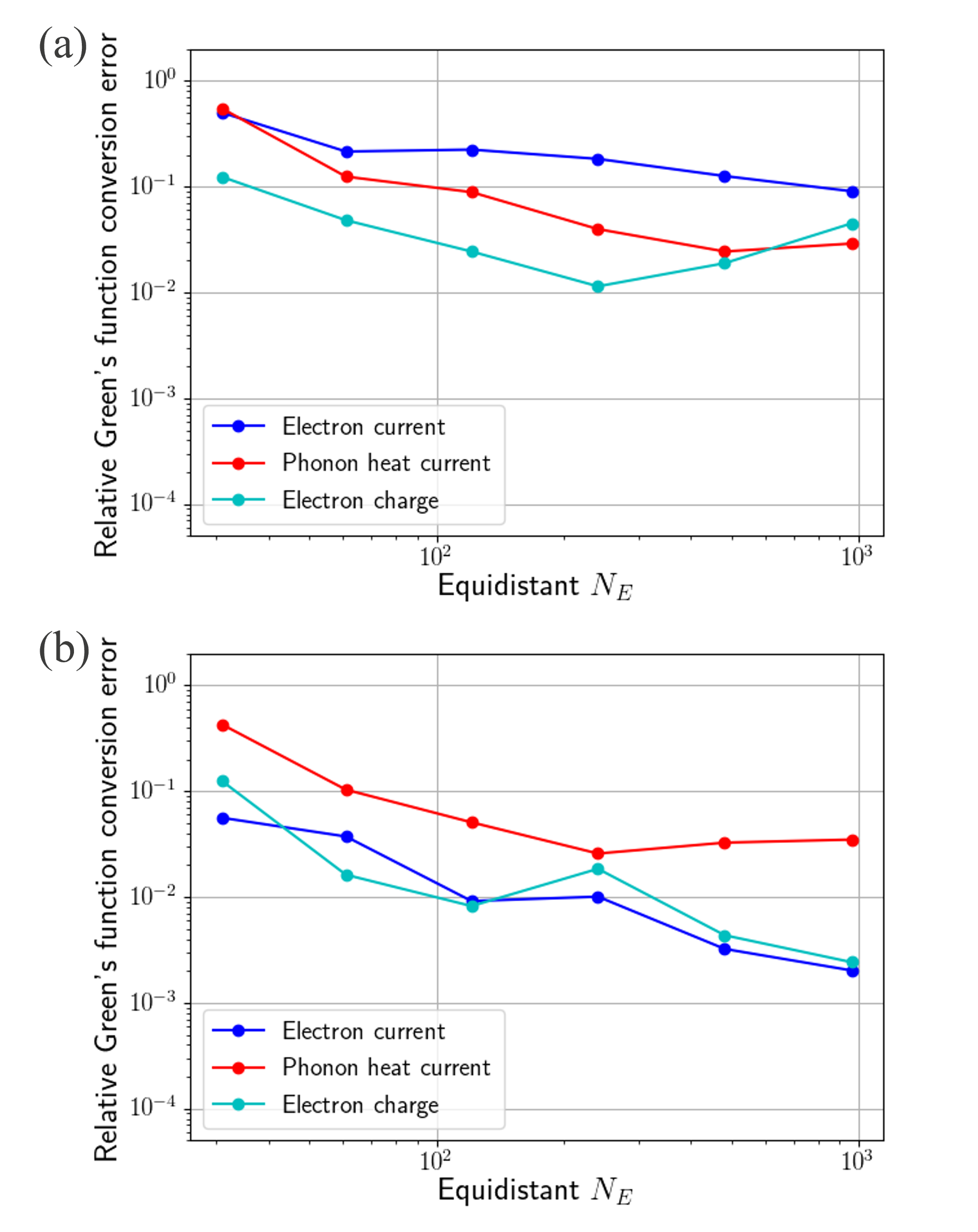}
    \caption{The relative Green's function conversion error on the electron current, phonon heat current and charge density for the system in Section \ref{secsec:device_simulation} as a function of the number of initial equidistant grid energy points. The error is calculated as the difference between the results when the calculation is calculated with the conventional convolution-based implementation using the full adaptive grid and the FFT-based implementation using a converted equidistant grid. In both cases the self-energies are only calculated on the initial equidistant grid and the self-energies for adaptively added points is obtained by interpolation. (a) shows the results for the leave-out strategy of Fig. \ref{fig:equidistant_mesh_err2} (a) and (b) shows the results for the averaged strategy of Fig. \ref{fig:equidistant_mesh_err2} (b).}
    \label{fig:Greens_func_conv_err}
\end{figure}

An estimate of the Green's function conversion errors are shown in Fig. \ref{fig:Greens_func_conv_err} (a) and (b) for the "leave-out" and "averaged" approach, respectively. Fig. \ref{fig:Greens_func_conv_err} (a) demonstrates that the "leave-out" approach shows no or very slow convergence of the macroscopic parameters with increasing initial equidistant grid sizes. For the "averaged" approach, the electronic properties of the system converge and reach a relative error below 1\% for $\sim$500 initial equidistant grid energy points. The phonon heat current, however, demonstrates no or very slow convergence, resulting in a relative error of a few percent at high numbers of initial equidistant grid energy points. Although unfortunate, this is not surprising nor does it invalidate the FFT-based implementation. 

As indicated by Fig. \ref{fig:equidistant_mesh_err2}, the "averaged" approach has the effect of shifting the mass of the spectra to different energies. The shift in energy is at most the distance between equidistant grid points. For the electron self-energy, this does not pose any problems as the shift in energy will be on the same order of magnitude as the electron grid, for the electron Green's function, or much smaller, for the phonon Green's function. For the phonon self-energy, however, this does introduce a slightly larger error as the shift in energy happens on the electron energy grid and can thus be significantly larger than the distance between phonon energy grid points. Additionally, a shift in energy also results in a shift in the energy of the phonons that are created in the system, which has a direct influence on the phonon heat current. The electronic properties, on the other hand, are less dependent on the energy of the electrons. 

\clearpage

\bibliography{main}

\begin{thebibliography}{30}%
\makeatletter
\providecommand \@ifxundefined [1]{%
 \@ifx{#1\undefined}
}%
\providecommand \@ifnum [1]{%
 \ifnum #1\expandafter \@firstoftwo
 \else \expandafter \@secondoftwo
 \fi
}%
\providecommand \@ifx [1]{%
 \ifx #1\expandafter \@firstoftwo
 \else \expandafter \@secondoftwo
 \fi
}%
\providecommand \natexlab [1]{#1}%
\providecommand \enquote  [1]{``#1''}%
\providecommand \bibnamefont  [1]{#1}%
\providecommand \bibfnamefont [1]{#1}%
\providecommand \citenamefont [1]{#1}%
\providecommand \href@noop [0]{\@secondoftwo}%
\providecommand \href [0]{\begingroup \@sanitize@url \@href}%
\providecommand \@href[1]{\@@startlink{#1}\@@href}%
\providecommand \@@href[1]{\endgroup#1\@@endlink}%
\providecommand \@sanitize@url [0]{\catcode `\\12\catcode `\$12\catcode `\&12\catcode `\#12\catcode `\^12\catcode `\_12\catcode `\%12\relax}%
\providecommand \@@startlink[1]{}%
\providecommand \@@endlink[0]{}%
\providecommand \url  [0]{\begingroup\@sanitize@url \@url }%
\providecommand \@url [1]{\endgroup\@href {#1}{\urlprefix }}%
\providecommand \urlprefix  [0]{URL }%
\providecommand \Eprint [0]{\href }%
\providecommand \doibase [0]{https://doi.org/}%
\providecommand \selectlanguage [0]{\@gobble}%
\providecommand \bibinfo  [0]{\@secondoftwo}%
\providecommand \bibfield  [0]{\@secondoftwo}%
\providecommand \translation [1]{[#1]}%
\providecommand \BibitemOpen [0]{}%
\providecommand \bibitemStop [0]{}%
\providecommand \bibitemNoStop [0]{.\EOS\space}%
\providecommand \EOS [0]{\spacefactor3000\relax}%
\providecommand \BibitemShut  [1]{\csname bibitem#1\endcsname}%
\let\auto@bib@innerbib\@empty
\bibitem [{\citenamefont {Schwierz}\ \emph {et~al.}(2015)\citenamefont {Schwierz}, \citenamefont {Pezoldt},\ and\ \citenamefont {Granzner}}]{Schwierz2015}%
  \BibitemOpen
  \bibfield  {author} {\bibinfo {author} {\bibfnamefont {F.}~\bibnamefont {Schwierz}}, \bibinfo {author} {\bibfnamefont {J.}~\bibnamefont {Pezoldt}},\ and\ \bibinfo {author} {\bibfnamefont {R.}~\bibnamefont {Granzner}},\ }\bibfield  {title} {\bibinfo {title} {Two-dimensional materials and their prospects in transistor electronics},\ }\href@noop {} {\bibfield  {journal} {\bibinfo  {journal} {Nanoscale}\ }\textbf {\bibinfo {volume} {7}},\ \bibinfo {pages} {8261} (\bibinfo {year} {2015})}\BibitemShut {NoStop}%
\bibitem [{\citenamefont {Logoteta}\ \emph {et~al.}(2014)\citenamefont {Logoteta}, \citenamefont {Zhang},\ and\ \citenamefont {Fiori}}]{Logoteta2014}%
  \BibitemOpen
  \bibfield  {author} {\bibinfo {author} {\bibfnamefont {D.}~\bibnamefont {Logoteta}}, \bibinfo {author} {\bibfnamefont {Q.}~\bibnamefont {Zhang}},\ and\ \bibinfo {author} {\bibfnamefont {G.}~\bibnamefont {Fiori}},\ }\bibfield  {title} {\bibinfo {title} {What can we really expect from 2d materials for electronic applications?},\ }in\ \href {https://doi.org/10.1109/drc.2014.6872357} {\emph {\bibinfo {booktitle} {72nd Device Research Conference}}}\ (\bibinfo  {publisher} {{IEEE}},\ \bibinfo {year} {2014})\BibitemShut {NoStop}%
\bibitem [{\citenamefont {Chhowalla}\ \emph {et~al.}(2016)\citenamefont {Chhowalla}, \citenamefont {Jena},\ and\ \citenamefont {Zhang}}]{Chhowalla2016}%
  \BibitemOpen
  \bibfield  {author} {\bibinfo {author} {\bibfnamefont {M.}~\bibnamefont {Chhowalla}}, \bibinfo {author} {\bibfnamefont {D.}~\bibnamefont {Jena}},\ and\ \bibinfo {author} {\bibfnamefont {H.}~\bibnamefont {Zhang}},\ }\bibfield  {title} {\bibinfo {title} {Two-dimensional semiconductors for transistors},\ }\href@noop {} {\bibfield  {journal} {\bibinfo  {journal} {Nature Reviews Materials}\ }\textbf {\bibinfo {volume} {1}},\ \bibinfo {pages} {16052} (\bibinfo {year} {2016})}\BibitemShut {NoStop}%
\bibitem [{\citenamefont {Kang}\ \emph {et~al.}(2014)\citenamefont {Kang}, \citenamefont {Cao}, \citenamefont {Xie}, \citenamefont {Sarkar}, \citenamefont {Liu},\ and\ \citenamefont {Banerjee}}]{Kang2014}%
  \BibitemOpen
  \bibfield  {author} {\bibinfo {author} {\bibfnamefont {J.}~\bibnamefont {Kang}}, \bibinfo {author} {\bibfnamefont {W.}~\bibnamefont {Cao}}, \bibinfo {author} {\bibfnamefont {X.}~\bibnamefont {Xie}}, \bibinfo {author} {\bibfnamefont {D.}~\bibnamefont {Sarkar}}, \bibinfo {author} {\bibfnamefont {W.}~\bibnamefont {Liu}},\ and\ \bibinfo {author} {\bibfnamefont {K.}~\bibnamefont {Banerjee}},\ }\bibfield  {title} {\bibinfo {title} {Graphene and beyond-graphene 2d crystals for next-generation green electronics},\ }in\ \href@noop {} {\emph {\bibinfo {booktitle} {Micro-and Nanotechnology Sensors, Systems, and Applications VI}}},\ Vol.\ \bibinfo {volume} {9083}\ (\bibinfo {organization} {International Society for Optics and Photonics},\ \bibinfo {year} {2014})\ p.\ \bibinfo {pages} {908305}\BibitemShut {NoStop}%
\bibitem [{\citenamefont {Afzalian}(2021)}]{Afzalian2020}%
  \BibitemOpen
  \bibfield  {author} {\bibinfo {author} {\bibfnamefont {A.}~\bibnamefont {Afzalian}},\ }\bibfield  {title} {\bibinfo {title} {Ab initio perspective of ultra-scaled {CMOS} from 2d-material fundamentals to dynamically doped transistors},\ }\bibfield  {journal} {\bibinfo  {journal} {npj 2D Materials and Applications}\ }\textbf {\bibinfo {volume} {5}},\ \href {https://doi.org/10.1038/s41699-020-00181-1} {10.1038/s41699-020-00181-1} (\bibinfo {year} {2021})\BibitemShut {NoStop}%
\bibitem [{\citenamefont {Duflou}\ \emph {et~al.}(2021)\citenamefont {Duflou}, \citenamefont {Houssa},\ and\ \citenamefont {Afzalian}}]{Duflou2021}%
  \BibitemOpen
  \bibfield  {author} {\bibinfo {author} {\bibfnamefont {R.}~\bibnamefont {Duflou}}, \bibinfo {author} {\bibfnamefont {M.}~\bibnamefont {Houssa}},\ and\ \bibinfo {author} {\bibfnamefont {A.}~\bibnamefont {Afzalian}},\ }\bibfield  {title} {\bibinfo {title} {Electron-phonon scattering in cold-metal contacted two-dimensional semiconductor devices},\ }in\ \href {https://doi.org/10.1109/SISPAD54002.2021.9592538} {\emph {\bibinfo {booktitle} {2021 International Conference on Simulation of Semiconductor Processes and Devices (SISPAD)}}}\ (\bibinfo {year} {2021})\ pp.\ \bibinfo {pages} {94--97}\BibitemShut {NoStop}%
\bibitem [{\citenamefont {Rhyner}\ and\ \citenamefont {Luisier}(2014)}]{Rhyner2014}%
  \BibitemOpen
  \bibfield  {author} {\bibinfo {author} {\bibfnamefont {R.}~\bibnamefont {Rhyner}}\ and\ \bibinfo {author} {\bibfnamefont {M.}~\bibnamefont {Luisier}},\ }\bibfield  {title} {\bibinfo {title} {Atomistic modeling of coupled electron-phonon transport in nanowire transistors},\ }\href {https://doi.org/10.1103/PhysRevB.89.235311} {\bibfield  {journal} {\bibinfo  {journal} {Phys. Rev. B}\ }\textbf {\bibinfo {volume} {89}},\ \bibinfo {pages} {235311} (\bibinfo {year} {2014})}\BibitemShut {NoStop}%
\bibitem [{\citenamefont {Afzalian}\ \emph {et~al.}(2021)\citenamefont {Afzalian}, \citenamefont {Akhoundi}, \citenamefont {Gaddemane}, \citenamefont {Duflou},\ and\ \citenamefont {Houssa}}]{Afzalian2021}%
  \BibitemOpen
  \bibfield  {author} {\bibinfo {author} {\bibfnamefont {A.}~\bibnamefont {Afzalian}}, \bibinfo {author} {\bibfnamefont {E.}~\bibnamefont {Akhoundi}}, \bibinfo {author} {\bibfnamefont {G.}~\bibnamefont {Gaddemane}}, \bibinfo {author} {\bibfnamefont {R.}~\bibnamefont {Duflou}},\ and\ \bibinfo {author} {\bibfnamefont {M.}~\bibnamefont {Houssa}},\ }\bibfield  {title} {\bibinfo {title} {Advanced dft–negf transport techniques for novel 2-d material and device exploration including hfs2/wse2 van der waals heterojunction tfet and wte2/ws2 metal/semiconductor contact},\ }\href {https://doi.org/10.1109/TED.2021.3078412} {\bibfield  {journal} {\bibinfo  {journal} {IEEE Transactions on Electron Devices}\ }\textbf {\bibinfo {volume} {68}},\ \bibinfo {pages} {5372} (\bibinfo {year} {2021})}\BibitemShut {NoStop}%
\bibitem [{\citenamefont {Duflou}\ \emph {et~al.}(2023)\citenamefont {Duflou}, \citenamefont {Houssa},\ and\ \citenamefont {Afzalian}}]{Duflou2023SISPAD}%
  \BibitemOpen
  \bibfield  {author} {\bibinfo {author} {\bibfnamefont {R.}~\bibnamefont {Duflou}}, \bibinfo {author} {\bibfnamefont {M.}~\bibnamefont {Houssa}},\ and\ \bibinfo {author} {\bibfnamefont {A.}~\bibnamefont {Afzalian}},\ }\bibfield  {title} {\bibinfo {title} {Ballistic heat transport in mos2 monolayers},\ }in\ \href@noop {} {\emph {\bibinfo {booktitle} {2023 International Conference on Simulation of Semiconductor Processes and Devices (SISPAD)}}}\ (\bibinfo {year} {2023})\BibitemShut {NoStop}%
\bibitem [{\citenamefont {Giustino}(2017)}]{Giustino2017}%
  \BibitemOpen
  \bibfield  {author} {\bibinfo {author} {\bibfnamefont {F.}~\bibnamefont {Giustino}},\ }\bibfield  {title} {\bibinfo {title} {Electron-phonon interactions from first principles},\ }\bibfield  {journal} {\bibinfo  {journal} {Reviews of Modern Physics}\ }\textbf {\bibinfo {volume} {89}},\ \href {https://doi.org/10.1103/revmodphys.89.015003} {10.1103/revmodphys.89.015003} (\bibinfo {year} {2017})\BibitemShut {NoStop}%
\bibitem [{\citenamefont {Marzari}\ \emph {et~al.}(2012)\citenamefont {Marzari}, \citenamefont {Mostofi}, \citenamefont {Yates}, \citenamefont {Souza},\ and\ \citenamefont {Vanderbilt}}]{Marzari2012}%
  \BibitemOpen
  \bibfield  {author} {\bibinfo {author} {\bibfnamefont {N.}~\bibnamefont {Marzari}}, \bibinfo {author} {\bibfnamefont {A.~A.}\ \bibnamefont {Mostofi}}, \bibinfo {author} {\bibfnamefont {J.~R.}\ \bibnamefont {Yates}}, \bibinfo {author} {\bibfnamefont {I.}~\bibnamefont {Souza}},\ and\ \bibinfo {author} {\bibfnamefont {D.}~\bibnamefont {Vanderbilt}},\ }\bibfield  {title} {\bibinfo {title} {Maximally localized wannier functions: Theory and applications},\ }\href {https://doi.org/10.1103/RevModPhys.84.1419} {\bibfield  {journal} {\bibinfo  {journal} {Rev. Mod. Phys.}\ }\textbf {\bibinfo {volume} {84}},\ \bibinfo {pages} {1419} (\bibinfo {year} {2012})}\BibitemShut {NoStop}%
\bibitem [{\citenamefont {Giustino}\ \emph {et~al.}(2007)\citenamefont {Giustino}, \citenamefont {Cohen},\ and\ \citenamefont {Louie}}]{Giustino2007}%
  \BibitemOpen
  \bibfield  {author} {\bibinfo {author} {\bibfnamefont {F.}~\bibnamefont {Giustino}}, \bibinfo {author} {\bibfnamefont {M.~L.}\ \bibnamefont {Cohen}},\ and\ \bibinfo {author} {\bibfnamefont {S.~G.}\ \bibnamefont {Louie}},\ }\bibfield  {title} {\bibinfo {title} {Electron-phonon interaction using wannier functions},\ }\href {https://doi.org/10.1103/PhysRevB.76.165108} {\bibfield  {journal} {\bibinfo  {journal} {Phys. Rev. B}\ }\textbf {\bibinfo {volume} {76}},\ \bibinfo {pages} {165108} (\bibinfo {year} {2007})}\BibitemShut {NoStop}%
\bibitem [{\citenamefont {Fetter}\ and\ \citenamefont {Walecka}(1971)}]{Fetterch3}%
  \BibitemOpen
  \bibfield  {author} {\bibinfo {author} {\bibfnamefont {A.~L.}\ \bibnamefont {Fetter}}\ and\ \bibinfo {author} {\bibfnamefont {J.~D.}\ \bibnamefont {Walecka}},\ }\href@noop {} {\emph {\bibinfo {title} {Quantum Theory of Many-Particle Systems}}}\ (\bibinfo  {publisher} {McGraw-Hill},\ \bibinfo {address} {Boston},\ \bibinfo {year} {1971})\ Chap.~\bibinfo {chapter} {3}\BibitemShut {NoStop}%
\bibitem [{\citenamefont {Maciejko}(2007)}]{Maciejko2007}%
  \BibitemOpen
  \bibfield  {author} {\bibinfo {author} {\bibfnamefont {J.}~\bibnamefont {Maciejko}},\ }\bibfield  {title} {\bibinfo {title} {An introduction to nonequilibrium many-body theory},\ }\href@noop {} {\bibfield  {journal} {\bibinfo  {journal} {Lecture Notes, Springer}\ } (\bibinfo {year} {2007})}\BibitemShut {NoStop}%
\bibitem [{\citenamefont {Svizhenko}\ \emph {et~al.}(2002)\citenamefont {Svizhenko}, \citenamefont {Anantram}, \citenamefont {Govindan}, \citenamefont {Biegel},\ and\ \citenamefont {Venugopal}}]{Svizhenko2002}%
  \BibitemOpen
  \bibfield  {author} {\bibinfo {author} {\bibfnamefont {A.}~\bibnamefont {Svizhenko}}, \bibinfo {author} {\bibfnamefont {M.}~\bibnamefont {Anantram}}, \bibinfo {author} {\bibfnamefont {T.}~\bibnamefont {Govindan}}, \bibinfo {author} {\bibfnamefont {B.}~\bibnamefont {Biegel}},\ and\ \bibinfo {author} {\bibfnamefont {R.}~\bibnamefont {Venugopal}},\ }\bibfield  {title} {\bibinfo {title} {Two-dimensional quantum mechanical modeling of nanotransistors},\ }\href@noop {} {\bibfield  {journal} {\bibinfo  {journal} {Journal of Applied Physics}\ }\textbf {\bibinfo {volume} {91}},\ \bibinfo {pages} {2343} (\bibinfo {year} {2002})}\BibitemShut {NoStop}%
\bibitem [{Sup()}]{Supp}%
  \BibitemOpen
  \href@noop {} {}\bibinfo {note} {See the Supplemental Material for a rigorous treatment of the interaction terms between the phonons in the leads and the device.}\BibitemShut {Stop}%
\bibitem [{\citenamefont {Lake}\ and\ \citenamefont {Pandey}(2006)}]{Lake2006}%
  \BibitemOpen
  \bibfield  {author} {\bibinfo {author} {\bibfnamefont {R.}~\bibnamefont {Lake}}\ and\ \bibinfo {author} {\bibfnamefont {R.}~\bibnamefont {Pandey}},\ }\bibfield  {title} {\bibinfo {title} {Non-equilibrium green functions in electronic device modeling},\ }\href@noop {} {\bibfield  {journal} {\bibinfo  {journal} {arXiv preprint cond-mat/0607219}\ } (\bibinfo {year} {2006})}\BibitemShut {NoStop}%
\bibitem [{\citenamefont {Datta}(2005{\natexlab{a}})}]{Dattasupp}%
  \BibitemOpen
  \bibfield  {author} {\bibinfo {author} {\bibfnamefont {S.}~\bibnamefont {Datta}},\ }\href@noop {} {\emph {\bibinfo {title} {Quantum transport: atom to transistor}}}\ (\bibinfo  {publisher} {Cambridge university press},\ \bibinfo {year} {2005})\ Chap.~\bibinfo {chapter} {A1}\BibitemShut {NoStop}%
\bibitem [{\citenamefont {Sancho}\ \emph {et~al.}(1985)\citenamefont {Sancho}, \citenamefont {Sancho}, \citenamefont {Sancho},\ and\ \citenamefont {Rubio}}]{Sancho1985}%
  \BibitemOpen
  \bibfield  {author} {\bibinfo {author} {\bibfnamefont {M.~P.~L.}\ \bibnamefont {Sancho}}, \bibinfo {author} {\bibfnamefont {J.~M.~L.}\ \bibnamefont {Sancho}}, \bibinfo {author} {\bibfnamefont {J.~M.~L.}\ \bibnamefont {Sancho}},\ and\ \bibinfo {author} {\bibfnamefont {J.}~\bibnamefont {Rubio}},\ }\bibfield  {title} {\bibinfo {title} {Highly convergent schemes for the calculation of bulk and surface green functions},\ }\href {https://doi.org/10.1088/0305-4608/15/4/009} {\bibfield  {journal} {\bibinfo  {journal} {Journal of Physics F: Metal Physics}\ }\textbf {\bibinfo {volume} {15}},\ \bibinfo {pages} {851} (\bibinfo {year} {1985})}\BibitemShut {NoStop}%
\bibitem [{\citenamefont {Mingo}\ and\ \citenamefont {Yang}(2003)}]{Mingo2003}%
  \BibitemOpen
  \bibfield  {author} {\bibinfo {author} {\bibfnamefont {N.}~\bibnamefont {Mingo}}\ and\ \bibinfo {author} {\bibfnamefont {L.}~\bibnamefont {Yang}},\ }\bibfield  {title} {\bibinfo {title} {Phonon transport in nanowires coated with an amorphous material: An atomistic green's function approach},\ }\href {https://doi.org/10.1103/PhysRevB.68.245406} {\bibfield  {journal} {\bibinfo  {journal} {Phys. Rev. B}\ }\textbf {\bibinfo {volume} {68}},\ \bibinfo {pages} {245406} (\bibinfo {year} {2003})}\BibitemShut {NoStop}%
\bibitem [{\citenamefont {Mingo}(2009)}]{Mingo2009}%
  \BibitemOpen
  \bibfield  {author} {\bibinfo {author} {\bibfnamefont {N.}~\bibnamefont {Mingo}},\ }\bibinfo {title} {Green's function methods for phonon transport through nano-contacts},\ in\ \href {https://doi.org/10.1007/978-3-642-04258-4_3} {\emph {\bibinfo {booktitle} {Thermal Nanosystems and Nanomaterials}}}\ (\bibinfo  {publisher} {Springer Berlin Heidelberg},\ \bibinfo {address} {Berlin, Heidelberg},\ \bibinfo {year} {2009})\ pp.\ \bibinfo {pages} {63--94}\BibitemShut {NoStop}%
\bibitem [{\citenamefont {Giannozzi}\ \emph {et~al.}(2009)\citenamefont {Giannozzi}, \citenamefont {Baroni}, \citenamefont {Bonini}, \citenamefont {Calandra}, \citenamefont {Car}, \citenamefont {Cavazzoni}, \citenamefont {Ceresoli}, \citenamefont {Chiarotti}, \citenamefont {Cococcioni}, \citenamefont {Dabo}, \citenamefont {Corso}, \citenamefont {de~Gironcoli}, \citenamefont {Fabris}, \citenamefont {Fratesi}, \citenamefont {Gebauer}, \citenamefont {Gerstmann}, \citenamefont {Gougoussis}, \citenamefont {Kokalj}, \citenamefont {Lazzeri}, \citenamefont {Martin-Samos}, \citenamefont {Marzari}, \citenamefont {Mauri}, \citenamefont {Mazzarello}, \citenamefont {Paolini}, \citenamefont {Pasquarello}, \citenamefont {Paulatto}, \citenamefont {Sbraccia}, \citenamefont {Scandolo}, \citenamefont {Sclauzero}, \citenamefont {Seitsonen}, \citenamefont {Smogunov}, \citenamefont {Umari},\ and\ \citenamefont {Wentzcovitch}}]{Giannozzi2009}%
  \BibitemOpen
  \bibfield  {author} {\bibinfo {author} {\bibfnamefont {P.}~\bibnamefont {Giannozzi}}, \bibinfo {author} {\bibfnamefont {S.}~\bibnamefont {Baroni}}, \bibinfo {author} {\bibfnamefont {N.}~\bibnamefont {Bonini}}, \bibinfo {author} {\bibfnamefont {M.}~\bibnamefont {Calandra}}, \bibinfo {author} {\bibfnamefont {R.}~\bibnamefont {Car}}, \bibinfo {author} {\bibfnamefont {C.}~\bibnamefont {Cavazzoni}}, \bibinfo {author} {\bibfnamefont {D.}~\bibnamefont {Ceresoli}}, \bibinfo {author} {\bibfnamefont {G.~L.}\ \bibnamefont {Chiarotti}}, \bibinfo {author} {\bibfnamefont {M.}~\bibnamefont {Cococcioni}}, \bibinfo {author} {\bibfnamefont {I.}~\bibnamefont {Dabo}}, \bibinfo {author} {\bibfnamefont {A.~D.}\ \bibnamefont {Corso}}, \bibinfo {author} {\bibfnamefont {S.}~\bibnamefont {de~Gironcoli}}, \bibinfo {author} {\bibfnamefont {S.}~\bibnamefont {Fabris}}, \bibinfo {author} {\bibfnamefont {G.}~\bibnamefont {Fratesi}}, \bibinfo {author} {\bibfnamefont {R.}~\bibnamefont {Gebauer}}, \bibinfo {author} {\bibfnamefont
  {U.}~\bibnamefont {Gerstmann}}, \bibinfo {author} {\bibfnamefont {C.}~\bibnamefont {Gougoussis}}, \bibinfo {author} {\bibfnamefont {A.}~\bibnamefont {Kokalj}}, \bibinfo {author} {\bibfnamefont {M.}~\bibnamefont {Lazzeri}}, \bibinfo {author} {\bibfnamefont {L.}~\bibnamefont {Martin-Samos}}, \bibinfo {author} {\bibfnamefont {N.}~\bibnamefont {Marzari}}, \bibinfo {author} {\bibfnamefont {F.}~\bibnamefont {Mauri}}, \bibinfo {author} {\bibfnamefont {R.}~\bibnamefont {Mazzarello}}, \bibinfo {author} {\bibfnamefont {S.}~\bibnamefont {Paolini}}, \bibinfo {author} {\bibfnamefont {A.}~\bibnamefont {Pasquarello}}, \bibinfo {author} {\bibfnamefont {L.}~\bibnamefont {Paulatto}}, \bibinfo {author} {\bibfnamefont {C.}~\bibnamefont {Sbraccia}}, \bibinfo {author} {\bibfnamefont {S.}~\bibnamefont {Scandolo}}, \bibinfo {author} {\bibfnamefont {G.}~\bibnamefont {Sclauzero}}, \bibinfo {author} {\bibfnamefont {A.~P.}\ \bibnamefont {Seitsonen}}, \bibinfo {author} {\bibfnamefont {A.}~\bibnamefont {Smogunov}}, \bibinfo {author}
  {\bibfnamefont {P.}~\bibnamefont {Umari}},\ and\ \bibinfo {author} {\bibfnamefont {R.~M.}\ \bibnamefont {Wentzcovitch}},\ }\bibfield  {title} {\bibinfo {title} {{QUANTUM} {ESPRESSO}: a modular and open-source software project for quantum simulations of materials},\ }\href {https://doi.org/10.1088/0953-8984/21/39/395502} {\bibfield  {journal} {\bibinfo  {journal} {Journal of Physics: Condensed Matter}\ }\textbf {\bibinfo {volume} {21}},\ \bibinfo {pages} {395502} (\bibinfo {year} {2009})}\BibitemShut {NoStop}%
\bibitem [{\citenamefont {Al-Hilli}\ and\ \citenamefont {Evans}(1972)}]{Alhilli1972}%
  \BibitemOpen
  \bibfield  {author} {\bibinfo {author} {\bibfnamefont {A.}~\bibnamefont {Al-Hilli}}\ and\ \bibinfo {author} {\bibfnamefont {B.}~\bibnamefont {Evans}},\ }\bibfield  {title} {\bibinfo {title} {The preparation and properties of transition metal dichalcogenide single crystals},\ }\href {https://doi.org/https://doi.org/10.1016/0022-0248(72)90129-7} {\bibfield  {journal} {\bibinfo  {journal} {Journal of Crystal Growth}\ }\textbf {\bibinfo {volume} {15}},\ \bibinfo {pages} {93} (\bibinfo {year} {1972})}\BibitemShut {NoStop}%
\bibitem [{\citenamefont {Grimme}\ \emph {et~al.}(2010)\citenamefont {Grimme}, \citenamefont {Antony}, \citenamefont {Ehrlich},\ and\ \citenamefont {Krieg}}]{Grimme2010}%
  \BibitemOpen
  \bibfield  {author} {\bibinfo {author} {\bibfnamefont {S.}~\bibnamefont {Grimme}}, \bibinfo {author} {\bibfnamefont {J.}~\bibnamefont {Antony}}, \bibinfo {author} {\bibfnamefont {S.}~\bibnamefont {Ehrlich}},\ and\ \bibinfo {author} {\bibfnamefont {H.}~\bibnamefont {Krieg}},\ }\bibfield  {title} {\bibinfo {title} {{A consistent and accurate ab initio parametrization of density functional dispersion correction (DFT-D) for the 94 elements H-Pu}},\ }\href {https://doi.org/10.1063/1.3382344} {\bibfield  {journal} {\bibinfo  {journal} {The Journal of Chemical Physics}\ }\textbf {\bibinfo {volume} {132}},\ \bibinfo {pages} {154104} (\bibinfo {year} {2010})},\ \Eprint {https://arxiv.org/abs/https://pubs.aip.org/aip/jcp/article-pdf/doi/10.1063/1.3382344/15684000/154104\_1\_online.pdf} {https://pubs.aip.org/aip/jcp/article-pdf/doi/10.1063/1.3382344/15684000/154104\_1\_online.pdf} \BibitemShut {NoStop}%
\bibitem [{\citenamefont {Fischetti}\ and\ \citenamefont {Vandenberghe}(2016)}]{Fischetti2016}%
  \BibitemOpen
  \bibfield  {author} {\bibinfo {author} {\bibfnamefont {M.~V.}\ \bibnamefont {Fischetti}}\ and\ \bibinfo {author} {\bibfnamefont {W.~G.}\ \bibnamefont {Vandenberghe}},\ }\bibfield  {title} {\bibinfo {title} {Mermin-wagner theorem, flexural modes, and degraded carrier mobility in two-dimensional crystals with broken horizontal mirror symmetry},\ }\href {https://doi.org/10.1103/PhysRevB.93.155413} {\bibfield  {journal} {\bibinfo  {journal} {Phys. Rev. B}\ }\textbf {\bibinfo {volume} {93}},\ \bibinfo {pages} {155413} (\bibinfo {year} {2016})}\BibitemShut {NoStop}%
\bibitem [{\citenamefont {Chu}\ \emph {et~al.}(2018)\citenamefont {Chu}, \citenamefont {Sarangapani}, \citenamefont {Charles}, \citenamefont {Klimeck},\ and\ \citenamefont {Kubis}}]{Chu2018}%
  \BibitemOpen
  \bibfield  {author} {\bibinfo {author} {\bibfnamefont {Y.}~\bibnamefont {Chu}}, \bibinfo {author} {\bibfnamefont {P.}~\bibnamefont {Sarangapani}}, \bibinfo {author} {\bibfnamefont {J.}~\bibnamefont {Charles}}, \bibinfo {author} {\bibfnamefont {G.}~\bibnamefont {Klimeck}},\ and\ \bibinfo {author} {\bibfnamefont {T.}~\bibnamefont {Kubis}},\ }\bibfield  {title} {\bibinfo {title} {Explicit screening full band quantum transport model for semiconductor nanodevices},\ }\href {https://doi.org/10.1063/1.5031461} {\bibfield  {journal} {\bibinfo  {journal} {Journal of Applied Physics}\ }\textbf {\bibinfo {volume} {123}},\ \bibinfo {pages} {244501} (\bibinfo {year} {2018})}\BibitemShut {NoStop}%
\bibitem [{\citenamefont {Afzalian}\ \emph {et~al.}(2011)\citenamefont {Afzalian}, \citenamefont {Colinge},\ and\ \citenamefont {Flandre}}]{Afzalian2011}%
  \BibitemOpen
  \bibfield  {author} {\bibinfo {author} {\bibfnamefont {A.}~\bibnamefont {Afzalian}}, \bibinfo {author} {\bibfnamefont {J.-P.}\ \bibnamefont {Colinge}},\ and\ \bibinfo {author} {\bibfnamefont {D.}~\bibnamefont {Flandre}},\ }\bibfield  {title} {\bibinfo {title} {Physics of gate modulated resonant tunneling (rt)-fets: Multi-barrier mosfet for steep slope and high on-current},\ }\href {https://doi.org/https://doi.org/10.1016/j.sse.2011.01.016} {\bibfield  {journal} {\bibinfo  {journal} {Solid-State Electronics}\ }\textbf {\bibinfo {volume} {59}},\ \bibinfo {pages} {50} (\bibinfo {year} {2011})}\BibitemShut {NoStop}%
\bibitem [{\citenamefont {Afzalian}\ \emph {et~al.}(2019)\citenamefont {Afzalian}, \citenamefont {Doornbos}, \citenamefont {Shen}, \citenamefont {Passlack},\ and\ \citenamefont {Wu}}]{Afzalian2019}%
  \BibitemOpen
  \bibfield  {author} {\bibinfo {author} {\bibfnamefont {A.}~\bibnamefont {Afzalian}}, \bibinfo {author} {\bibfnamefont {G.}~\bibnamefont {Doornbos}}, \bibinfo {author} {\bibfnamefont {T.-M.}\ \bibnamefont {Shen}}, \bibinfo {author} {\bibfnamefont {M.}~\bibnamefont {Passlack}},\ and\ \bibinfo {author} {\bibfnamefont {J.}~\bibnamefont {Wu}},\ }\bibfield  {title} {\bibinfo {title} {A high-performance inas/gasb core-shell nanowire line-tunneling tfet: An atomistic mode-space negf study},\ }\href {https://doi.org/10.1109/JEDS.2018.2881335} {\bibfield  {journal} {\bibinfo  {journal} {IEEE Journal of the Electron Devices Society}\ }\textbf {\bibinfo {volume} {7}},\ \bibinfo {pages} {88} (\bibinfo {year} {2019})}\BibitemShut {NoStop}%
\bibitem [{\citenamefont {Datta}(2005{\natexlab{b}})}]{Dattach8}%
  \BibitemOpen
  \bibfield  {author} {\bibinfo {author} {\bibfnamefont {S.}~\bibnamefont {Datta}},\ }\href@noop {} {\emph {\bibinfo {title} {Quantum transport: atom to transistor}}}\ (\bibinfo  {publisher} {Cambridge university press},\ \bibinfo {year} {2005})\ Chap.~\bibinfo {chapter} {8}\BibitemShut {NoStop}%
\bibitem [{\citenamefont {Wagner}(1991)}]{Wagner1991}%
  \BibitemOpen
  \bibfield  {author} {\bibinfo {author} {\bibfnamefont {M.}~\bibnamefont {Wagner}},\ }\bibfield  {title} {\bibinfo {title} {Expansions of nonequilibrium green’s functions},\ }\href@noop {} {\bibfield  {journal} {\bibinfo  {journal} {Physical Review B}\ }\textbf {\bibinfo {volume} {44}},\ \bibinfo {pages} {6104} (\bibinfo {year} {1991})}\BibitemShut {NoStop}%
\end{thebibliography}%

\end{document}


\preprint{PRB/123-QED}

\title{Supplemental Material for Fully coupled electron-phonon transport in two-dimensional-material-based devices using efficient FFT-based self-energy calculations}

\author{Rutger Duflou}
\email{rutger.duflou@imec.be}
\author{Gautam Gaddemane}
\author{Michel Houssa}
\author{Aryan Afzalian}
\affiliation{
imec, Kapeldreef 75, Leuven, 3001,Belgium}
\affiliation{Semiconductor Physics Laboratory, KU Leuven, Celestijnenlaan 200 D, Leuven, 3001, Belgium}

\date{\today}

\maketitle


\section{Influence of the leads on the phonon Green's function}

To compute the steady-state occupation of the Green's functions in the NEGF formalism, we generally subdivide the system into a central device and the leads. The Hamiltonian governing the dynamics of the system,

\begin{align}
    \hat{H} =
    &\sum_{nn'\kb_{\perp}}
\bar{h}_{\substack{nn'\\\kb_{\perp}}}
 \cbc_{\kb_{\perp}n}\cb_{\kb_{\perp}n'} + \sum_{\nu\nu'\qb_{\perp}}
 \bar{d}_{\substack{\nu\nu'\\\qb_{\perp}}}
     \abc_{\qb_{\perp}\nu}\ab_{\qb_{\perp}\nu'}.
\label{eq:hamiltonian_mix}\\ 
 &+N_{\perp}^{-\frac{1}{2}}\sum_{\substack{n n'\nu\\\kb_{\perp}\qb_{\perp}}}
    \bar{g}_{\substack{nn'\nu\\\kb_{\perp}\qb_{\perp}}}
    \cbc_{\kb_{\perp}+\qb_{\perp}
    n}\cb_{\kb_{\perp}n'}(\ab_{\qb_{\perp}\nu}+\abc_{-\qb_{\perp}\nu}),\nonumber
\end{align}

is then generally subdivided into parts governing the device and lead sub-blocks, $\hat{H}_0$, parts connecting those sub-blocks, $\hat{H}_P$, and parts governing the electron-phonon interactions, $\hat{H}_I$, which are handled by Appendix C in the main text and will be neglected here. For the electrons we thus have

\begin{align}
\hat{H}^{el}_0 =& \sum_{nn'\kb_{\perp}}^{regions}
\bar{h}_{\substack{nn'\\\kb_{\perp}}}
 \cbc_{\kb_{\perp}n}\cb_{\kb_{\perp}n'}, \\ 
\hat{H}^{el}_P =& \sum_{nn'\kb_{\perp}}^{interfaces}
\bar{h}_{\substack{nn'\\\kb_{\perp}}}
 \cbc_{\kb_{\perp}n}\cb_{\kb_{\perp}n'}.
\end{align}

The matrix elements of $\hat{H}^{el}_P$ can be grouped into a matrix $\mathbf{U}_{\kb_{\perp}}$, which arises in the Dyson equation,

\begin{align}
&\Gb_{\kb_{\perp}}(t,t') = \Gb^0_{\kb_{\perp}}(t,t')
    + \int_C dt_1\Gb^0_{\kb_{\perp}}(t,t_1)\mathbf{U}_{\kb_{\perp}}
    \Gb_{\kb_{\perp}}(t_1,t').
\end{align}

We want a similar expression for the phonons,

\begin{align}
&\Db_{\qb_{\perp}}(t,t') = \Db^0_{\qb_{\perp}}(t,t')
    + \int_C dt_1\Db^0_{\qb_{\perp}}(t,t_1)\mathbf{V}_{\qb_{\perp}}
    \Db_{\qb_{\perp}}(t_1,t').\label{eq:Dyson_D}
\end{align}

However, contrary to the electrons and contrary to the simplistic denotation in Fig. 2 of the main text, merely assigning the matrix elements $\bar{d}_{\substack{\nu\mu\\\qb_{\perp}}}$ to either $\hat{H}^{ph}_0$ or $\hat{H}^{ph}_P$ depending on whether they are in a sub-block or connect sub-blocks is insufficient to obtain \eqref{eq:Dyson_D}. As we will show here, we require our Hamiltonians to have the form

\begin{align}
	\hat{H}^{ph}_0 = 
	\sum_{\substack{\nu\mu\\\qb_{\perp}}}&
	(\bar{d}_{\substack{\nu\mu\\\qb_{\perp}}}-v_{\substack{\nu\mu\\\qb_{\perp}}})
	\abc_{\qb_{\perp}\nu}\ab_{\qb_{\perp}\mu}\\
	 &- \frac{v_{\qb_{\perp}\nu\mu}}{2}\big(
	\abc_{\qb_{\perp} \nu}\abc_{-\qb_{\perp} \mu}
	+\ab_{-\qb_{\perp} \nu}\ab_{\qb_{\perp} \mu}\big),\nonumber
\end{align}

\begin{align}
\begin{split}
	\hat{H}^{ph}_P = \sum_{\substack{\nu\mu\\\qb_{\perp}}} &v_{\substack{\nu\mu\\\qb_{\perp}}} \big(\abc_{\qb_{\perp} \nu}\ab_{\qb_{\perp} \mu}\\
	&\quad+\frac{1}{2}\big(
	\abc_{\qb_{\perp} \nu}\abc_{-\qb_{\perp} \mu}
	+\ab_{-\qb_{\perp} \nu}\ab_{\qb_{\perp} \mu}
	\big)\big).\label{eq:perturb_term}
\end{split}
\end{align}

The procedure is very similar to Appendix C in the main text. The perturbation expansion of the phonon Green's function,

\begin{align}
    iD_{\substack{\nu,\mu\\\qb_{\perp}}}(t,t') =& \frac{1}{\hb}\big<T_c\big[e^{\frac{-i}{\hb}\int_C \hat{H}_{P}(t_1) dt_1}(\ab_{\qb_{\perp}\nu}(t)+\abc_{-\qb_{\perp}\nu}(t))\nonumber\\
    &\qquad\qquad(\abc_{\qb_{\perp}\mu}(t')+\ab_{-\qb_{\perp}\mu}(t'))\big]\big>,
    \label{eq:Greens_func_int}
\end{align}

is limited to a first-order approximation,
\begin{widetext}

\begin{align}
\begin{split}
    iD_{\substack{\nu,\mu\\\qb_{\perp}}}&(t,t') = \frac{1}{\hb}\braket{T_c\left[(\ab_{\qb_{\perp}\nu}(t)+\abc_{-\qb_{\perp}\nu}(t))(\abc_{\qb_{\perp}\mu}(t')+\ab_{-\qb_{\perp}\mu}(t'))\right]} \\
&+ 
    \braket{T_c\left[\frac{-i}{\hb^2}\int_C \hat{H}_{P}(t_1) dt_1(\ab_{\qb_{\perp}\nu}(t)+\abc_{-\qb_{\perp}\nu}(t))(\abc_{\qb_{\perp}\mu}(t')+\ab_{-\qb_{\perp}\mu}(t'))\right]}.
\end{split}
\label{eq:ph_D_perturb}
\end{align}

The first term in \eqref{eq:ph_D_perturb} gives rise to the first non-interacting phonons Green's function term in \eqref{eq:Dyson_D}. The second term in \eqref{eq:ph_D_perturb} can be decomposed using Wick's theorem.

\begin{align}
    \frac{-i}{\hb^2}&
    \sum_{\substack{\nu_1 \mu_1\\\qb_{\perp,1}}}
    v_{\substack{\nu_1\mu_1\\\qb_{\perp,1}}}\int_C dt_1\\
\big(&\braket{
    \ab_{\qb_{\perp,1}\mu_1}(t_1)
    \abc_{\qb_{\perp,1}\nu_1}(t_1)}
    \braket{
    (\ab_{\qb_{\perp}\nu}(t)+\abc_{-\qb_{\perp}\nu}(t))
    (\abc_{\qb_{\perp}\mu}(t')+\ab_{-\qb_{\perp}\mu}(t'))}\nonumber\\
+&\braket{
    (\ab_{\qb_{\perp}\nu}(t)+\abc_{-\qb_{\perp}\nu}(t))
    \ab_{\qb_{\perp,1}\mu_1}(t_1)}
    \braket{
    \abc_{\qb_{\perp,1}\nu_1}(t_1)
    (\abc_{\qb_{\perp}\mu}(t')+\ab_{-\qb_{\perp}\mu}(t'))}\nonumber\\
+&\braket{
    (\ab_{\qb_{\perp}\nu}(t)+\abc_{-\qb_{\perp}\nu}(t))
    \abc_{\qb_{\perp,1}\nu_1}(t_1)}
    \braket{
    \ab_{\qb_{\perp,1}\mu_1}(t_1)
    (\abc_{\qb_{\perp}\mu}(t')+\ab_{-\qb_{\perp}\mu}(t'))}\nonumber\\
+&\frac{1}{2}\braket{
    (\ab_{\qb_{\perp}\nu}(t)+\abc_{-\qb_{\perp}\nu}(t))
    \abc_{-\qb_{\perp,1}\mu_1}(t_1)}
    \braket{
    \abc_{\qb_{\perp,1}\nu_1}(t_1)
    (\abc_{\qb_{\perp}\mu}(t')+\ab_{-\qb_{\perp}\mu}(t'))}\nonumber\\
+&\frac{1}{2}\braket{
    (\ab_{\qb_{\perp}\nu}(t)+\abc_{-\qb_{\perp}\nu}(t))
    \abc_{\qb_{\perp,1}\nu_1}(t_1)}
    \braket{
    \abc_{-\qb_{\perp,1}\mu_1}(t_1)
    (\abc_{\qb_{\perp}\mu}(t')+\ab_{-\qb_{\perp}\mu}(t'))}\nonumber\\
+&\frac{1}{2}\braket{
    (\ab_{\qb_{\perp}\nu}(t)+\abc_{-\qb_{\perp}\nu}(t))
    \ab_{\qb_{\perp,1}\mu_1}(t_1)}
    \braket{
    \ab_{-\qb_{\perp,1}\nu_1}(t_1)
    (\abc_{\qb_{\perp}\mu}(t')+\ab_{-\qb_{\perp}\mu}(t'))}\nonumber\\
+&\frac{1}{2}\braket{
    (\ab_{\qb_{\perp}\nu}(t)+\abc_{-\qb_{\perp}\nu}(t))
    \ab_{-\qb_{\perp,1}\nu_1}(t_1)}
    \braket{
    \ab_{\qb_{\perp,1}\mu_1}(t_1)
    (\abc_{\qb_{\perp}\mu}(t')+\ab_{-\qb_{\perp}\mu}(t'))}\nonumber\big)
\end{align}

The first term corresponds to a disconnected diagram and can be neglected. Applying momentum conservation for the remaining terms, we obtain

\begin{align}
    \frac{-i}{\hb^2}&
    \sum_{\nu_1 \mu_1}
    \int_C dt_1\\
\big(&v_{\substack{\nu_1\mu_1\\-\qb_{\perp}}}\braket{
    (\ab_{\qb_{\perp}\nu}(t)+\abc_{-\qb_{\perp}\nu}(t))
    \ab_{-\qb_{\perp}\mu_1}(t_1)}
    \braket{
    \abc_{-\qb_{\perp}\nu_1}(t_1)
    (\abc_{\qb_{\perp}\mu}(t')+\ab_{-\qb_{\perp}\mu}(t'))}\nonumber\\
+&v_{\substack{\nu_1\mu_1\\\qb_{\perp}}}\braket{
    (\ab_{\qb_{\perp}\nu}(t)+\abc_{-\qb_{\perp}\nu}(t))
    \abc_{\qb_{\perp}\nu_1}(t_1)}
    \braket{
    \ab_{\qb_{\perp}\mu_1}(t_1)
    (\abc_{\qb_{\perp}\mu}(t')+\ab_{-\qb_{\perp}\mu}(t'))}\nonumber\\
+&\frac{v_{\substack{\nu_1\mu_1\\-\qb_{\perp}}}}{2}\braket{
    (\ab_{\qb_{\perp}\nu}(t)+\abc_{-\qb_{\perp}\nu}(t))
    \abc_{\qb_{\perp}\mu_1}(t_1)}
    \braket{
    \abc_{-\qb_{\perp}\nu_1}(t_1)
    (\abc_{\qb_{\perp}\mu}(t')+\ab_{-\qb_{\perp}\mu}(t'))}\nonumber\\
+&\frac{v_{\substack{\nu_1\mu_1\\\qb_{\perp}}}}{2}\braket{
    (\ab_{\qb_{\perp}\nu}(t)+\abc_{-\qb_{\perp}\nu}(t))
    \abc_{\qb_{\perp}\nu_1}(t_1)}
    \braket{
    \abc_{-\qb_{\perp}\mu_1}(t_1)
    (\abc_{\qb_{\perp}\mu}(t')+\ab_{-\qb_{\perp}\mu}(t'))}\nonumber\\
+&\frac{
    v_{\substack{\nu_1\mu_1\\-\qb_{\perp}}}}{2}\braket{(\ab_{\qb_{\perp}\nu}(t)+\abc_{-\qb_{\perp}\nu}(t))
    \ab_{-\qb_{\perp}\mu_1}(t_1)}
    \braket{
    \ab_{\qb_{\perp}\nu_1}(t_1)
    (\abc_{\qb_{\perp}\mu}(t')+\ab_{-\qb_{\perp}\mu}(t'))}\nonumber\\
+&\frac{v_{\substack{\nu_1\mu_1\\\qb_{\perp}}}}{2}\braket{
    (\ab_{\qb_{\perp}\nu}(t)+\abc_{-\qb_{\perp}\nu}(t))
    \ab_{-\qb_{\perp}\nu_1}(t_1)}
    \braket{
    \ab_{\qb_{\perp}\mu_1}(t_1)
    (\abc_{\qb_{\perp}\mu}(t')+\ab_{-\qb_{\perp}\mu}(t'))}\nonumber\big).
\end{align}

We now impose the symmetry requirement $v_{\substack{\nu_1\mu_1\\\qb_{\perp}}} = v_{\substack{\mu_1\nu_1\\-\qb_{\perp}}}$, which implies that the third and fourth, and fifth and sixth term are equal when summing over all $\nu_1$ and $\mu_1$, eliminating the factor of two in the denominator. We thus obtain

\begin{align}
    \frac{-i}{\hb^2}&
    \sum_{\nu_1 \mu_1}v_{\substack{\nu_1\mu_1\\\qb_{\perp}}}
    \int_C dt_1\\
\big(&\braket{
    (\ab_{\qb_{\perp}\nu}(t)+\abc_{-\qb_{\perp}\nu}(t))
    \ab_{-\qb_{\perp}\nu_1}(t_1)}
    \braket{
    \abc_{-\qb_{\perp}\mu_1}(t_1)
    (\abc_{\qb_{\perp}\mu}(t')+\ab_{-\qb_{\perp}\mu}(t'))}\nonumber\\
+&\braket{
    (\ab_{\qb_{\perp}\nu}(t)+\abc_{-\qb_{\perp}\nu}(t))
    \abc_{\qb_{\perp}\nu_1}(t_1)}
    \braket{
    \ab_{\qb_{\perp}\mu_1}(t_1)
    (\abc_{\qb_{\perp}\mu}(t')+\ab_{-\qb_{\perp}\mu}(t'))}\nonumber\\
+&\braket{
    (\ab_{\qb_{\perp}\nu}(t)+\abc_{-\qb_{\perp}\nu}(t))
    \abc_{\qb_{\perp}\nu_1}(t_1)}
    \braket{
    \abc_{-\qb_{\perp}\mu_1}(t_1)
    (\abc_{\qb_{\perp}\mu}(t')+\ab_{-\qb_{\perp}\mu}(t'))}\nonumber\\
+&\braket{
    (\ab_{\qb_{\perp}\nu}(t)+\abc_{-\qb_{\perp}\nu}(t))
    \ab_{-\qb_{\perp}\nu_1}(t_1)}
    \braket{
    \ab_{\qb_{\perp}\mu_1}(t_1)
    (\abc_{\qb_{\perp}\mu}(t')+\ab_{-\qb_{\perp}\mu}(t'))}\nonumber\big).
\end{align}

Summing everything gives us

\begin{align}
    \frac{-i}{\hb^2}&
    \sum_{\nu_1 \mu_1}v_{\substack{\nu_1\mu_1\\\qb_{\perp}}}
    \int_C dt_1\\
\big(&\braket{
    (\ab_{\qb_{\perp}\nu}(t)+\abc_{-\qb_{\perp}\nu}(t))
    (\abc_{\qb_{\perp}\nu_1}(t_1)+\ab_{-\qb_{\perp}\nu_1}(t_1))}
    \braket{
    (\ab_{\qb_{\perp}\mu_1}(t_1)+\abc_{-\qb_{\perp}\mu_1}(t_1))
    (\abc_{\qb_{\perp}\mu}(t')+\ab_{-\qb_{\perp}\mu}(t'))}\big),\nonumber
\end{align}

\end{widetext}

which is equal to

\begin{equation}
 -i\sum_{\nu_1 \mu_1}v_{\substack{\nu_1\mu_1\\\qb_{\perp}}}
    \int_C dt_1 iD^0_{\substack{\nu,\nu_1\\\qb_{\perp}}}(t,t_1)
    iD^0_{\substack{\mu_1,\mu\\\qb_{\perp}}}(t_1,t').
\end{equation}

Grouping the elements into matrices, inserting in \eqref{eq:ph_D_perturb} and converting to a Dyson equation for higher-order terms then gives us \eqref{eq:Dyson_D}.